\documentclass{aastex631}
\usepackage{graphicx,graphics,aas_macros,amsmath}
\pdfoutput=1
\usepackage{booktabs}
\usepackage{graphicx}
\usepackage{longtable}
\usepackage{rotating}
\usepackage{wrapfig}
\usepackage{xspace}
\usepackage{geometry}
\usepackage[normalem]{ulem}
\usepackage{array} \newcolumntype{z}{>{\centering\arraybackslash}p{1.25cm}}

\usepackage{threeparttable}

\newcommand{\mone}{^{-1}}
\newcommand{\mtwo}  {^{-2}}

\newcommand{\chan}    {{Chandra}\xspace}

\newcommand{\cmmtwo}  {\,\mathrm{cm\mtwo}}

\newcommand{\pflux}   {\,\mathrm{photon\,cm\mtwo\,s\mone}}
\newcommand{\elum}    {\,\mathrm{erg\,s\mone}}
\newcounter{xion}     \newcommand{\eli}[2]  {\setcounter{xion}{#2}#1{~\sc\roman{xion}}}
\newcommand{\hetg}    {{HETG}\xspace}

\newcommand{\kms}     {\,\mathrm{km\,s\mone}}

\newcommand{\mang}    {\,\mathrm{{\mbox{\AA}}}\xspace}
\newcommand{\mk}      {\,\mathrm{MK}}

\newcommand{\msun}  {\,M_\odot}

\usepackage{verbatim}

\title{ob-stars}

\begin{document}

\title{Survey of X-rays from Massive Stars Observed at High Spectral Resolution with {\em Chandra}}

\author[0000-0002-1131-3059]{Pragati Pradhan}\thanks{pradhanp@erau.edu}
\affiliation{Massachusetts Institute of Technology , 77 Massachusetts
  Ave., Cambridge, MA 02139, USA}
  
 \affiliation{Embry Riddle Aeronautical University, Department of
 Physics \& Astronomy, 3700 Willow Creek Road
Prescott, AZ 86301, USA}

\author[0000-0002-3860-6230]{David P.\ Huenemoerder}\thanks{dph@mit.edu}
\affiliation{Massachusetts Institute of Technology , 77 Massachusetts
  Ave., Cambridge, MA 02139, USA}

\author[0000-0002-7204-5502]{Richard Ignace} \affiliation{Department
  of Physics \& Astronomy, East Tennessee State University, Johnson
  City, TN 37614 USA}

\author[0000-0003-3298-7455]{Joy S.\ Nichols}
\affiliation{Center for Astrophysics | Harvard \& Smithsonian , 60 Garden
  Street, Cambridge, MA 02138, USA }

\author[0000-0002-6737-538X]{A.M.T. Pollock}
\affiliation{Department of Physics and Astronomy, University of
  Sheffield, Hounsfield Road, Sheffield S3 7RH, UK}

\newcommand*{\pp}{\textcolor[rgb]{0,0.5,0}} 

\begin{abstract}

Identifying trends between observational data and the range of
physical parameters of massive stars is a critical step to the
still-elusive full understanding of the source, structure, and
evolution of X-ray emission from the stellar winds, requiring a
substantial sample size and systematic analysis methods.  The \emph{Chandra}
data archive as of 2022 contains 37 high resolution spectra of O, B,
and WR stars, observed with the \emph{Chandra}/HETGS and of sufficient
quality to fit the continua and emission line profiles.  Using a
systematic approach to the data analysis, we explore morphological
trends in the line profiles (i.e., O, Ne, Mg, Si) and find that the
centroid offsets of resolved lines versus wavelength can be separated
in three empirically-defined groups based on the amount of line
broadening and centroid offset. Using \ion{Fe}{17} (15.01 \AA, 17.05
\AA) and \ion{Ne}{10} $\alpha$ (12.13 \AA) lines which are prevalent
among the sample stars, we find a well-correlated linear trend of
increasing Full Width Half Maximum (FWHM) with faster wind terminal
velocity. The H-like/He-like total line flux ratio for strong lines
displays different trends with spectral class depending on ion
species.  Some of the sources in our sample have peculiar properties
(e.g., magnetic and $\gamma$ Cas-analogue stars) and we find that
these sources stand out as outliers from more regular trends. Finally,
our spectral analysis is presented summarily in terms of X-ray
spectral energy distributions in specific luminosity for each source,
plus tables of line identifications and fluxes.

\end{abstract}

\section{Introduction}

X-ray studies of massive star winds have been central to the ongoing
understanding of their structured and time-variable flows
\citep[e.g.,][]{2009A&ARv..17..309G, 2016AdSpR..58..739O}.  Studies
tend to fall into two categories: intensive observations of individual
or binary stars and coarse observations for larger samples of objects,
for example characterizing basic X-ray properties for stars in a
cluster.  A good example of the former would include $\zeta$~Pup, with
around a $\sim 1$~Msec of data from {\em XMM-Newton} \citep{naze2012}
and a similar amount from {\em Chandra}
\citep{huenemoerder:al:2020,cohen2020,Nichols2021,CohenOverdorff2022}.
The star draws this continued attention because of its relative
proximity, X-ray brightness, and high mass-loss rate
($2.5\times$10$^{-6}~{M_{\odot}}$/yr; \citealt{2020MNRAS.499.6044C})
and high-speed wind $v_{\infty}$ of $2250\,\mathrm{km/s}$
\citep{puls2006,cohen2010}.  There are numerous examples for X-ray
studies of massive star populations in cluxters
\citep[e.g.,][]{oskinova2005,rauw2016}.

Fewer investigations have sought to frame massive stars in terms of an
X-ray classification based on high-resolution spectral data.
Certainly, there have been efforts to group high-resolution data in
relation to stellar winds \citep{2006MNRAS.372..313O,
  2007ApJ...668..456W, 2017A&A...608A..54C}, magnetic stars
\citep{2011MNRAS.416.1456O, 2014ApJS..215...10N,2016AdSpR..58..680U},
or studies of individual colliding-wind binaries like
{$\eta$}~Carinae, WR~147, WR~140, WR~48a, WR~25
\citep{Corcoran2001,zhekov2010,pollock:al:2005,zhekov2014,pradhan2021}. \citealt{2009ApJ...703..633W}
explored trends for massive stars based on \emph{Chandra} grating spectra and
found a trend of X-ray hardness correlated with spectral subtype in O
and early B stars indicating a correlation between X-ray plasma
temperature and stellar effective temperature. Some have argued that
this variation in the X-ray hardness with spectral subtype among O
stars is possibly an absorption effect \citep{leut2010,cohen2021}.


There remain outstanding questions concerning X-ray production from
early-type massive-star winds.  A key mechanism long invoked for
explaining the presence of multi-million degree gas in the winds of OB
stars, as well as WR stars, is rooted in the line-deshadowing
instability (LDI) endemic to the wind-driving physics
\citep[e.g.,][]{1997A&A...322..878F}.  The radiative line-driving
force through spectral lines \citep{1980ApJ...241..300L} relates to
the velocity gradient of the flow, which can have positive
re-enforcement in the presence of perturbations that result in
producing shock structures distributed throughout the wind
\citep{1998Ap&SS.260..113F}.  Indeed, not only does LDI predict the
presence of X-ray producing hot plasma, it can naturally account for
the presence of structured outflow seen in other diagnostics,
structure normally grouped under the heading of ``clumping''
\citep{1991A&A...247..455H, 1995A&A...295..725B, 1998A&A...335.1003H,
  2006AA...454..625P}.

While LDI is qualitatively capable of accounting for X-ray production
and structured flow in OB~star winds, it has faced challenges to
account quantitatively for the observations.  There have been concerns
about whether LDI can produce the requisite level of wind emission
measure implied by observed X-ray luminosities
\citep{1997A&A...322..878F, 2007ApJ...668..456W}.  Also, there is
uncertainty whether the LDI mechanism is capable of accounting for the
well-known trend between X-ray and bolometric luminosities, with $L_X
/ L_{\rm bol} \sim 10^{-7}$ observed among O and early-B stars
\citep{1980ApJ...239L..65L, 1997A&A...322..167B, 2009A&ARv..17..309G,
  2009A&A...506.1055N}.  \cite{2013MNRAS.429.3379O} have examined this
scaling relation and generally found that LDI combined with
photoabsorption of X-rays by the wind itself can account for the
observed behavior with bolometric luminosities.  There have also been
concerns for whether LDI could produce X-rays at low wind velocities
as implied, but this now seems to have been addressed
\citep{2013MNRAS.428.1837S}.

To understand better the underlying physics and properties for X-ray
production in massive star winds requires a sufficient sample of stars
to relate observations -- spectral energy distributions and line
shapes -- to the heating of the gas \citep{2014ApJ...788...90G,
  2014MNRAS.444.3729C}.  An analysis of a representative sample is
complicated by the fact that early-type stars are relatively faint
X-ray sources which limits the number of sources with high counts,
plus the heterogeneous nature of early-type stars.  Concerning the
latter, there are two issues worth highlighting.  The first is that
binarity is common \citep{2012Sci...337..444S,2013ApJ...764..166D},
and magnetism is not too unusual \citep{2016MNRAS.456....2W}.  Both
can alter the X-ray production in relation to predictions for the LDI
mechanism operating in single-star winds \citep{1992ApJ...389..635U,
  1992ApJ...386..265S,2008MNRAS.388.1047P,
  2008AA...490..793H,2015A&A...582A..45F, 2009ApJ...692L..76W,
  2016MNRAS.456....2W}.  The second issue with the heterogeneous
sample is the nature of wind photoabsorption that can limit the escape
of X-rays from the wind.  The issue arises in two forms.  One is
simply that some early-type stars have high wind opacity that results
in wavelength-dependent photoabsorption \citep[e.g.,
][]{2006ApJ...648..565O}; two is that structured winds (i.e.,
porosity) can alter the effectiveness of the wind photoabsorption
\citep{2010MNRAS.405.2391C, 2011BSRSL..80...54O, 2014MNRAS.439..908C}.

Understanding the LDI mechanism and disentangling its influence from
other contributing mechanisms (i.e., wind collision or magnetism)
continues to be a topic of active study.  \cite{2021MNRAS.503..715C}
have presented a summary of high-resolution grating {\em Chandra}
spectra for six O~stars.  In their focus on O~stars, they provide an
analysis of spectral trends and line profile fitting with spherically
symmetric wind models for nominally single non-magnetic stars.  Among
their conclusions, chief is that wind absorption of X-rays is an
important consideration, and that around 90\% of X-rays produced in
distributed wind shocks (i.e., via the LDI mechanism) are absorbed in
the wind itself, with only 10\% of the emission escaping.  They
attribute the empirical relation of $L_X \approx 10^{-7}\,L_{\rm bol}$
as primarily related to wind absorption effects.

Here we continue and extend the efforts of \cite{2007ApJ...668..456W},
\cite{2009ApJ...703..633W}, and \cite{2021MNRAS.503..715C} by
considering a sample of early-type massive stars with high-resolution
{\em Chandra} spectra.  For our approach we adopt three main criteria
for inclusion in the sample: (1) sources must be massive stars, (2)
sources must have HETG spectra, and (3) the spectra, possibly from
multiple pointings, must have a minimum number of counts to facilitate
line profile analysis.  The application of these criteria leads to
about 40 sources consisting of a mix of types -- putative single stars
(or stars with distant companions), binary stars, magnetic stars, and
a range of spectral types B, O, and Wolf-Rayet (WR).  Section
\ref{sec:obs} describes the observations, Section \ref{sec:sample}
details the sample of stars, Section \ref{sec:fitting} details the
fitting procedures. The results, discussion and summary are outlined
in Sections \ref{sec:res}, \ref{sec:disc} and \ref{sec:sum},
respectively, followed by line measurement tables in
Appendix~\ref{sec:linemeas} and spectrum plots as luminosity density
in Appendix~\ref{sec:specplots}.

\section{Observations}
\label{sec:obs}

Data for this project were taken from archival \emph{Chandra} High Energy
Transmission Grating Spectrometer (HETGS) observations.  The HETGS
covers the band from \eli{Fe}{25} ($1.85\mang$) to \eli{N}{7}
($24.78\mang$) with resolving powers up to about 1000 and effective
area up to about $150\,\mathrm{cm^2}$.  For details on the instrument,
see \citet{canizares2005} or the \emph{Chandra} Proposers' Observatory
Guide\footnote{For a detailed description of the spectrometer
parameters and performance, see
\url{https://cxc.harvard.edu/proposer/POG/html/chap8.html}}.

The data are primarily comprised of the targets proposed by the
principal investigators of the observations.  In some cases
(HD~150135, Cyg OB2-9, and $\zeta\,$Ori B), we identified
serendipitous sources in the field by searching for overlap between
OB-star catalogs and all HETGS pointings. We accepted only candidate
objects which were within $3\,\mathrm{arcmin}$ of the aim-point, since
the resolving power rapidly degrades off-axis beyond
$2\,\mathrm{arcmin}$.

For high-resolution spectroscopy, a relatively large number of counts
is necessary to achieve sufficient signal-to-noise (S/N) ratios for
analysis of lines in terms of line width and location of peak line
emission.  For merged spectra of unique sources, we adopted a minimum
number of counts in a dispersed spectrum of about 1800. Among the
sources that we examined, the highest number of counts was over
150,000 for $\gamma\,$Cas (the next brightest was 62,000).  Slightly
more than half the sources had more than 10,000 counts.  The count
rates per second ranged from $0.012$ to $3.0$. Observational information for the sample is given in Table~\ref{tbl:obsid}.
%
%
\begin{table}
\centering
\caption{The list of sources along with their observed fluxes
  (uncorrected for absorption), count rates, exposures,
  percentage uncertainty, number of observations, and a list
  of the observation-identifiers used in this work. The
  uncertainty is based on the square root of total counts and
  pertains as well to the integrated flux, $f_\mathrm{x}$.
}
\begin{tabular}{|cc|c|r|c|c|c|p{0.47\columnwidth}|}
\hline
Index&
Object& 
$10^{11}f_x$&
\multicolumn{1}{c|}{Rate}&
Exp.&
Err.&
\#&
\multicolumn{1}{c|}{ObsIDs}\\
&
&
\multicolumn{1}{c|}{\footnotesize{(erg cm$^{-2}$ s$^{-1}$)}}&
\multicolumn{1}{c|}{\footnotesize{(ct ks$^{-1}$)}}&
\multicolumn{1}{c|}{(ks)}&
(\%)&
&
\\\hline 
1&	 $9\,$Sgr&	  0.24&	 40.39&	 146&	 1.3&        2&	 5398 6285 \\
2&	 $\zeta\,$Oph&	  0.52&	 89.90&	 84&	 1.2&        2&	 2571 4367 \\
3&	 M17 Cen 1B&	  0.10&	 17.03&	 172&	 1.8&        2&	 9948 10680 \\
4&	 M17 Cen 1A&	  0.25&	 40.27&	 172&	 1.2&        2&	 9948 10680 \\
5&	 HD~191612&	  0.10&	 16.88&	 338&	 1.3&        10&	 16653 16654 16655 17489 17655 17694 18743 18753 18754 18821 \\
6&	 Cyg~OB2-12&	  0.20&	 46.80&	 138&	 1.2&        1&	 16659 \\
7&	 Cyg~OB2-9&	  0.16&	 28.73&	 65&	 2.3&        1&	 2572 \\
8&	 Cyg~OB2-8A&	  0.52&	 146.80& 65&	 1.0&        1&	 2572 \\
9&	 HD~206267&	  0.13&	 19.84&	 212&	 1.5&        5&	 1888 1889 16309 16532 16533 \\
10&	$\gamma\,$Cas&	  17.65& 3000.00&51&	 0.3&        1&	 1895 \\
11&	$\xi\,$Per&	  0.28&	 44.84&	 159&	 1.2&        1&	 4512 \\
12&	$15\,$Mon&	  0.22&	 19.88&	 100&	 2.2&        3&	 5401 6247 6248 \\
13&	$\delta\,$Ori&	  0.94&	 99.49&	 528&	 0.4&        5&	 639 14567 14568 14569 14570 \\
14&	$\epsilon\,$Ori&  0.76&	 89.93&	 92&	 1.1&        1&	 3753 \\
15&	V640 Mon&	  0.32&	 52.97&	 320&	 0.8&        6&	 17730 17731 17732 18751 18752 19955 \\
16&	$\zeta\,$Ori~A&	  1.22&	 143.70& 367&	 0.4&        7&	 610 1524 13460 13461 14373 14374 14375 \\
17&	$\zeta\,$Ori~B&	  0.11&	 8.05&	 205&	 2.5&        3&	 13460 14373 14374 \\
18&	$\sigma\,$Ori~Aa& 0.21&	 25.44&	 91&	 2.1&        1&	 3738 \\
19&	$\theta^1\,$Ori~C&3.09&	 546.00& 2035&	 0.1&        68&	 3 4 2567 2568 7407 7408 7409 7410 8568 8589 8895 8896 8897 22334 22335 22336 22337 22338 22339 22340 22341 22342 22343 22892 22893 22904 22993 22994 22995 22996 22997 22998 22999 23000 23001 23002 23003 23004 23005 23006 23007 23008 23009 23010 23011 23012 23087 23097 23104 23114 23115 23120 23206 23207 23208 23233 24622 24623 24624 24829 24830 24831 24832 24834 24842 24873 24874 24906 \\
20&	$\iota\,$Ori&	  0.85&	 117.10& 50&	 1.3&        2&	 599 2420 \\
21&	$\kappa\,$Ori&	  0.28&	 25.84&	 234&	 1.3&        4&	 9939 9940 10839 10846 \\
22&	WR~6&	          0.15&  33.55& 440&	 0.8&	     3&  14533 14534 14535 \\
23&	$\tau\,$CMa&	  0.14&	 12.27&	 286&	 1.7&        5&	 2525 2526 17441 17442 17593 \\
24&	HD~42054&	  0.27&	 50.20&	 150&	 1.2&        2&	 11021 12226 \\
25&	$\zeta\,$Pup&	  1.41&	 178.60& 881&	 0.3&        22&	 640 20154 20155 20156 20157 20158 21111 21112 21113 21114 21115 21116 21659 21661 21673 21898 22049 22076 22278 22279 22280 22281 \\
26&	$\gamma^2\,$Vel&  1.50&	 359.80& 65&	 0.7&        1&	 629 \\
27&	RCW 38 IRS 2&	  0.11&	 21.96&	 185&	 1.6&        5&	 12351 15706 16535 16536 17638 \\
28&	HD~93129A&	  0.13&	 33.75&	 138&	 1.5&        7&	 5397 7201 7202 7203 7204 7228 7229 \\
29&	HD~93250&	  0.21&	 46.32&	 194&	 1.1&        5&	 5399 5400 7189 7341 7342 \\
30&	WR~25&	          1.00&	 215.10& 85&	 0.7&        2&	 18616 19867 \\
31&	$\theta\,$Car&	  0.23&	 15.34&	 120&	 2.3&        2&	 15730 16537 \\
32&	HD 110432&	  3.80&	 451.70& 140&	 0.4&        1&	 9947 \\
33&	$\beta\,$Cru&	  0.28&	 28.53&	 74&	 2.2&        1&	 2575 \\
34&	HD~148937&	  0.38&	 86.46&	 99&	 1.1&        1&	 10982 \\
35&	HD~150135&	  0.10&	 19.07&	 358&	 1.2&        3&	 2569 14598 14599 \\
36&	HD~150136&	  0.62&	 137.70& 358&	 0.5&        3&	 2569 14598 14599 \\
37&	$\tau\,$Sco&	  1.60&	 343.80& 72&	 0.6&        2&	 638 2305 \\
\hline
\end{tabular}
    \label{tbl:obsid}
\end{table}

\section{Sample of stars} 
\label{sec:sample}


\begin{deluxetable}{cccCCCCCC}  
  \tabletypesize{\scriptsize}
  \tablecaption{Stellar Parameters}
  \tablehead{
    \colhead{Index}&
    \colhead{Star}&
    \colhead{Sp.\ Type}&
    \colhead{$d$}&
    \colhead{$T_\mathrm{eff}$}&
    \colhead{$v_{\infty}$}&
    \colhead{$\log \dot M$}&
    \colhead{$\log L_x$}&
    \colhead{$\log N_\mathrm{H}$}  \\
    &
    &
    &
            (\mathrm{pc})&
            (\mathrm{K)}&
            ($\kms$)&
            \mathrm{dex} ($M_\odot\,\mathrm{yr}^{-1}\,$)&
                 \mathrm{dex}  ($\mathrm{erg\,s^{-1}}$)&
               \mathrm{dex}    ($\cmmtwo$)
  }
  \startdata
 1 &          $9\,$Sgr &                 O4 V((f))\textsuperscript {2} &  1156\textsuperscript {1} &  43000\textsuperscript {1} &  2750\textsuperscript {1} &  -6.32\textsuperscript {1} & 32.96 \pm 0.07 &   21.29 $\pm$ 0.07\textsuperscript {41} \\
 2 &      $\zeta\,$Oph &                 O9.2 IVnn\textsuperscript {5} &  182\textsuperscript {29} & 32500\textsuperscript {19} & 1470\textsuperscript {33} & -7.03\textsuperscript {36} & 31.41 \pm 0.02 &  20.69 $\pm$ 0.10\textsuperscript {41}  \\
 3 &        M17 Cen 1B &                   O2-4 V\textsuperscript {30} & 1630\textsuperscript {30} &                    \nodata &              3331 \,(850) &                    \nodata & 32.97 \pm 0.09 & 22.30 $\pm$ 0.05\textsuperscript {39x}  \\
 4 &        M17 Cen 1A &                  O2-4 Vp\textsuperscript {30} & 1630\textsuperscript {30} &                    \nodata &              2932 \,(974) &                    \nodata & 33.32 \pm 0.08 & 22.30 $\pm$ 0.05\textsuperscript {39x}  \\
 5 &         HD~191612 &                 O6-8 f?p\textsuperscript {13} & 2010\textsuperscript {29} & 36000\textsuperscript {21} & 2400\textsuperscript {21} & -7.90\textsuperscript {21} &          33.13 &               21.65\textsuperscript {y} \\
 6 &        Cyg~OB2-12 &                 B3-4 Ia+\textsuperscript {16} & 1750\textsuperscript {16} &                    \nodata &  400\textsuperscript {16} & -5.52\textsuperscript {16} & 33.64 \pm 0.03 & 22.31 $\pm$ 0.01\textsuperscript {16x}  \\
 7 &         Cyg~OB2-9 &      O4 If + O5.5 III(f)\textsuperscript {13} & 1709\textsuperscript {29} &  40000\textsuperscript {1} &              1685 \,(414) &  -5.64\textsuperscript {1} &          33.05 &             22.09 \textsuperscript {1x} \\
 8 &        Cyg~OB2-8A & O6 Ib(fc) + O4.5 III(fc)\textsuperscript {13} & 1804\textsuperscript {29} &  37000\textsuperscript {1} &              2216 \,(267) &  -5.87\textsuperscript {1} &          33.72 &             21.93\textsuperscript {1x}  \\
 9 &         HD~206267 &     O6 V(n)((fc)) + B0 V\textsuperscript {13} & 1093\textsuperscript {28} & 40900\textsuperscript {19} & 2680\textsuperscript {33} &                    \nodata &          33.29 &             21.64 \textsuperscript {y}  \\
10 &     $\gamma\,$Cas &                  B0.5 IVe\textsuperscript {6} &   168\textsuperscript {7} &                    \nodata & 1800\textsuperscript {46} &                    \nodata & 32.78 \pm 0.00 &  20.17 $\pm$ 0.09\textsuperscript {41}  \\
11 &        $\xi\,$Per &          O7.5 III((f))(n)\textsuperscript {5} & 1134\textsuperscript {29} & 35000\textsuperscript {19} & 2330\textsuperscript {33} & -6.66\textsuperscript {37} & 32.85 \pm 0.04 &   21.05 $\pm$ 0.08\textsuperscript {41} \\
12 &         $15\,$Mon &               O7 V((f))z\textsuperscript {13} &  719\textsuperscript {30} &  37000\textsuperscript {1} & 2055\textsuperscript {33} &  -7.95\textsuperscript {1} & 32.18 \pm 0.01 &  20.32 $\pm$ 0.10\textsuperscript {41}  \\
13 &     $\delta\,$Ori &              O9.5 II Nwk\textsuperscript {13} &   212\textsuperscript {7} & 31000\textsuperscript {19} & 1995\textsuperscript {33} & -6.40\textsuperscript {43} & 31.73 \pm 0.01 &  20.18 $\pm$ 0.07\textsuperscript {41}  \\
14 &   $\epsilon\,$Ori &                     B0 Ib\textsuperscript {4} &   606\textsuperscript {7} & 27000\textsuperscript {10} & 1910\textsuperscript {33} & -6.35\textsuperscript {10} & 32.59 \pm 0.02 &  20.49 $\pm$ 0.11\textsuperscript {41}  \\
15 &          V640 Mon &        O8 Iabf + O8.5:fp\textsuperscript {13} & 1478\textsuperscript {29} & 35100\textsuperscript {23} & 2410\textsuperscript {33} & -6.22\textsuperscript {23} & 33.07 \pm 0.05 &  21.18 $\pm$ 0.11\textsuperscript {41}  \\
16 &    $\zeta\,$Ori~A &              O9.2 Ib Nwk\textsuperscript {13} &   226\textsuperscript {7} & 31000\textsuperscript {19} & 1860\textsuperscript {33} & -6.47\textsuperscript {37} & 31.93 \pm 0.01 &   20.40 $\pm$ 0.09\textsuperscript {41} \\
17 &    $\zeta\,$Ori~B &                O9.7 IIIn\textsuperscript {13} &   226\textsuperscript {7} &                    \nodata &                728 \,(94) &                    \nodata & 30.87 \pm 0.01 &   20.40 $\pm$ 0.09\textsuperscript {41} \\
18 &  $\sigma\,$Ori~Aa &       O9.5 V + B0.2 V(n)\textsuperscript {13} &  388\textsuperscript {12} & 33000\textsuperscript {22} & 1060\textsuperscript {22} & -7.00\textsuperscript {22} & 31.66 \pm 0.01 &  20.55 $\pm$ 0.07\textsuperscript {41}  \\
19 & $\theta^1\,$Ori~C &                     O7 Vp\textsuperscript {5} &  373\textsuperscript {29} &  37000\textsuperscript {1} &   580\textsuperscript {1} &  -7.94\textsuperscript {1} & 32.96 \pm 0.11 &   21.54 $\pm$ 0.11\textsuperscript {41} \\
20 &      $\iota\,$Ori &        O8.5 III + B0.2 V\textsuperscript {13} &  412\textsuperscript {28} & 32900\textsuperscript {19} & 2195\textsuperscript {33} & -9.49\textsuperscript {37} & 32.27 \pm 0.01 &   20.20 $\pm$ 0.10\textsuperscript {41} \\
21 &     $\kappa\,$Ori &                  B0.5 Ia\textsuperscript {17} &   198\textsuperscript {7} & 25700\textsuperscript {18} & 1525\textsuperscript {33} &                    \nodata & 31.22 \pm 0.02 &   20.61 $\pm$ 0.08\textsuperscript {41} \\
22 &              WR~6 &                     WN4b\textsuperscript {26} & 2270\textsuperscript {26} &                    \nodata & 1700\textsuperscript {25} & -4.20\textsuperscript {25} &          33.09 &              21.2\textsuperscript {47x} \\
23 &       $\tau\,$CMa &                     O9 II\textsuperscript {5} & 1500\textsuperscript {28} & 31000\textsuperscript {19} & 1960\textsuperscript {33} &                    \nodata & 32.72 \pm 0.03 &  20.80 $\pm$ 0.08\textsuperscript {41}  \\
24 &          HD~42054 &                     B5 Ve\textsuperscript {6} &  334\textsuperscript {29} & 17860\textsuperscript {15} &              1684 \,(557) &                    \nodata &          31.56 &           19.78 \textsuperscript {15x}  \\
25 &      $\zeta\,$Pup &                 O4 I(n)fp\textsuperscript {5} &   332\textsuperscript {7} & 39000\textsuperscript {32} & 2485\textsuperscript {33} & -5.07\textsuperscript {32} & 32.28 \pm 0.00 &  19.95 $\pm$ 0.08\textsuperscript {41}  \\
26 &   $\gamma^2\,$Vel &         WC8 + O7.5 III-V\textsuperscript {26} &  340\textsuperscript {26} & 35000\textsuperscript {24} & 1415\textsuperscript {33} &  -4.5\textsuperscript {24} & 32.32 \pm 0.00 &  19.74 $\pm$ 0.06\textsuperscript {41}  \\
27 &      RCW 38 IRS 2 &                  O5.5 V \textsuperscript {42} & 1700\textsuperscript {42} &                    \nodata &             3462 \,(1144) &                    \nodata & 32.93 \pm 0.09 & 22.15 $\pm$ 0.06\textsuperscript {40x}  \\
28 &         HD~93129A &                  O2 If*+\textsuperscript {31} & 2430\textsuperscript {30} & 52000\textsuperscript {34} & 3150\textsuperscript {33} &  -4.7\textsuperscript {34} &          33.22 &              21.44\textsuperscript {1x} \\
29 &          HD~93250 &                 O4 IV(fc)\textsuperscript {3} &  2555\textsuperscript {1} &  42000\textsuperscript {1} & 3160\textsuperscript {33} &  -5.98\textsuperscript {1} & 33.47 \pm 0.12 &  21.39 $\pm$ 0.15\textsuperscript {41}  \\
30 &             WR~25 &        O2.5 If*/WN6 + OB\textsuperscript {26} & 1985\textsuperscript {26} &                    \nodata & 2480\textsuperscript {25} & -4.60\textsuperscript {25} & 33.87 \pm 0.08 &  21.55 $\pm$ 0.13\textsuperscript {41}  \\
31 &     $\theta\,$Car &                B0.5 Vp +\textsuperscript {35} &   140\textsuperscript {7} & 31000\textsuperscript {38} &                482 \,(85) &                    \nodata & 30.77 \pm 0.01 &  20.28 $\pm$ 0.08\textsuperscript {41}  \\
32 &         HD 110432 &                B0.5 IVpe\textsuperscript {44} &  416\textsuperscript {29} & 20350\textsuperscript {14} &              1121 \,(357) &                    \nodata &          32.93 &              21.52\textsuperscript {y}  \\
33 &      $\beta\,$Cru &                     B1 IV\textsuperscript {4} &    85\textsuperscript {7} & 27000\textsuperscript {11} &  420\textsuperscript {11} & -9.00\textsuperscript {11} &          30.40 &            19.54\textsuperscript {11x}  \\
34 &         HD~148937 &                   O6 f?p+\textsuperscript {5} & 1102\textsuperscript {29} & 41000\textsuperscript {20} & 2215\textsuperscript {33} &                    \nodata & 33.09 \pm 0.11 &  21.60 $\pm$ 0.10\textsuperscript {41}  \\
35 &         HD~150135 &              O6.5 V((f))z\textsuperscript {5} & 1118\textsuperscript {29} & 37900\textsuperscript {19} & 2455\textsuperscript {27} &                    \nodata &          32.77 &              21.60\textsuperscript {y}  \\
36 &         HD~150136 &    O3.5/4 III(f*) + O6 IV\textsuperscript {5} & 1049\textsuperscript {29} &  46500\textsuperscript {9} &  3500\textsuperscript {9} &  -6.00\textsuperscript {9} &          33.59 &             21.61 \textsuperscript {y}  \\
37 &       $\tau\,$Sco &                     B0 IV\textsuperscript {4} &  215\textsuperscript {29} &  32000\textsuperscript {8} & 1000\textsuperscript {45} &  -9.3\textsuperscript {45} & 31.99 \pm 0.01 &   20.48 $\pm$ 0.09\textsuperscript {41} \\
  \enddata
  \tablecomments{Objects are ordered by increasing Galactic
    longitude.
   $L_\mathrm{x}$ was empirically determined from
    the flux-corrected HETG spectra, using the given values of $d$ and
    $N_\mathrm{H}$, with uncertainties being from the systematic error
    values on $N_\mathrm{H}.$ Uncertainties from counting statistics
    are given as percentages in Table~\ref{tbl:obsid} and scale directly
    to $L_\mathrm{x}$.  Values of $v_\infty$ with uncertainties in
    parentheses are newly determined values from this work, estimated
    from the widths of \eli{Ne}{10} and \eli{Fe}{17}
    lines. Superscripts indicate the references as follows (and for
    each reference, see references therein).  ``x" indicates the
    $N_\mathrm{H}$ value was derived from X-ray data and may include
    wind absorption as well as interstellar absorption.  ''y"
    indicates the $N_\mathrm{H}$ value was derived from
    $N_\mathrm{H}=0.83\times E(B-V) 10^{22}\,\mathrm{cm^{-2}}$
    \citep{liszt:2014a,liszt:2014b}.
  }
  \tablerefs{
    1: \citet{2018AA...620A..89N};
    2: \citet{2016yCat..51520031A};
    3: \citet{2016ApJS..224....4M};
    4: \citet{2009AA...501..297Z};
    5: \citet{2011ApJS..193...24S};
    6: \citet{1982ApJS...50...55S};
    7: \citet{2007AA...474..653V};
    8: \citet{2014AA...566A...7N};
    9: \citet{2012AA...540A..97M};
    10: \citet{2016MNRAS.456.2907P};
    11: \citet{2008MNRAS.386.1855C};
    12: \citet{2016AJ....152..213S};
    13: \citet{2019AA...626A..20M};
    14: \citet{2005AA...441..235Z};
    15: \citet{2018AA...619A.148N};
    16: \citet{2017ApJ...845...39O};
    17: \citet{1968ApJS...17..371L};
    18: \citet{2018AA...614A..91H};
    19: \citet{2005AA...436.1049M};
    20: \citet{2008AJ....135.1946N};
    21: \citet{2013MNRAS.431.2253M};
    22: \citet{2008ApJ...683..796S};
    23: \citet{2008AA...489..713L};
    24: \citet{2000AA...358..187D};
    25: \citet{2019AA...625A..57H};
    26: \citet{2020MNRAS.493.1512R};
    27: \citet{2005MNRAS.361..191S};
    28: \citet{2020AA...636A..28M};
    29: \citet{2018AJ....156...58B};
    30: \citet{2020AA...643A.138M};
    31: \citet{2017MNRAS.464.3561M};
    32: \citet{2006AA...454..625P};
    33: \citet{1990ApJ...361..607P};
    34: \citet{2019AA...621A..63G};
    35: \citet{1969ApJ...157..313H};
    36: \citet{2021MNRAS.503..715C};
    37: \citet{2014MNRAS.439..908C};
    38: \citet{2008AA...488..287H};
    39: \citet{2007ApJS..169..353B};
    40: \citet{2002ApJ...580L.161W};
    41: \citet{1994ApJS...93..211D};
    42: \citet{2010MNRAS.401..275W};
    43: \citet{2015ApJ...809..135S};
    44: \citet{2022MNRAS.510.2286N};
    45: \citet{2011MNRAS.416.1456O};
    46: \citet{1999ApJ...517..866S};
    47: \citet{2015ApJ...815...29H}.
  }
  \label{stellar-parameters}
\end{deluxetable}

The properties of our sample stars are summarized in
Table~\ref{stellar-parameters}.  The sample is comprised of a wide
range of X-ray luminosities, as derived from this work, from a few
$10^{30}$ up to $1.6\times10^{34}\elum$, which approximately covers
the known range of X-ray luminosities encountered for massive stars
\citep{gomez-moran:al:2018}. The luminosities and
  uncertainties listed in Table~\ref{stellar-parameters} were computed
  from the flux-calibrated high resolution spectra after correction
  for interstellar absorption.

There are 8 probable isolated stars and 27 stars in known binary
systems.  Eight stars in the sample are known or suspected magnetic
stars; three are $\gamma\,$ Cas and analogs; and three in the Cyg OB2
cluster display peculiar X-ray properties.  The normal and atypical
stellar groups are listed in Table~\ref{tbl:peculiar}.
\begin{deluxetable}{cccc}
 	 \tablecaption{Single and Multiple Normal and Peculiar OB stars}
 	 \tablehead{\colhead{Normal OB stars}&
 	            \colhead{magnetic stars and candidates}&
 	            \colhead{$\gamma$ Cas and analogs}&
 	            \colhead{Cyg OB2 stars}
 	}
 	  \startdata
 	  \hline
 	  \multicolumn{4}{c}{Apparently single stars} \\
 	  \hline
 	  $\zeta$ Oph & $\tau$ Sco && \\
 	  $\xi$ Per &&& \\
 	  $\epsilon$ Ori &&& \\
 	  $\zeta$ Ori B &&& \\
 	  $\kappa$ Ori &&& \\
 	  $\zeta$ Pup &&& \\
 	  $\beta$ Cru &&& \\
 	  \hline
 	  \multicolumn{4}{c}{Confirmed binary or multiple systems} \\
 	  \hline
 	  M17 Cen 1A & 9 Sgr & $\gamma$ Cas & Cyg OB2 8A \\
 	  M17 Cen 1B & $\theta^1$ Ori C & HD 110432 & Cyg OB2 9 \\
 	  HD 206267 & HD 191612 & HD 42054 & \\
 	  15 Mon & V640 Mon && \\
 	  $\delta$ Ori & $\zeta$ Ori A && \\
 	  $\sigma$ Ori Aa & $\theta$ Car && \\
 	  $\iota$ Ori & HD 148937 && \\
 	  $\tau$ CMa &&& \\
 	  $\gamma^2$ Vel &&& \\
 	  RCW 38 IRS 2 &&& \\
 	  HD 93129A &&& \\
 	  HD 93250 &&& \\
 	  WR 25 &&& \\
 	  HD 150135 &&& \\
 	  HD 150136 &&& \\
 	  \hline
 	  \multicolumn{4}{c}{Systems of uncertain status} \\
 	  \hline
 	  WR 6 &&& Cyg OB2 12 \\
 	  \enddata
 	  \end{deluxetable} \label{tbl:peculiar}

Stellar parameters for the stars studied were collected from the
literature available before 2022 April.  A reference for each value is
noted in the table.  Preference was given to more recent work.
Because we are using these values to evaluate trends across the
spectral types surveyed, we did not collect the primary references for
all values.  If GAIA parallaxes were available for a source and
appropriate priors were used by the respective authors in the
literature references to make distance estimates, these distance
estimates are preferred in Table~\ref{stellar-parameters}.

\section{Spectral Extraction and Fitting} 
\label{sec:fitting}

For most observations, we retrieved data from the \chan archive and
reprocessed with CIAO \citep{CIAO:2006}, using \texttt{TGCat} scripts
\citep{Huenemoerder2011}.  We modified the detection defaults to
search in a small box centered on the known source position.  For
these extractions, we used CIAO versions 4.11-4.13, depending on the
date of production.  For a few sources (WR~6, WR~25, and
$\zeta\,$Pup), we already possessed data products from our previously
published work
\citep{huenemoerder:al:2015,huenemoerder:al:2020,pradhan:al:2021}, and
adopted those files for this study.  In those cases CIAO versions 4.5,
4.8, and 4.10 had been used.  In all cases we used the matched
calibration database (CALDB) versions as updated after the
observations so that time-dependent quantities (like the detector
filter contamination) would be from interpolations, not
extrapolations.

Some sources with small angular separations required modification of
the cross-dispersion extraction region.  $\zeta\,$ Ori A and B are
quite close ($2.4\,\mathrm{arcsec}$ separation); in two observations
we used a narrow region, and in two observations we had to ignore
component B altogether because of overlap with the much brighter
component A.  For HD~150135, one observation had an unfavorable roll,
for which we reduced the extraction width to avoid the dispersed
spectrum of HD~150136.  The CIAO response computations account for
this reduced width.


Our goal is to provide emission-line parameters for the sample of
stars in a source-model-free manner.  We adopt a simple approach for
line profile fitting consisting of a sum of Gaussian emission lines
plus a continuum model.  While emission lines are theoretically not
necessarily described by a Gaussian profile, this shape is typically
adequate for characterizing the line profiles, given the wind
broadening typical of the lines combined with the level of resolving
power for the HETG.  For some long-wavelength lines where wind
absorption is stronger and resolving power is highest,
asymmetrically-shaped lines may be expected \citep[e.g.,
][]{2016AdSpR..58..694I}.  Even in these cases, a Gaussian is often
still sufficient to characterize the lines relative to others and to
other stars on a uniform basis.

Line fluxes are tabulated in Appendix~\ref{sec:linemeas}, both as the
model-independent as-observed values, and as corrected for absorption
using column densities, $N_\mathrm{H}$, given in
Table~\ref{stellar-parameters}.  Judgement must be applied when using
the absorption-corrected values, since the column density comes from a
variety of sources.  If scaled from $E(B-V)$, we used the relation
from \citet{liszt:2014a}, which is an average.  If from X-ray data, it
may already include some component from wind absorption, which can be
significant, and which may require stellar wind modeling or
independent determinations of the intrinsic and interstellar
components \citep[see for example][Figure 11]{huenemoerder:al:2015}. While we give uncertainties on $N_\mathrm{H}$ from the
  literature in Table~\ref{stellar-parameters}, this is a systematic
  and not a random error when considering line fluxes; line fluxes for
  any object will be correlated with variation in assumed absorption.
  
Fitting the lines requires a well determined continuum level.  In
parts of the spectrum, however, the line density and wind broadening
creates a pseudo-continuum which can be very different from the true
continuum.  In order to remove a large part of the systematic
uncertainty due to continuum placement, we fit absorbed plasma models
to coarsely binned spectra and then used the model evaluated without
the emission line contributions as a model continuum.  The binning at
a resolution well above the instrumental value removes dependence on
the details of line shapes and centroids, and allows line fluxes to
determine dominant temperatures and normalizations.  Such a model is
not adequate for high quality line flux determinations, but it is good
for informing fits of what emission lines are likely present and what
order of magnitude their relative strengths are. This provisional
model is used to guide the multi-component Gaussian fitting.  While we
fit a plasma temperature and an absorption column in this continuum
model estimation, we do not trust these as meaningful physical
parameters and so do not here report them as interesting results. The
temperature components and absorption are quite degenerate; physically
reliable values would require detailed modeling of each source's
spectrum, which is beyond the scope of this primarily empirical study.
    
\begin{figure}[htb]
  \centering\leavevmode
  \includegraphics*[width=0.75\columnwidth, viewport= 0 0 515 300]
                   {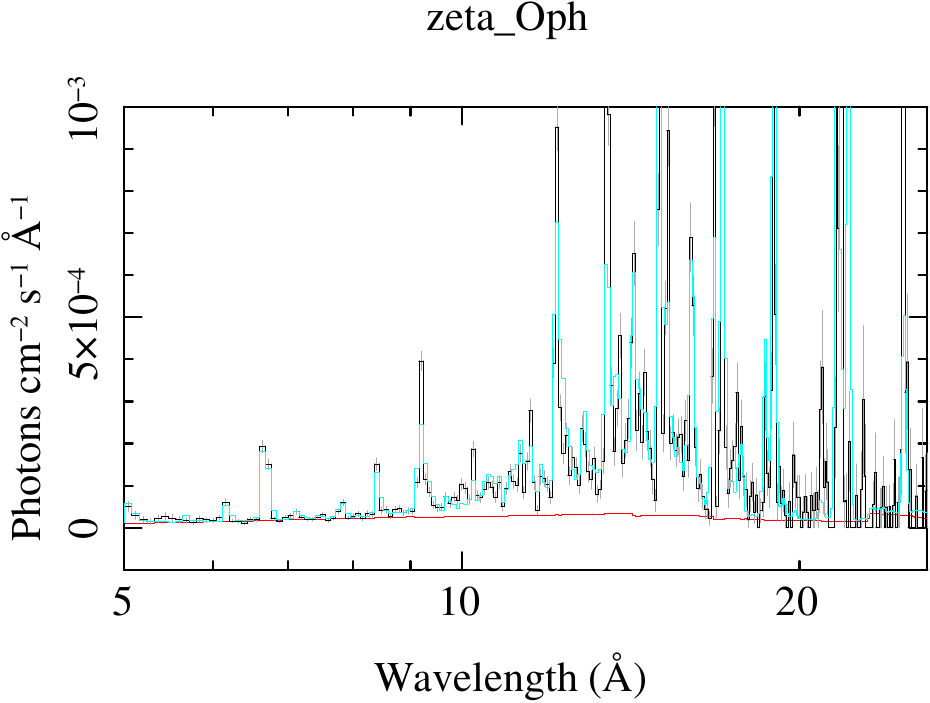} \caption{The black
                   histogram shows the flux spectrum for
                     $\zeta\,$Oph.  The light-colored curve overlaying it (cyan) is a plasma model
                     including line plus continuum emission.  Below that (red) is the
                     continuum contribution only, which is used as 
                     the baseline for Gaussian line fitting.}
                   \label{fig:contin}
\end{figure}

We made no use of zeroth order spectra since the CCD imaging
resolution is too low to aid in continuum level determination, and
also because the zeroth order images are generally saturated,
suffering from heavy photon pileup which is very difficult to
mitigate.  Figure \ref{fig:contin} shows an example of a counts
spectrum, the plasma model for lines plus continuum, and the
continuum-only evaluation.  Between about $10$ and $20\mang$ the true
continuum is well below the pseudo-continuum.

We adopt standard X-ray spectral fitting methods that employ iterative
forward-folding, in which one takes a model flux, convolves it by the
instrument response, computes a statistic, and then iterates
parameters to minimize the statistic.  This is required because the
count spectrum cannot be uniquely inverted, and because inversion can
amplify noise.  The forward folding also maintains rigorous handling
of Poisson counting statistics.  Our fitting handles the four first
orders of high energy grating (HEG) $\pm1$ and medium energy grating
(MEG) $\pm1$ simultaneously.  Since the wind-broadened lines are
resolved at the lower MEG resolving power, we matched HEG grids to MEG
and combine data dynamically during the fit --- each order is still
folded by the individual responses, but model counts are combined
before computing the statistic.  This is mainly a convenience because
it means models only need to be evaluated once on a common grid,
saving computation time (which can be significant when evaluating
confidence limits), and for visualization of a single, combined
result.  There is no quantitative statistical advantage because we are
using a maximum-likelihood statistic.  We used a Cash statistic and an
amoeba-subplex minimization method \citep{rowan:1990}.\footnote{The
ISIS implementation of the amoeba-subplex method is from
\url{https://www.netlib.org/opt/subplex.tgz.}}  All spectral fitting
was performed using the Interactive Spectral Interpretation System
\citep[ISIS software\footnote{For information on the Interactive
  Spectral Interpretation System (ISIS software package), see
  \url{https://space.mit.edu/cxc/ISIS/index.html}},][]{houck2000},
which also provides interfaces to AtomDB \citep{smith2001,foster2012}
as well as to {\tt xspec} \citep{Arnaud:1996} models.

Emission lines were fitted in groups of overlapping or close features,
using sums of Gaussian profiles folded through the instrument
response. If features were too blended or too weak for unconstrained
fits, they were restricted accordingly, either by constraining its
parameters to a stronger line's parameters or by freezing the position
or width and fitting the flux. For example, for the weak \ion{Fe}{25}
line (1.858 \AA) present in some stars, we froze the wavelength to the
theoretical value and the Gaussian $\sigma$ to the typical value for
massive stars, so that we could constrain only the flux. The $r$, $i$
lines (3.949 \AA, 3.968 \AA) of \ion{Ar}{17} cannot be resolved with
HETG and therefore both lines were fitted with a single Gaussian while
the $f$ line (3.994 \AA) was tied in wavelength offset and Gaussian
$\sigma$ to the ($r+i$) line's value. The \ion{S}{15} triplets (5.039
\AA, 5.065 \AA, 5.102 \AA) and \ion{Si}{13} triplets (6.648 \AA, 6.687
\AA, 6.740 \AA) were fitted with three Gaussians with the wavelength
offsets and Gaussian $\sigma$ of the $i$ and $f$ line tied to the $r$
line. Being blended with \ion{Ne}{10} (9.291 \AA), the \ion{Mg}{11}
region was fitted with six Gaussians (9.169 \AA, 9.230 \AA, 9.291\AA,
9.314 \AA, 9.362 \AA, 9.481 \AA) with the wavelength offset and
Gaussian $\sigma$ of five lines fixed to the strongest \ion{Mg}{11}
$r$-line. In some cases, we also froze the Gaussian $\sigma$ of the
\ion{Ne}{10} $\beta$ (10.239 \AA) and \ion{Ne}{10} $\gamma$ (9.708
\AA) lines since this region is heavily blended with Fe making
independent constraints difficult. The Gaussian $\sigma$ and
wavelength offsets of \ion{Fe}{17} (12.266 \AA) and \ion{Fe}{21}
(12.393 \AA) were tied to the main \ion{Ne}{10} $\mathrm{Ly}
\alpha$-like line (12.135 \AA).  The \ion{Ne}{9} region was fitted
with six lines (13.447 \AA, 13.497 \AA, 13.552 \AA, 13.699
\AA, 13.795 \AA, 13.825 \AA) with the wavelength offset and Gaussian
$\sigma$ of five lines tied to the strongest \ion{Ne}{9} $r$
line. The \ion{Fe} {17} region was fitted with four Gaussian lines
(15.014 \AA, 15.079 \AA, 15.176 \AA, 15.261 \AA) with the wavelength
offsets tied to the 15.014 \AA\ line. The \ion{O}{8} region was fitted
with four Gaussian lines (16.006 \AA, 16.071 \AA, 16.110 \AA, 16.159
\AA) at with the wavelength offsets tied to the 16.006 \AA\ line. The
\ion{Fe}{17} region was fitted with two lines (17.051 \AA, 17.096 \AA)
with the normalization of the latter line fixed at 0.86 times the
first line as guided by plasma models, and the wavelength offset of
17.096 \AA\ line tied to the 17.051 \AA\ line. The \ion{O}{7} region
were fitted with four Gaussian lines (21.170\AA, 21.601 \AA, 21.802
\AA, 22.098 \AA), with the wavelength offset tied to the strongest
\ion{O}{7} (r) line. All the isolated H-lines \ion{S}{16} (4.730 \AA),
\ion{Si}{13} $\beta$ (5.681 \AA), \ion{Si}{14} (6.183 \AA),
\ion{Mg}{12} (8.422 \AA), \ion{Ne}{10} $\gamma$ (9.708 \AA),
\ion{Ne}{10} $\beta$ (10.239 \AA), \ion{Ne}{10} (12.135 \AA),
\ion{Fe}{17} (16.780 \AA), \ion{O}{7} $\beta$ (18.627 \AA), \ion{O}{8}
(18.970 \AA), \ion{N}{7} (24.782) were freely fitted. Emission line
fit results are given in the tables in Appendix~\ref{sec:linemeas};
for tied parameters in constrained fits, uncertainties have null
values. Note that for certain stars (e.g., $\zeta$~Pup), there were
additional lines present, and we refer the reader to the tables in
Appendix~\ref{sec:linemeas} for complete information of the line
measurements.

\section{Results}\label{sec:res}

%

Fundamental diagnostics of stellar winds are found in the line
centroid, which can be blue-shifted from the rest wavelength by wind
absorption, and the line width, which is proportional to the wind
velocity.  These are the primary empirical products of the emission
line fitting.

\begin{figure}[htb]
    \centering\leavevmode
    \includegraphics*[width=0.65\columnwidth]{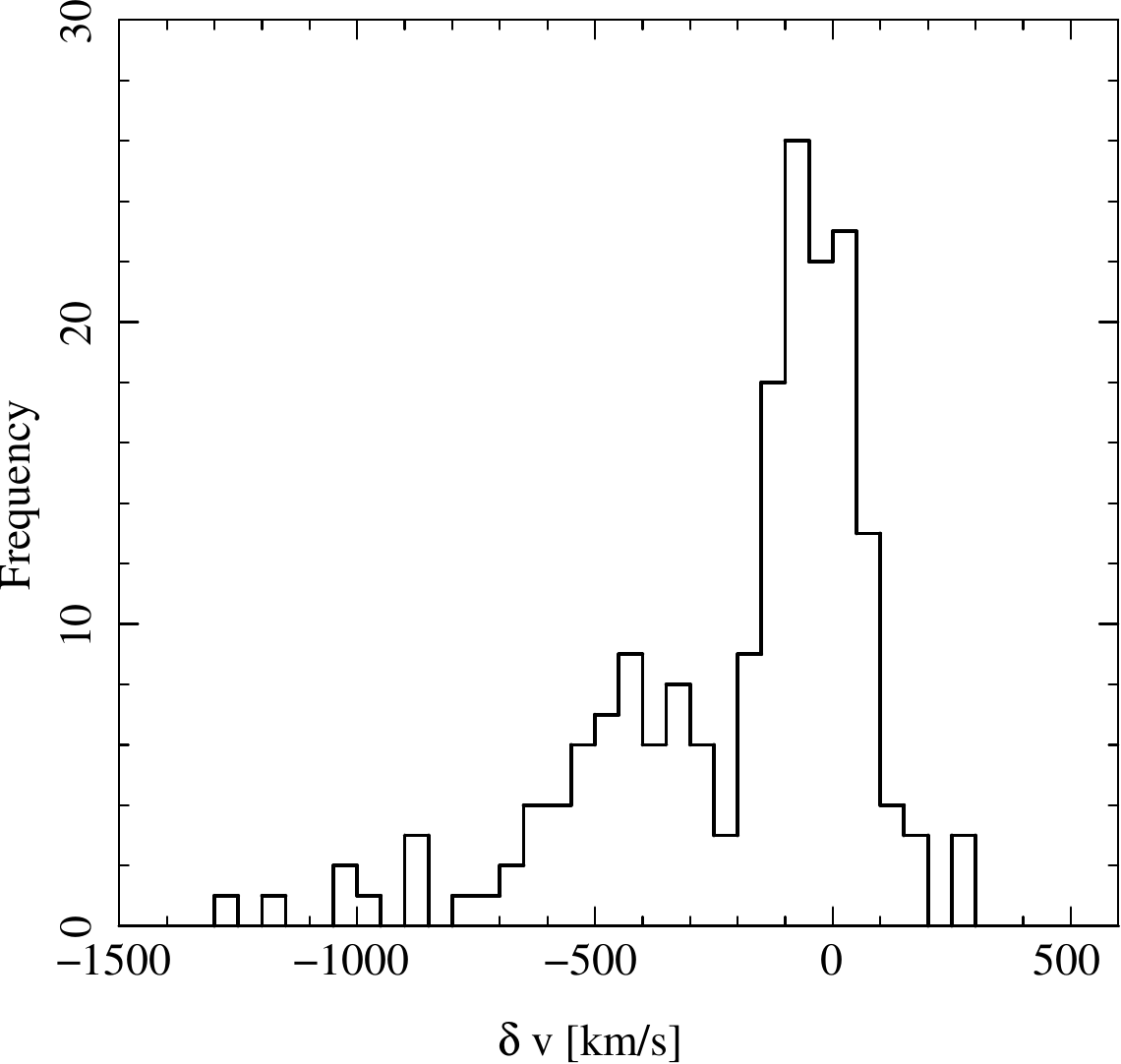} \caption{The
    frequency histogram of line centroid velocity offsets.  Two
    distinct groups are present, centered near $0\kms$ and $-400\kms$,
    with a long tail to large negative velocities.}
    \label{fig:dvhist}
\end{figure}

\begin{figure}[htb]
  \centering\leavevmode
  \includegraphics[width=0.65\columnwidth]{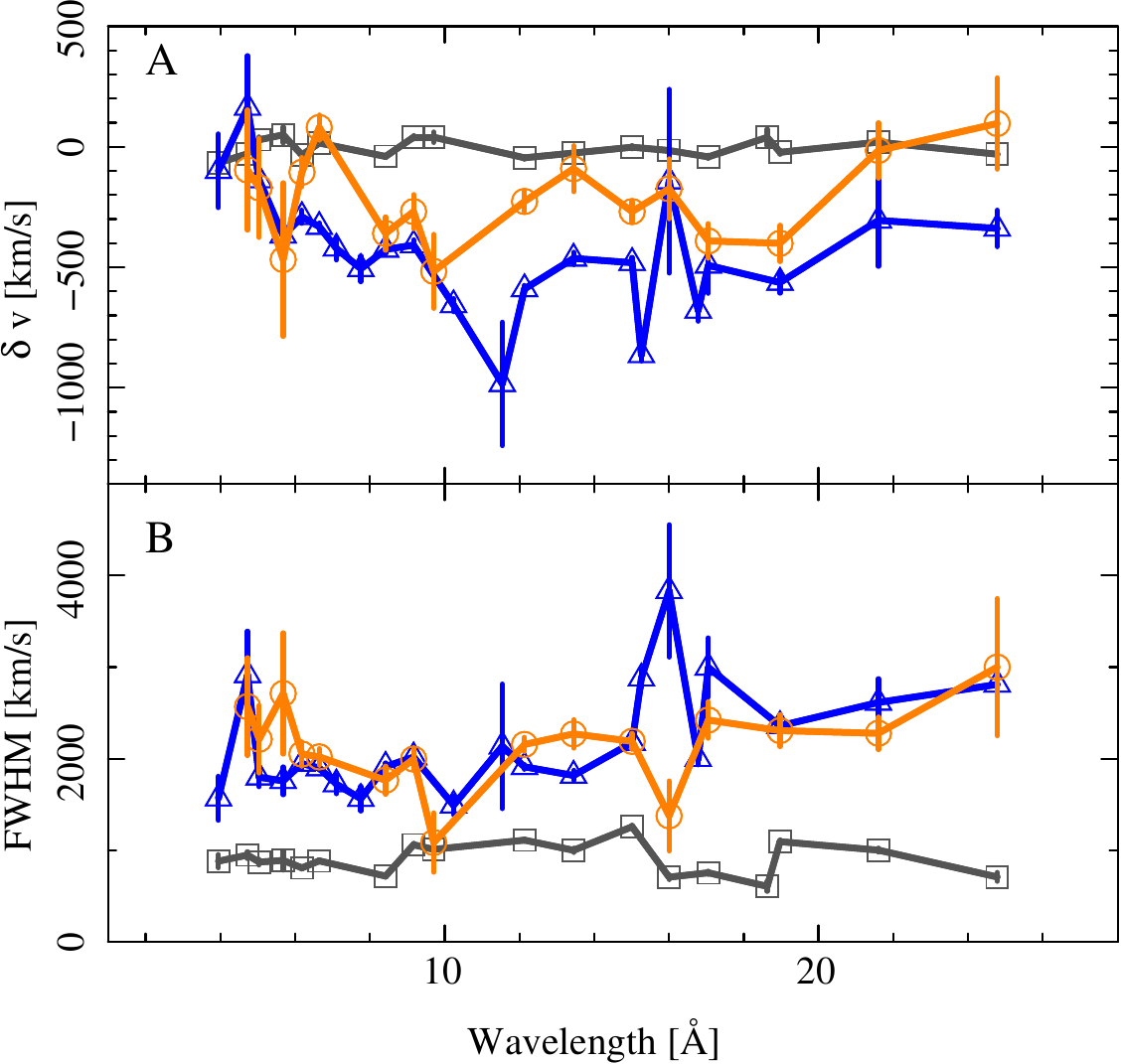}
  \caption{Error-weighted mean line centroid offsets vs line
    wavelength (top panel, A), and error-weighted-mean line widths vs
    line wavelength (bottom panel, B).  Classes of objects were
    defined from the velocity offset histogram group (see
    Figure~\ref{fig:dvhist} and are denoted as squares (gray) for zero
    offset, circles (orange) for intermediate offset and triangles
    (blue) for large negative offsets.}
    \label{fig:dv_fwhm_vs_wave}
\end{figure}

In Figure~\ref{fig:dvhist} we show the line centroid offset frequency
histogram for lines in the $7$--$18\mang$ region where the wind
continuum absorption is moderate.
We see two distinct groups, one near zero offset, and another near
$-400\kms$, as well as a long tail to large negative velocities.  We
have used this distribution to sort objects into ``low",
``intermediate", or ``large" velocity offset classes. In
Figure~\ref{fig:dv_fwhm_vs_wave} we show the weighted mean offsets
against wavelength for stars in these groups.

\begin{figure}[htb]
  \centering\leavevmode \includegraphics*[width=0.65\columnwidth,
  viewport= 0 0 585 525]{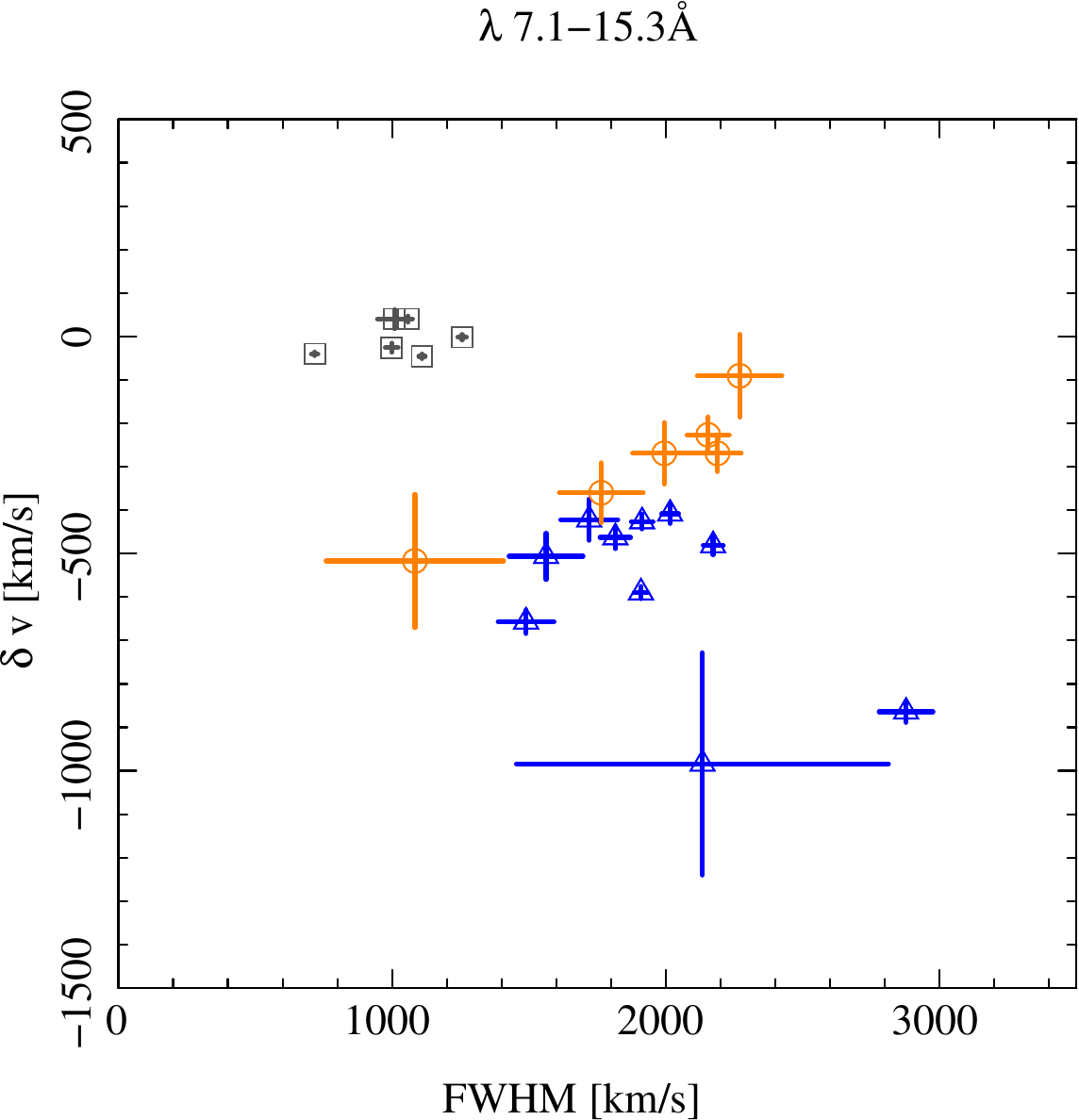}
  \caption{The correlation of centroid offset vs line width for the
    groupings defined by centroid offset vs wavelength shown in
    Figures~\ref{fig:dvhist} and \ref{fig:dv_fwhm_vs_wave}, for lines
    in the $7.1\mang$--$15.3\mang$ range. Symbols and colors are as in
    Figure~\ref{fig:dv_fwhm_vs_wave}.}
    \label{fig:dv_vs_fwhm}
\end{figure}
Each point represents an error-weighted-mean for a particular emission
line for each class.  In Figure~\ref{fig:dv_fwhm_vs_wave} we show the
line widths for these same groupings, and now see only two clear loci
-- those with low width, and thoses with higher width, which increases
with wavelength.  If we look at the correlation of offset and width
(Figure~\ref{fig:dv_vs_fwhm}), we see there is a clear separation of
stars with unshifted narrow lines from the others. The most-shifted
group (largest absolute $\Delta v$) is comprised of 9 objects
($\zeta\,$Pup, Cyg~OB2-9, HD~150136, HD~206267, HD~93129A, M17~Cen~1A,
RCW~38, WR~25, and WR~6), the intermediate group has 6 objects
($\tau\,$CMa, $\xi\,$Per, 9~Sgr, Cyg~OB2-8A, HD~150135, and HD~93250),
and the remaining 22 are in the narrow, unshifted group.

\begin{figure}[htb]
    \centering\leavevmode
    \includegraphics*[width=0.65\columnwidth]{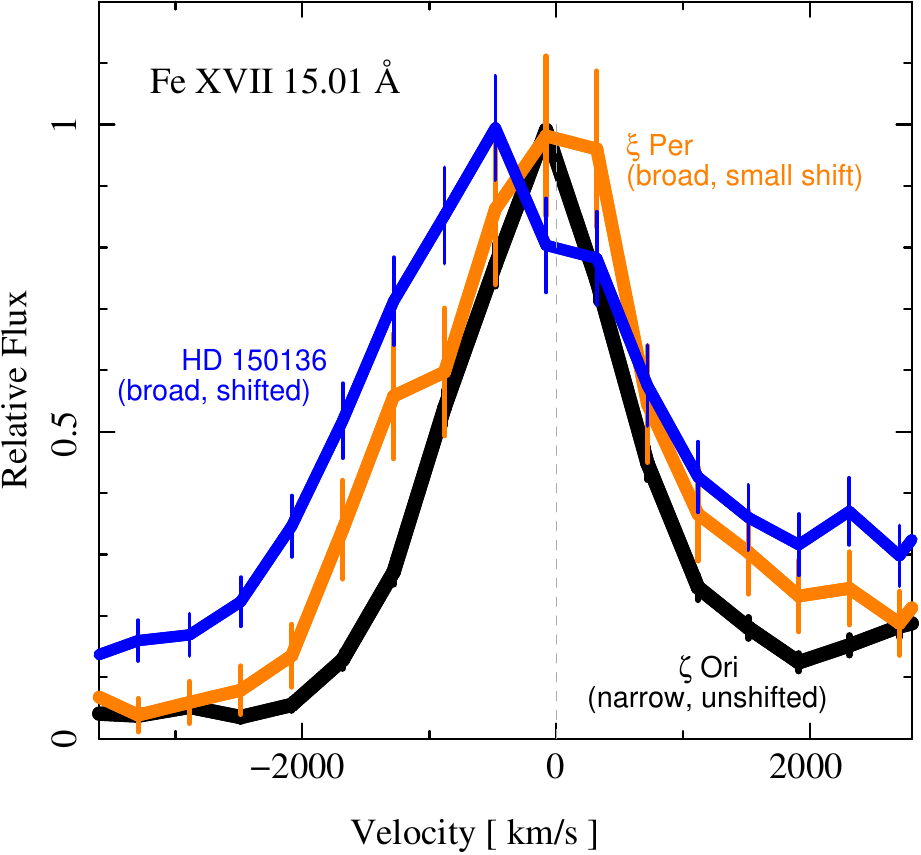}
    \caption{The \eli{Fe}{17} $15\mang$ velocity profile for stars from each of the three velocity-offset selected groups. Colors are as in Figure~\ref{fig:dv_fwhm_vs_wave}.}
    \label{fig:lines_detail}
\end{figure}

In Figure~\ref{fig:lines_detail} we show the velocity profile of the
\eli{Fe}{17} $15\mang$ line for three well-exposed spectra for a star
from each group.
We can easily see the difference between the unshifted and
most-shifted cases in centroid and width.  The middle case is more
subtle, the apparent shift being more due to the width and asymmetry
than the position of maximum line flux.

\begin{figure}[htb]
    \centering\leavevmode
    \includegraphics*[width=0.75\columnwidth]{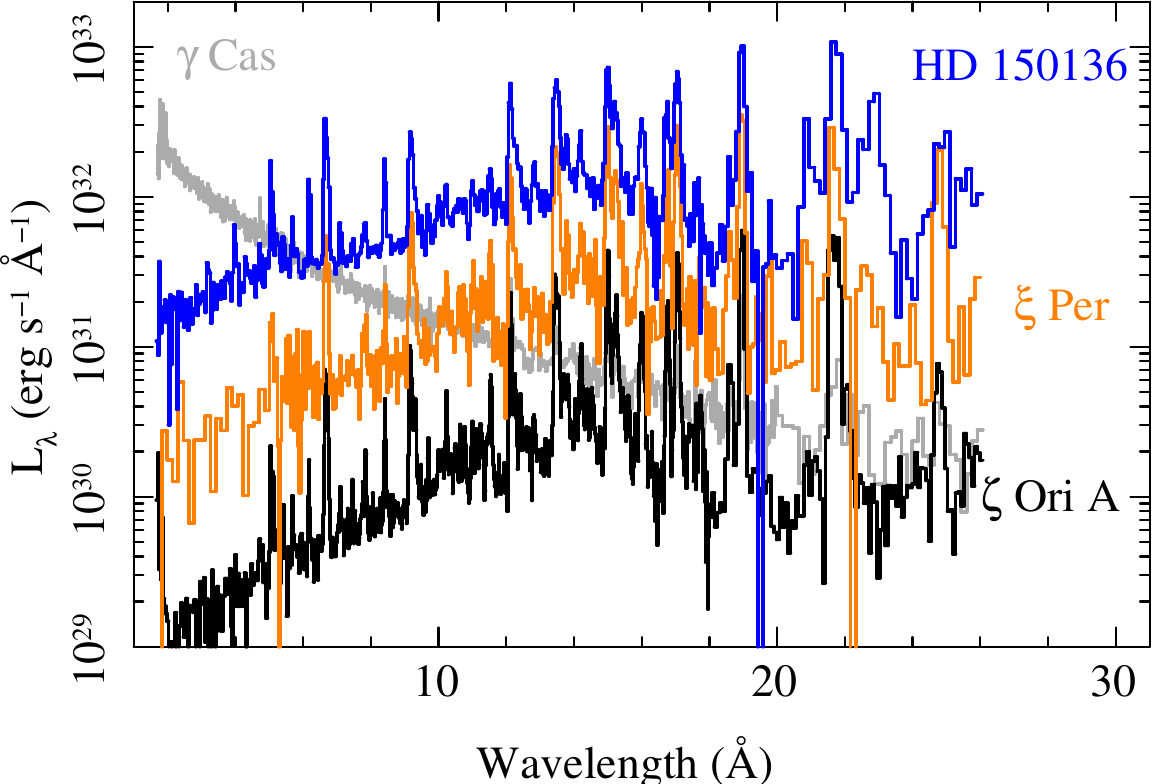} \caption{The
    luminosity density for stars from each of the three
    velocity-offset selected groups, plus the very different
    $\gamma\,$Cas. Colors are as in Figure~\ref{fig:dv_fwhm_vs_wave};
    $\gamma\,$Cas is shown in gray. Spectra have been
    corrected for absorption using the nominal interstellar values
    from Table~\ref{stellar-parameters}.}
    \label{fig:spec_examples}
\end{figure}

In Figure~\ref{fig:spec_examples} we show a broader band view of the
same three stars in units of luminosity density, along with the very
different $\gamma\,$Cas spectrum. In this view, the OB-wind-shock
systems all look quite similar, with perhaps slightly different
spectral slopes and a spectral break near $15\mang$. At the longest
wavelength( $>20\mang$), there is an apparent flattening or even rise
in flux; this is artificial, and due to reaching the noise floor in
the continuum. This is due to division of constant background counts
by a falling effective area. If we were rigorously fitting broad-band
spectral models, this would be handled by including background spectra
or models in the analysis, but results here are not affected by
background.

\begin{figure}[htb]
    \centering\leavevmode \includegraphics*[viewport=0 0 565 560,
    width=0.65\columnwidth]{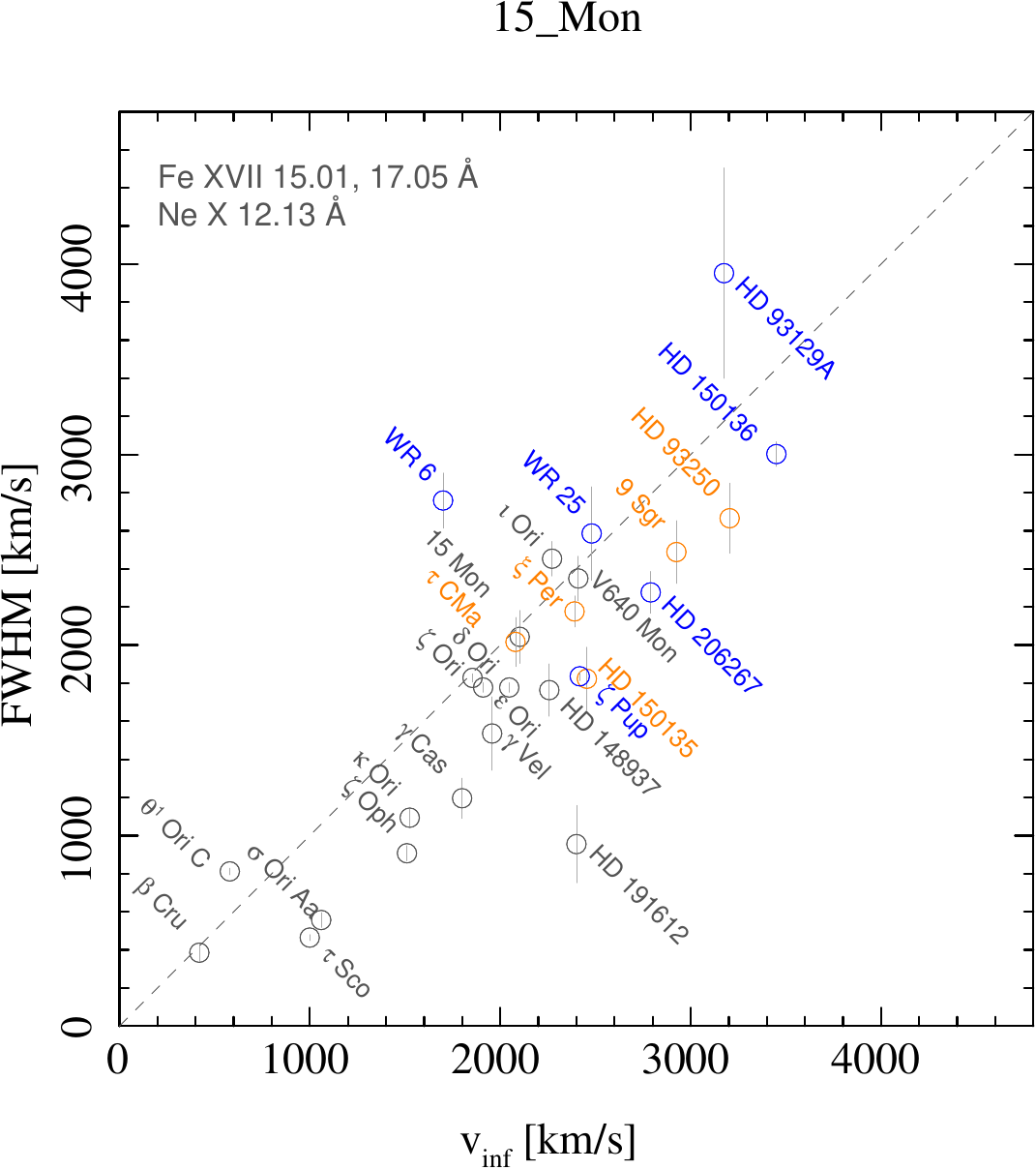} \caption{For
    each star with good measurements of the line width, we show their
    error-weighted-mean $FWHM$ over \eli{Ne}{10} $12\mang$,
    \eli{Fe}{17} $15\mang$ and $17\mang$. Colors are as in
    Figure~\ref{fig:dv_fwhm_vs_wave}. The dashed diagonal line is the
    $1:1$ dependence.}
    \label{fig:fwhm_vs_vinf}
\end{figure}

The weighted means, however, can hide significant outliers, if a mean
is dominated by group of measurements with small errorbars.  Hence, it
is still important to look at correlations among individual objects.
In Figure~\ref{fig:fwhm_vs_vinf} we show our measured line widths
against $v_\infty$ from the literature.  Here we can see that while
there is a group of objects with $FWHM\sim1000\kms$ from the
``unshifted, narrow'' group (gray circles), there are also members of
this group with $FWHM\sim2000\kms$.  The $FWHM$ generally correlates
with $v_\infty$, but there are significant deviations from the $1:1$
relationship.  Since the mean X-ray $FWHM$ seems a reasonable proxy
for $v_\infty$, we have entered the mean X-ray $FWHM$ into
Table~\ref{stellar-parameters} as the value for $v_\infty$ in the case
of objects with no value available in the literature.  The
individually fit values for each feature for each star are given in
tables in Appendix~\ref{sec:linemeas}.

The spectra for each object are shown in Appendix~\ref{sec:specplots},
Figures~\ref{fig:lx_9_Sgr}--\ref{fig:lx_tau_Sco}.  In these figures,
we show the absorption-corrected specific luminosity using the
distances and $N_\mathrm{H}$ values from
Table~\ref{stellar-parameters} as well as the spectra uncorrected for
absorption, since in some cases this is a large and uncertain term, as
can be seen in some of the spectra (e.g., Cyg OB2-12,
Figure~\ref{fig:lx_Cyg_OB2-12}).  The logarithmic scales all have the
same range and allow inter-comparison of the strength and shape of
X-ray emission.  The linearly-scaled plots emphsize the emission lines
and allow quick assessment of plasma characteristics, such as seen
between the magnetic $\theta^1\,$Ori~C and non-magnetic $\iota\,$Ori
(Figures~\ref{fig:lx_tet01_Ori_C}--\ref{fig:lx_iota_Ori}).

%
\begin{figure}
    \centering
      \includegraphics[width=0.48\columnwidth, viewport= 0 135 575 700]{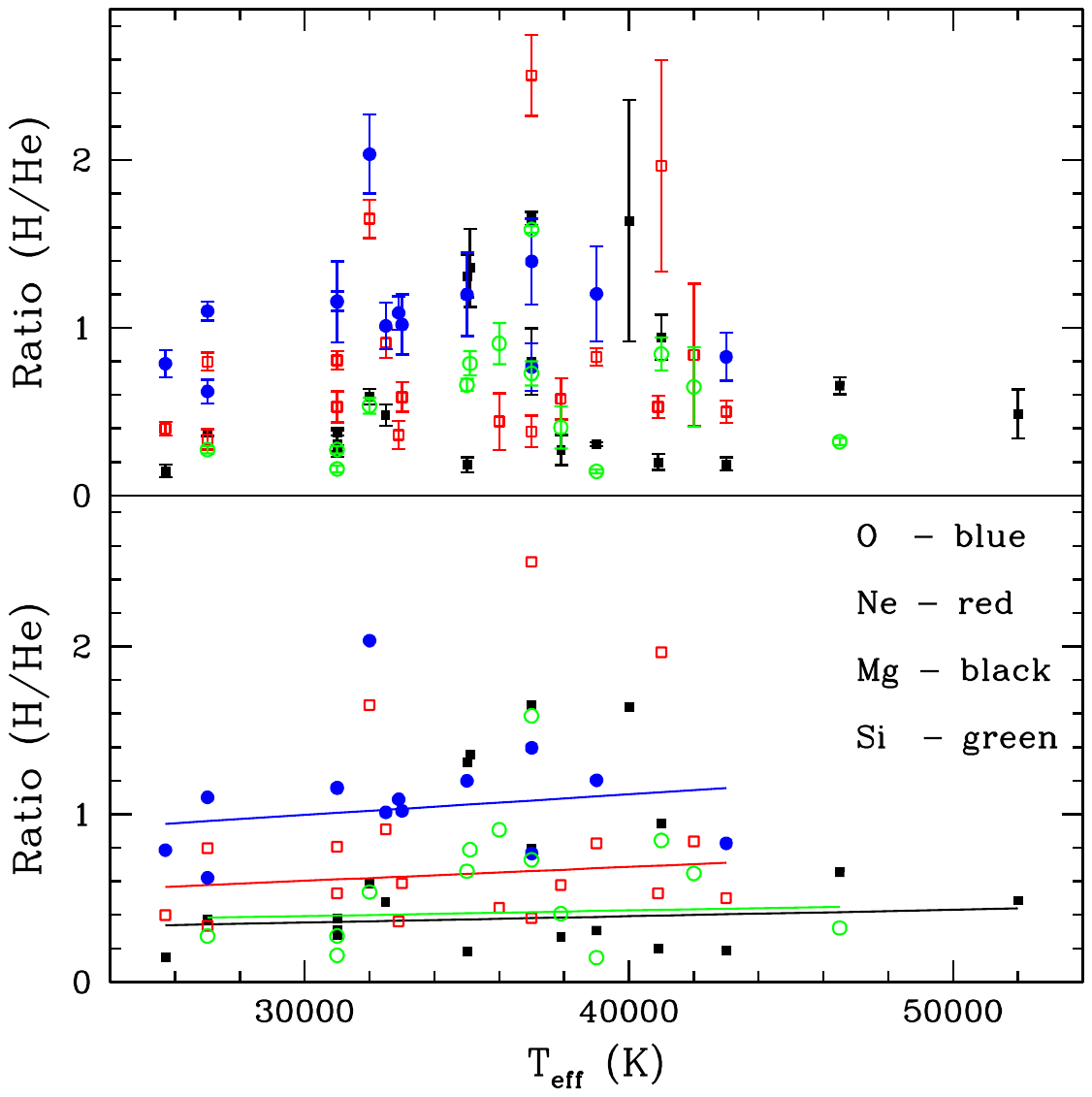}
      \includegraphics[width=0.48\columnwidth, viewport= 0 135 575 700]{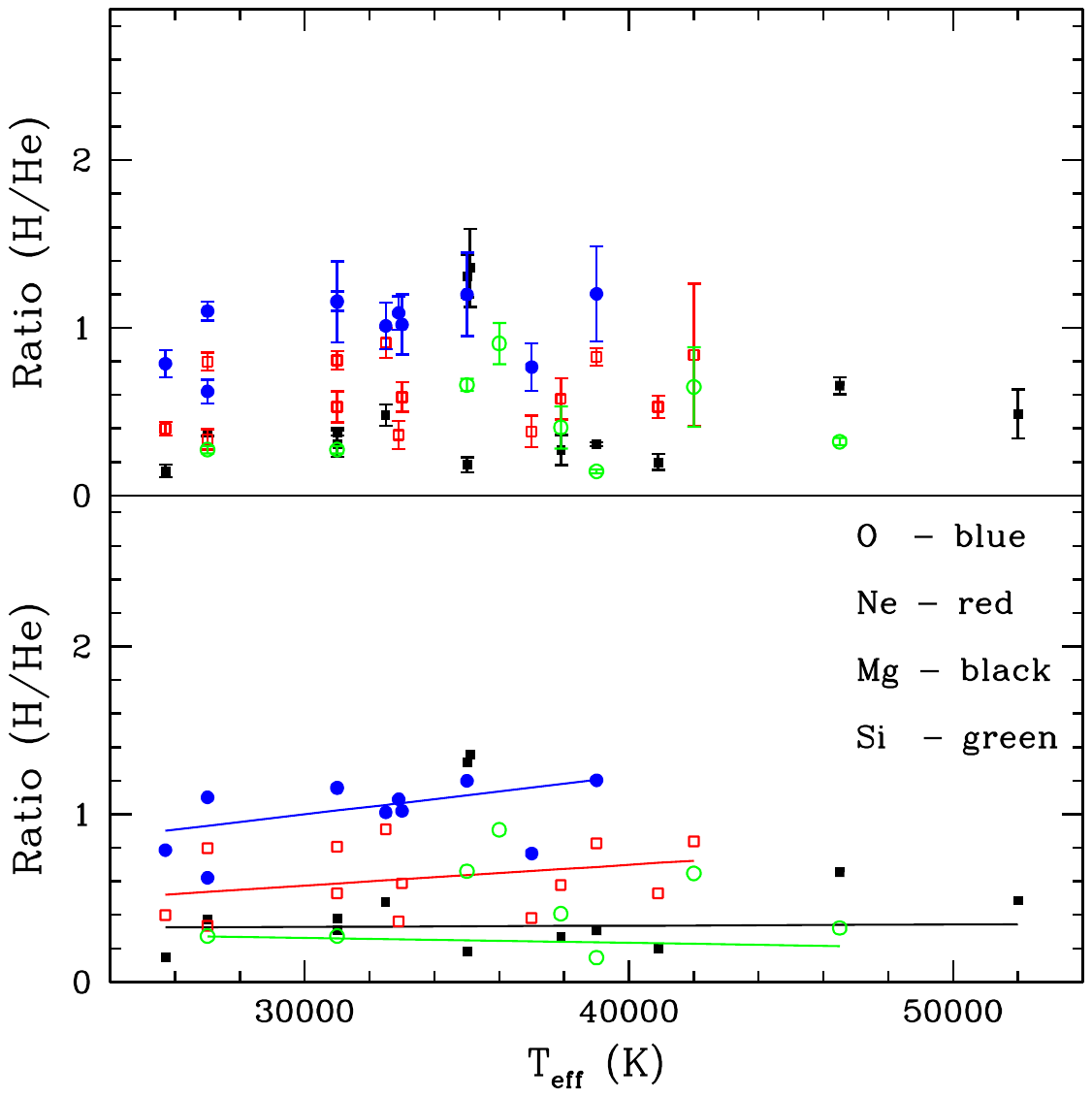}
    \caption{H/He line ratios involving O, Ne, Mg, and Si, as
    indicated, against source effective temperature.  The left side
    shows all stars in our sample; right side shows only the normal
    stars.  The upper panel displays ratios with flux measurement
    uncertainties.  The lower panel displays linear regressions to the
    respective ratio data sets.  The different lengths of line fits
    indicate the domain of temperature for which there are data.
    Line fluxes were corrected for the nominal interstellar
    absorption values given in Table~\ref{stellar-parameters}.}
   \label{fig:ratio-spectral_type}
\end{figure}
 

\section{Discussion}
\label{sec:disc}

The sample of stars in this paper more than doubles the
\cite{2009ApJ...703..633W} sample. Using \chan\ ACIS \hetg\ existing
spectra at the time, \cite{2009ApJ...703..633W} fit the \eli{Ne}{9}
and \eli{Ne}{10} lines for the 18 stars in their study.  These lines
were selected because the Ne H/He ratio is expected to show strong
trends throughout the temperature regime of O stars. They found that
stars of later O spectral type, and thus lower $T_\mathrm{eff}$,
showed lower X-ray ionization temperatures.  In comparing our work to
\cite{2009ApJ...703..633W}, we note some differences in fitting the
emission lines.  First, \cite{2009ApJ...703..633W} used a linear
continuum of constant value for each line fit, while we used a global
plasma model applied to all emission lines of a star.  Also,
\cite{2009ApJ...703..633W} used a single Gaussian fit to the
\eli{Ne}{9} triple rather than fit each component of the triplet, as
we did.  See Section \ref{sec:fitting} for a complete description of
the current analysis.
 
\cite{2009ApJ...703..633W} noted \eli{S}{15} emission essentially
disappears for spectral types later than O4. We confirm that trend
with our larger dataset, and can add that the only sources in our
study with measurable \eli{S}{15} emission later than O4 are either
magnetic stars, $\gamma$ Cas and analogs, or the Cyg OB2 stars.
Also, for ratios of H-like to He-like fluxes,
\cite{2009ApJ...703..633W} found that the ratios diminish
``drastically'' in later O spectral types for
\eli{Si}{14}/\eli{Si}{13} and \eli{Mg}{12}/\eli{Mg}{11}.  While we
confirm the general trend of lower ratios with later O spectral type,
with our larger dataset we find an approximate linear decline from
O3.5 to B0, with the dispersion about that linear relationship
decreasing with later spectral type (smaller $T_\mathrm{eff}$).  As
with the \eli{S}{15} results, only $\gamma$ Cas and analogs, and the
three Cyg OB2 stars, as well as some of the magnetic stars, have
anomalously large values compared to this trend.
 
These conclusions are demonstrated in
Figure~\ref{fig:ratio-spectral_type}.  In the left panel of
Figure~\ref{fig:ratio-spectral_type} the absorption-corrected
H-like/He-like ratios of Si, Mg, Ne, and O for all 37 stars in our
study are plotted vs. $T_\mathrm{eff}$ (each ion is color-coded on the
plot).  The upper plot shows the data points with error bars, and the
lower plot shows the data points without error bars, but with a linear
regression line for each ion.  For comparison, we show the same
representation in the right panel, but we have restricted this plot to
only the ``normal'' OB stars.  The ``atypical'' OB stars that we
excluded from the right panel were selected based on known attributes
(i.e., $\gamma$ Cas and analogs, magnetic stars) or based on empirical
evidence of disparate measurement values compared to other OB stars of
their spectral type (Cyg OB2 stars). Note that binary status of the
stars was not a selection criterion, so binaries exist in both the
normal (9 stars) and atypical (6 stars) groups.  See Table
\ref{tbl:peculiar} for list of stars in each group, and see
Table~\ref{trendfit} for slope and intercept values of the linear
fits.

A physical explanation of why these atypical stars have different
spectral parameters compared to our normal group of OB stars is beyond
the scope of this paper.  We found no systematic difference in the
emission line ratios with luminosity class for the normal stars.  The
H/He ratios can be an approximation to X-ray temperatures, so our
results are consistent with the lower temperatures and lower
ionization for later O-type spectral classes in normal stars.

\begin{figure}
    \centering
      \includegraphics*[width=0.65\columnwidth, viewport=0 0 495 510]{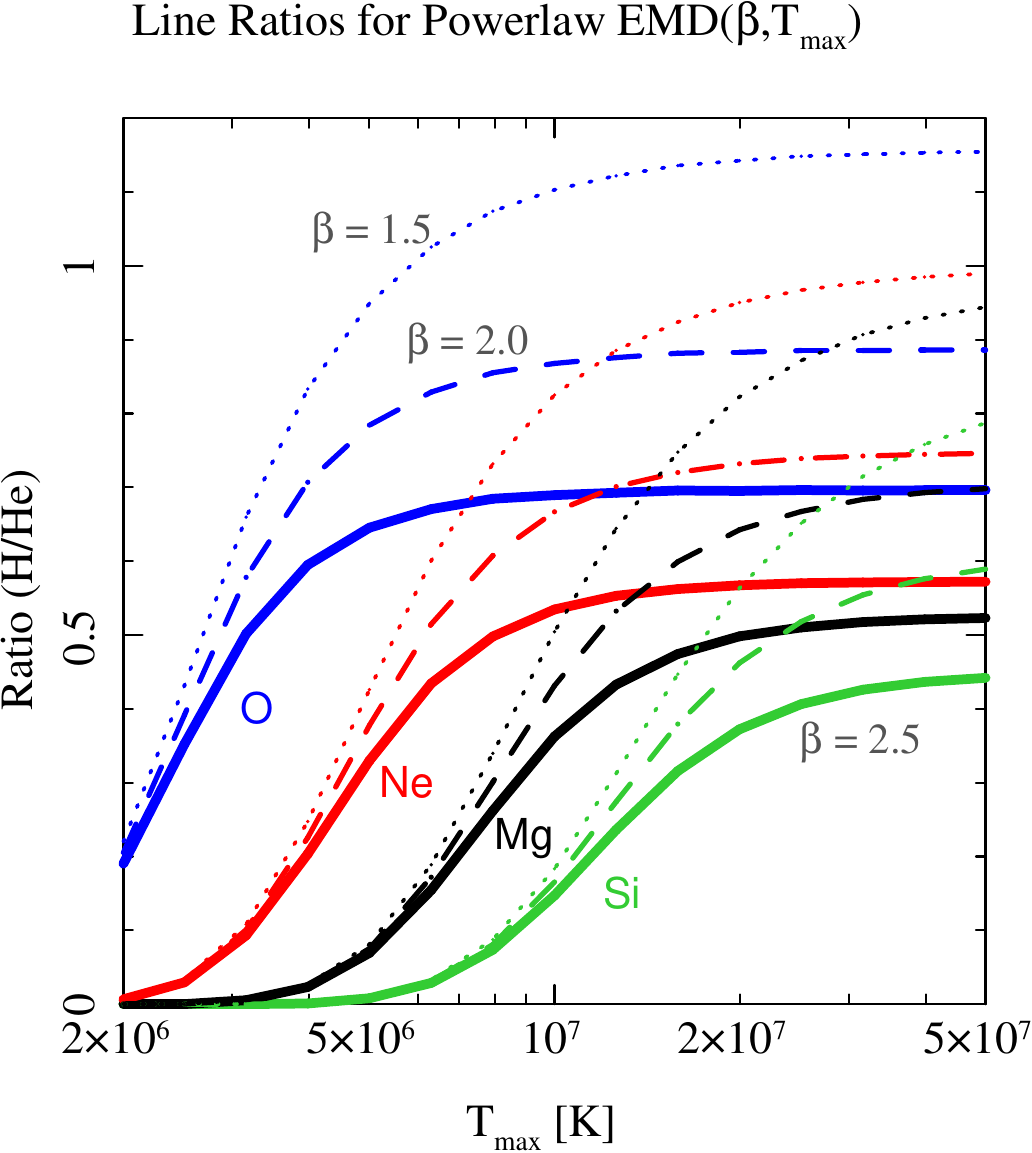}
    \caption{The H- to He-like photon flux ratios ($y$-axis) for several elements
    as labeled, for powerlaw emission measure distributions vs the
    high-temperature cutoff ($x$-axis) for three values of the
    exponent on temperature, $\beta$ (solid lines: $2.5$, dashed:
    $2.0$, and dotted: $1.5$, with larger values meaning steeper, or
    relatively less high-temperature plasma).}
   \label{fig:hheratios}
\end{figure}

It is tempting to use line ratios to determine a plasma temperature,
since there is a unique relationship between the H- to He-like ratio
and temperature for an isothermal plasma.  However, we generally have
emission measure distributions, which can greatly change the these
ratios.  As a qualitative aid, we adopt the powerlaw emission measure
distribution defined by \citet{huenemoerder:al:2020} to describe the
$\zeta\,$Pup X-ray spectrum.  The model has three primary parameters:
the exponent $\beta$ (unsigned, such that the emission measure is
$\propto T^{-\beta}$), the maximum temperature cutoff, and the minimum
temperature.  In Figure~\ref{fig:hheratios} we show theoretical H:He
photon flux ratios against the high temperature cutoff for three
values of $\beta$, using the AtomDB emissivities.

Here, the minimum temperature was kept at $1\mk$.  The solid curves
show the ratios for $\beta = 2.5$, about the value derived for
$\zeta\,$Pup. The ratios rise steeply at low $T_\mathrm{max}$ and then
quickly plateau.  This is because beyond a certain maximum
temperature, there is no longer any additional contribution from the
hotter H-like line, once $T_\mathrm{max}$ has exceeded the line's
temperature of maximum emissivity.  For shallower emission measure
distributions (smaller $\beta$; dashed and dotted lines), the ratios
are higher at any $T_\mathrm{max}$, since there is now relatively more
contribution from the H-like line at higher $T$.

Comparison of these trends to the observed ratios in
Figure~\ref{fig:ratio-spectral_type} is not direct.  We expect that
effective temperature defines the radiation field driving the wind so
higher $T_\mathrm{eff}$ implies, in general, a greater terminal
velocity, $v_\infty$.  This, in turn, relates to the maximum embedded
wind shock velocity, and thereby the $T_\mathrm{max}$ of the emission
measure distribution.  Hence, $T_\mathrm{eff}$ is a qualitative proxy
for $T_\mathrm{max}$.  So it is reasonable to compare the trends in
each plot, and we do indeed find them similar if we restrict to the
``normal'' stars (excluding known magnetic, colliding wind, or
$\gamma\,$Cas stars): the ratios rise slightly with temperature,
values are $\lesssim1$, and they are in the same order vs element.

The power-law exponents used are reasonable for OB-stars.  In addition
to the fit of the model to $\zeta\,$Pup, the emission measure
distribution analysis of \citet{wojdowski:schulz:2005} and
\citet{cohen2021} show slopes in this range.  The outliers in
Figure~\ref{fig:ratio-spectral_type} tend to be points for O and Ne.
These are the lines first affected by a decrease in relative emission
measure at low temperatures. In the powerlaw emission measure model,
if we were to increase the minimum temperature a bit, to say $3\mk$,
then the ratios for O and Ne would be significantly raised, by factors
of about 5 and 2, respectively.  For magnetically confined plasmas we
might even have a more ``coronal'' distribution which rises with
temperature, as is the case found for $\theta^1\,$Ori~C
\citep{wojdowski:schulz:2005}; it is indeed one of the outliers with
H:He ratios of about $3.0$ for O and Ne.

\begin{table}[ht] \label{trendfit}
\begin{center}
\caption{Fit Parameters for H/He Trends}
\begin{tabular}{|l|rr|rr|}
\hline\hline Species & \multicolumn{2}{c|}{All Stars} & \multicolumn{2}{c|}{Normal Stars} \\
   &  Slope ($K^{-1}$) & Intercept & Slope ($K^{-1}$) & Intercept \\ \hline
O  &  $123\times 10^{-7}$ & 0.628 & $220\times 10^{-7}$ & 0.312 \\
Ne &  $83\times 10^{-7}$ & 0.353 & $124\times 10^{-7}$ & 0.202 \\
Mg &  $38\times 10^{-7}$  & 0.238 & $7\times 10^{-7}$   & 0.308 \\
Si &  $34\times 10^{-7}$  & 0.292 & $-29\times 10^{-7}$ & 0.350 \\ \hline
\end{tabular}
\end{center}
\end{table}

\section{Summary}
\label{sec:sum}

In an effort to maximize the number of stars from which to assess
statistical trends of massive star X-ray properties from a uniform
analysis approach, we extracted a sample of hot massive star X-ray
spectra taken with the {\em Chandra} HETGS.  A selection criterion
requiring a minimum of approximately 1800 total counts, possibly from
multiple exposures, was adopted to ensure sufficient data quality for
fitting of resolved wind-broadened line profiles.  We did not limit
the categories of stars in terms of being single, binary, or magnetic.
We did however identify these separate categories with our analysis.
Our sample consisted of 37 WR, O, and B star sources (see
Tab.~\ref{stellar-parameters}).

We identified the stronger emission line features in the spectrum of
each source and fit them using Gaussian profiles.  The $FWHM$ of the
Gaussian and the peak of line center were fit parameters.  We were
able to accumulate an ensemble of line measurements in terms of line
width, location of peak emission, and line flux for each source, and
to assign uncertainties for each measure.

The primary results found from these spectra can be summarized as
follows:

\begin{enumerate}

\item For normal OB stars, our measurements of X-ray parameters such
  as H/He ratios of the X-ray emission lines and the $L_\mathrm{x}$
  values determined herein are, in general, well described by the
  optically-determined spectral types and their associated
  $T_\mathrm{eff}$.  This finding reinforces and extends the
  conclusions of \cite{2009ApJ...703..633W} and provides a framework
  for identifying atypical OB stars.

\item Three subgroups of stars were identified in terms of trends for
  line width (FWHM) and offset of peak line emission ($\Delta v$), as
  shown in Figure \ref{fig:dv_vs_fwhm}.  One group was identified as
  having little to no shift and low to modest line broadening.
  Another group had broad lines and the largest shifts in peak line
  emission.  The final group was intermediate, with modest to large
  line broadenings and low to modest line peak offsets.  The groupings
  generally correlated with wind speed.

\item X-ray emission line $FWHM$ for lines at $\lambda \gtrsim
  12\mang$ is a reasonable proxy for wind terminal velocity,
  $v_\infty$ (see, Fig.~\ref{fig:fwhm_vs_vinf}). This trend could be
  useful for wind modeling of heavily obscured sources where UV or
  optical diagnostics are not available.

\item The H/He line ratio versus the effective stellar temperature for
  O, Ne, Mg, Si are shown in Figure~\ref{fig:ratio-spectral_type}

\item For the benefit of future analyses, tabulations of line
  measurements, including the velocity offsets, the full-width half
  maxima (FWHMs) and total flux, are provided for each star in
  Appendix~\ref{sec:linemeas}.

\item A gallery of X-ray spectra the form of specific luminosity vs
  wavelength (linear and logarithmic versions) are provided in
  Appendix~\ref{sec:specplots}, placing all sources of our sample on a
  common footing for comparison.  Determination of specific
  luminosities were calculated using distances and interstellar
  absorption values from the literature.

\end{enumerate}

In this paper, we have performed a comprehensive spectral analysis
focusing on the emission line properties of massive stars. We were
able to investigate the individual emission lines in great details,
thanks to the impeccable spectral resolution of \chan grating
spectrometers that make such high resolution studies possible. With
the future of high resolution spectral studies heading towards micro
calorimeter measurements \citep{micro2021}, some of our findings are
natural segue into understanding the plasma properties of massive
stars to be investigated with XRISM and Athena/XIFU
\citep{xrism2020,pajot2018}. For instance, the empirical relation
between the line widths of emission lines and terminal wind velocity
of massive stars we present here (Fig.~\ref{fig:fwhm_vs_vinf}) can be
useful to measure terminal wind velocity in stars found in star
clusters where direct measurements with UV are not feasible due to
extinction. Additionally, XRISM will also provide some complementary
view of the hard X-ray properties of where \chan gratings are not
covered. This is especially useful to study line emission from hot and
unusual stars in our sample.

\begin{acknowledgements}
Support for this work was in part provided by NASA through the
Smithsonian Astrophysical Observatory (SAO) contract SV3-73016 to MIT
for Support of the \emph{Chandra} X-Ray Center (CXC) and Science Instruments,
and by \emph{Chandra} Award Number AR8-19001 (A, B, and C) issued by the CXC.
JSN acknowledges the support of NASA \emph{Chandra} contract NAS8-03060.  The
CXC is operated by the Smithsonian Astrophysical Observatory for and
on behalf of NASA under contract NAS8-03060.
\end{acknowledgements}

\appendix

\section{Notes on some individual stars}

As discussed above, massive stars are usually found in binary or
higher order multiple systems that have the potential to generate
X-rays through the interactions of colliding winds that can rival or
outshine the intrinsic emission of the winds of individual
stars. Establishing system multiplicity is therefore an important
consideration. Of the current sample of stars, the sometimes complex
multiplicities of HD~191612, Cyg OB2-9, Cyg OB2-8A, HD~206267, 15~Mon,
$\delta$~Ori, V460~Mon, $\zeta$~Ori, $\sigma$~Ori, $\theta^1$~Ori~C
and $\iota$~Ori were discussed by \citet{2019AA...626A..20M}; and
those of $\zeta$~Ori, $\iota$~Ori, 15~Mon, $\tau$~CMa, HD~206267 and
HD~93129A by \citet{2020AA...636A..28M}.

\begin{description}

\item[9~Sgr alias HD~164794] The two \emph{Chandra} HETG observations
  analysed here were taken 216 and 245 d after the 2004 October
  periastron passage of the most recent definition of the SB2 binary
  orbit of the 8.9-year period established by
  \citet{2021A&A...651A.119F}. The change in MEG count rate after the
  gap of about a month was small at $1.5\pm1.4$\%. The four XMM-Newton
  observations performed an orbital cycle later
  \citep{2016A&A...589A.121R} showed that most of the X-ray emission
  is intrinsic to the individual winds of the $53 \msun$ primary and
  $39 \msun$ secondary stars that dominate a relatively weak variable
  colliding-wind component that accounts for the increase of about
  20\% from apastron seen about a week before the 2013 September
  periastron.


\item[M17 Cen 1A \& 1B alias ALS~19613 A \& B] According to
  \citet{2020A&A...643A.138M} and references therein, these two bright
  X-ray sources separated by 1.65\arcsec\,at the center of the
  \ion{H}{2} region M17 are both spectroscopic binaries. They have
  also been resolved in \emph{Chandra} images following
  \citet{2007ApJS..169..353B} and appear with moderate pile-up
  warnings in the latest release 2.0.1 of the \emph{Chandra} Source Catalog.

\item[HD~191612] The 10 HETG observations of the magnetic star
  HD~191612 \citep{2016ApJ...831..138N} were scheduled in 2 groups
  according to the maximum and minimum phases of its primary's 537.2
  day rotational period. The two sets differed by about 20\% in count
  rate but with no obvious change in line shape so we combined all the
  data for analysis.

\item[Cyg OB2-12] According to \citet{2019A&A...627A..99N}, the
  regular 108-d X-ray variability of this B-type hypergiant is not
  obviously of binary origin.

\item[Cyg OB2-9 alias ALS~11422] This pair of supergiant and giant
  early O stars is one of the clearest examples of an X-ray
  colliding-wind binary system \citep{naze2012}: over the course of
  its 2.35-yr period, the luminosity varies inversely with the binary
  separation of its highly eccentric orbit. The single HETG spectrum
  available was obtained 186 days after the 2002 January periastron
  during an early \emph{Chandra} observation targeted at its neighbor
  CygOB2-8A that lies 3.8\arcmin~away. At this orbital phase, the
  luminosity of Cyg OB2-9 should already have been close to the
  minimum expected at apastron at a count rate consistent with the
  value observed.

\item[Cyg OB2-8A alias ALS~11423] The one HETG observation of this
  close double-lined spectroscopic binary was obtained near the X-ray
  minimum of the 25\% amplitude variations over the 21.9-day orbit. It
  was one of the brightest stars in the survey.


\item[$\gamma$~Cas] Following \citet{2022MNRAS.510.2286N}, the
  prototype of this class and two analogs are classified as binary
  systems in Table~\ref{tbl:peculiar}.





\item[V640~Mon alias HD~47129, Plaskett's Star] When phased by the
  14.4-d binary orbital period \citep{2008AA...489..713L}, the X-ray
  light curve of the 6 HETG observations of this SB2 shows a decrease
  of about 25\% at an epoch reasonably consistent with an expected
  stellar conjunction. This merits further exploration.





\item[$\iota$~Ori alias HD~37043] The tidally-induced photometric
  variations observed in the optical close to the periastron of the
  29.1-day orbital period of this eccentric binary
  \citep{2017MNRAS.467.2494P} classify $\iota$~Ori as the most massive
  of the heartbeat stars. The 2 HETG exposures were taken a day apart
  3 days before apastron early in the \emph{Chandra} mission and a long time
  before recognition of the system's heartbeat status.

\item[$\kappa$~Ori] The 234 ks HETG exposure time of this early B
  supergiant was accumulated in 4 observations scheduled over 10 days
  including one long exposure responsible for nearly half the
  time. The mean count rates show significant variability of $9\pm2$\%
  between the brightest and faintest. In common with stars of similar
  type, the fractional amplitude of X-ray variability is about an
  order of magnitude greater than that observed on similar timescales
  in the optical through TESS photometry.




\item[$\zeta$~Pup alias Naos, HD~66811] After the discovery with the
  XMM-Newton RGS of strong \ion{N}{6} and \ion{N}{7} lines in its
  spectrum, the nearby, bright, single early O-star supergiant
  $\zeta$~Pup has continued to be observed on a roughly annual basis
  by XMM-Newton for long-wavelength calibration purposes. While an
  initial astronomical assessment by \citet{naze2012} of these data
  mistakenly cited instrumental issues, it is now well established
  that the X-rays generated in the wind of $\zeta$~Pup are subject to
  slow stochastic variability at the few percent level as well as more
  coherent changes of similar amplitude at the same period of 1.78
  days seen in optical photometry \citep{Nichols2021}.

\item[$\gamma^2$~Vel alias WR~11, HD~68273] $\gamma^2$~Velorum is a
  first magnitude star, the nearest and brightest Wolf-Rayet star and
  a binary system. The orbital period is 78.53~d
  \citep{1997A&A...328..219S} and its complex but repeatable phased
  X-ray light curve with a dynamic range of about 5 is still best
  captured by the 1991-1993 ROSAT PSPC observations
  \citep{1995A&A...298..549W}. The combination of these data with a
  few subsequent XMM-Newton measurements shows a steady increase in
  intensity over 9 days after periastron. The single \emph{Chandra} HETG
  spectrum was obtained at about $\frac{2}{3}$ of the observed maximum
  with little evidence of any internal variability that might have
  been expected.

\item[RCW 38 IRS 2] The binary nature of this central ionizing system
  of the \ion{H}{2} region RCW 38 was discovered by
  \citet{2009AJ....138...33D}.



\item[WR~25 alias HD~93162] Colliding winds account for most of the
  X-rays from this very massive Wolf-Rayet binary system of 208-d
  period observed with the HETG near maximum light between 10 and 12 d
  before periastron by \citet{pradhan2021}. The level of intrinsic
  emission from the wind of the Wolf-Rayet primary star may be judged
  by comparison with WR~24, a neighboring single star in Carina of
  similar type. The 6 separate observations of WR~24 reported in the
  \emph{Chandra} Source Catalog define a consistent picture of an X-ray
  imaging count rate about a factor of 20 lower than that of the
  single imaging observation of WR~25 that was taken a few days before
  apastron.

\item[$\theta$~Car alias HD~93030] The star has an accurately defined
  spectroscopic binary period of 2.20 d although recent TESS
  photometry \citep{2021MNRAS.505.5725K} suggested that rotation could
  also be relevant. The HETG data were taken in two observations 10
  days apart and partially overlaped in phase, covering in total about
  half the period. The HETG spectrum resolves the narrow lines
  identified in the RGS spectrum early in the XMM-Newton mission
  \citep{2008AA...490..801N}. As reported in Table 1, the measured
  mean X-ray line width implies a wind terminal velocity of $482\pm85
  \kms$ that is consistent, for example, with the velocity profile of
  \ion{Si}{4} $\lambda\lambda~1393.8,1402.8$ absorption in the IUE
  spectrum.







%
%

\end{description}


\clearpage
\section{Line measurements tables}\label{sec:linemeas}

We present line measurements for each star in the following tables.
The first column gives the ion names, the second column $\lambda_0$ is
the theoretical wavelength of the line, the third column $\delta v$ is
the offset of the measured wavelength from the theoretical wavelength
in velocity units, the fourth column $\sigma_{\delta v}$ is the error
on this offset followed by the full-width, half-maximum (FWHM) of the
line and the error on the FWHM is $\sigma_{FW}$.  The next two columns
are the model-independent fluxes $f_x$ and the error on the flux,
$\sigma_{f_x}$, uncorrected for interstellar absorption.  The last two
columns $f_0$, and $\sigma_{f_0}$, are fluxes, $f_x$, and bounds
corrected for interstellar absorption using $N_\mathrm{H}$ and its
uncertainty from Table~\ref{stellar-parameters}.  (For a visual
estimate of the effect of absorption, see the plots in
Appendix~\ref{sec:specplots}.)  For those lines where the Gaussian
line center or line width were not determined by the fit, we only
provide fluxes.  Since the interstellar absorption term's
  bounds are systematic uncertainties, any change (as per the
  uncertainties on $N_\mathrm{H}$ listed in
  Table~\ref{stellar-parameters}) would be correlated over all lines
  for any given star.  Hence, we have not propagated the statistical
  flux uncertainty with the systematic absorption uncertainty.  We
  provide the latter to give some idea of the possible effects of
  uncertain absorption.  The fractional statistical uncertainties on
  $f_0$ are equal to those of $f_x$.

\clearpage

\restartappendixnumbering

\begin{deluxetable}{lrrccccccc} 
 	 \tablecaption{ $9\,$Sgr} 
 	 \tablehead{ 
 	 \colhead{Ion}& 
 	 \colhead{$\lambda_0$}& 
 	 \colhead{$\delta v$}& 
 	 \colhead{$\sigma_{\delta v}$}& 
 	 \colhead{$ FWHM$}& 
 	 \colhead{$\sigma_{FW}$}& 
 	   \colhead{$ f_x$}& 
 	 \colhead{$\sigma_{f_x}$}& 
 	   \colhead{$ f_0$}& 
 	 \colhead{$\sigma_{f_0}$}\\ 
 	 & 
 	 \colhead{(\AA)} & 
 	 \colhead{($\kms$)}  & 
 	 \colhead{($\kms$)}  & 
 	 \colhead{($\kms$)}  & 
 	  \colhead{($\kms$)}  & 
 	  \multicolumn{4}{c}{($ 10^{-5} \pflux$) } 
 } 
 \startdata
        Si XIV&       6.183&        \nodata&         \nodata&        \nodata&         \nodata&        0.09&        0.03&        0.10&        0.00 \\
       Si XIII&       6.648&           5&         214&        2774&         329&        0.69&        0.12&        0.77&        0.01 \\
       Si XIII&       6.740&           \nodata &           \nodata&        \nodata&        \nodata&        0.28&        0.05&        0.30&        0.00 \\
        Mg XII&       8.422&         -55&         244&        2757&         479&        0.36&        0.06&        0.43&        0.01 \\
     Mg XI (r)&       9.169&         260&         155&        3641&         315&        1.79&        0.20&        2.26&        0.08 \\
Ne IX $\delta$&       9.481&         \nodata&           \nodata&        \nodata&        \nodata&        0.16&        0.06&        0.20&        0.01 \\
  Ne X $\beta$&      10.239&        -627&         265&        1629&         754&        0.24&        0.11&        0.32&        0.02 \\
          Ne X&      12.135&        -418&         103&        2662&         228&        3.92&        0.28&        6.21&        0.47 \\
\enddata 
 \end{deluxetable} \label{tbl:9_Sgr}

\begin{deluxetable}{lrrccccccc} 
 	 \tablecaption{ $\zeta\,$Oph} 
 	 \tablehead{ 
 	 \colhead{Ion}& 
 	 \colhead{$\lambda_0$}& 
 	 \colhead{$\delta v$}& 
 	 \colhead{$\sigma_{\delta v}$}& 
 	 \colhead{$ FWHM$}& 
 	 \colhead{$\sigma_{FW}$}& 
 	   \colhead{$ f_x$}& 
 	 \colhead{$\sigma_{f_x}$}& 
 	   \colhead{$ f_0$}& 
 	 \colhead{$\sigma_{f_0}$}\\ 
 	 & 
 	 \colhead{(\AA)} & 
 	 \colhead{($\kms$)}  & 
 	 \colhead{($\kms$)}  & 
 	 \colhead{($\kms$)}  & 
 	  \colhead{($\kms$)}  & 
 	  \multicolumn{4}{c}{($ 10^{-5} \pflux$) } 
 } 
 \startdata
           S XV&       5.039&        \nodata &         \nodata &        \nodata &        \nodata &        0.45&        0.12&        0.45&        0.00 \\
         Si XIV&       6.183&        \nodata &        \nodata &         \nodata &         \nodata &        0.35&        0.06&        0.36&        0.00 \\
        Si XIII&       6.648&          23&          39&         875&          90&        1.27&        0.12&        1.31&        0.01 \\
        Si XIII&       6.687&          \nodata &           \nodata &         \nodata &        \nodata &        0.52&        0.08&        0.54&        0.00 \\
        Si XIII&       6.740&          \nodata &           \nodata &         \nodata &        \nodata &        0.70&        0.08&        0.72&        0.00 \\
         Mg XII&       8.422&           6&          53&         802&         123&        1.13&        0.13&        1.18&        0.01 \\
      Mg XI (r)&       9.169&           5&          32&         968&          67&        2.33&        0.17&        2.47&        0.03 \\
      Mg XI (i)&       9.230&           \nodata &           \nodata &         \nodata &        \nodata &        1.52&        0.14&        1.61&        0.02 \\
\enddata 
 \end{deluxetable} \label{tbl:zeta_Oph}

\begin{deluxetable}{lrrccccccc} 
 	 \tablecaption{   M17 Cen 1B} 
 	 \tablehead{ 
 	 \colhead{Ion}& 
 	 \colhead{$\lambda_0$}& 
 	 \colhead{$\delta v$}& 
 	 \colhead{$\sigma_{\delta v}$}& 
 	 \colhead{$ FWHM$}& 
 	 \colhead{$\sigma_{FW}$}& 
 	   \colhead{$ f_x$}& 
 	 \colhead{$\sigma_{f_x}$}& 
 	   \colhead{$ f_0$}& 
 	 \colhead{$\sigma_{f_0}$}\\ 
 	 & 
 	 \colhead{(\AA)} & 
 	 \colhead{($\kms$)}  & 
 	 \colhead{($\kms$)}  & 
 	 \colhead{($\kms$)}  & 
 	  \colhead{($\kms$)}  & 
 	  \multicolumn{4}{c}{($ 10^{-5} \pflux$) } 
 } 
 \startdata
      Fe XXV&       1.858&        \nodata&         \nodata&         \nodata&         \nodata&        0.38&        0.13&        0.39&        0.00 \\
       S XVI&       4.730&          \nodata&          \nodata&         \nodata&          \nodata&        0.20&        0.08&        0.31&        0.01 \\
        S XV&       5.039&          \nodata&          \nodata&          \nodata&            \nodata&        0.16&        0.06&        0.26&        0.01 \\
      Si XIV&       6.183&        -113&         290&        3331&         850&        0.34&        0.07&        0.81&        0.07 \\
   Mg XI (i)&       9.230&        \nodata&            \nodata&         \nodata&         \nodata&        0.17&        0.05&        1.95&        0.47 \\
        Ne X&       9.362&        \nodata&            \nodata&         \nodata&         \nodata&        0.08&        0.04&        0.98&        0.25 \\
\enddata 
 \end{deluxetable} \label{tbl:M17_Cen_1B}

\begin{deluxetable}{lrrccccccc} 
 	 \tablecaption{ M17 Cen 1A} 
 	 \tablehead{ 
 	 \colhead{Ion}& 
 	 \colhead{$\lambda_0$}& 
 	 \colhead{$\delta v$}& 
 	 \colhead{$\sigma_{\delta v}$}& 
 	 \colhead{$ FWHM$}& 
 	 \colhead{$\sigma_{FW}$}& 
 	   \colhead{$ f_x$}& 
 	 \colhead{$\sigma_{f_x}$}& 
 	   \colhead{$ f_0$}& 
 	 \colhead{$\sigma_{f_0}$}\\ 
 	 & 
 	 \colhead{(\AA)} & 
 	 \colhead{($\kms$)}  & 
 	 \colhead{($\kms$)}  & 
 	 \colhead{($\kms$)}  & 
 	  \colhead{($\kms$)}  & 
 	  \multicolumn{4}{c}{($ 10^{-5} \pflux$) } 
 } 
 \startdata
      Fe XXV&       1.858&        \nodata&         \nodata&        \nodata&        \nodata&        0.82&        0.23&        0.85&        0.00 \\
       S XVI&       4.730&        \nodata&         \nodata&        \nodata&        \nodata&        0.49&        0.14&        0.76&        0.03 \\
        S XV&       5.039&         158&         523&        2782&        1111&        0.40&        0.17&        0.65&        0.03 \\
        S XV&       5.102&         \nodata&         \nodata&        \nodata&        \nodata&       0.33&        0.12&        0.55&        0.03 \\
      Si XIV&       6.183&           8&         185&        3052&         356&        0.66&        0.08&        1.60&        0.14 \\
     Si XIII&       6.648&         -47&         200&        3276&         294&        0.69&        0.08&        2.02&        0.22 \\
     Si XIII&       6.740&         \nodata&         \nodata&        \nodata&        \nodata&        0.57&        0.08&        1.59&        0.16 \\
      Mg XII&       8.422&        -444&         390&        3218&         840&        0.26&        0.06&        1.67&        0.31 \\
\enddata 
 \end{deluxetable} \label{tbl:M17_Cen_1A}

\begin{deluxetable}{lrrccccccc} 
 	 \tablecaption{ HD~191612} 
 	 \tablehead{ 
 	 \colhead{Ion}& 
 	 \colhead{$\lambda_0$}& 
 	 \colhead{$\delta v$}& 
 	 \colhead{$\sigma_{\delta v}$}& 
 	 \colhead{$ FWHM$}& 
 	 \colhead{$\sigma_{FW}$}& 
 	   \colhead{$ f_x$}& 
 	 \colhead{$\sigma_{f_x}$}& 
 	   \colhead{$ f_0$}& 
 	 \colhead{$\sigma_{f_0}$}\\ 
 	 & 
 	 \colhead{(\AA)} & 
 	 \colhead{($\kms$)}  & 
 	 \colhead{($\kms$)}  & 
 	 \colhead{($\kms$)}  & 
 	  \colhead{($\kms$)}  & 
 	  \multicolumn{4}{c}{($ 10^{-5} \pflux$) } 
 } 
 \startdata
       Fe XXV&       1.858&         \nodata&        \nodata&        \nodata&        \nodata&        0.23&        0.09&        0.23&  \nodata \\
  Ar XVII (i)&       3.968&        \nodata&        \nodata&        \nodata&        \nodata&         0.06&        0.02&        0.07&  \nodata \\
        S XVI&       4.730&          \nodata&        \nodata&        \nodata&        \nodata&         0.13&        0.03&        0.15&  \nodata \\
         S XV&       5.039&        \nodata&        \nodata&        \nodata&        \nodata&         0.17&        0.04&        0.19&  \nodata \\
         S XV&       5.102&        \nodata&        \nodata&        \nodata&        \nodata&        0.10&        0.03&        0.11&  \nodata \\
       Si XIV&       6.183&         -37&          39&         556&         143&        0.34&        0.03&        0.42&  \nodata \\
      Si XIII&       6.648&           7&          54&        1031&         146&        0.36&        0.04&        0.46&  \nodata \\
      Si XIII&       6.687&         \nodata&        \nodata&        \nodata&        \nodata&        0.12&        0.02&        0.15&  \nodata \\
\enddata 
 \end{deluxetable} \label{tbl:HD_191612}

\begin{deluxetable}{lrrccccccc} 
 	 \tablecaption{ Cyg~OB2-12} 
 	 \tablehead{ 
 	 \colhead{Ion}& 
 	 \colhead{$\lambda_0$}& 
 	 \colhead{$\delta v$}& 
 	 \colhead{$\sigma_{\delta v}$}& 
 	 \colhead{$ FWHM$}& 
 	 \colhead{$\sigma_{FW}$}& 
 	   \colhead{$ f_x$}& 
 	 \colhead{$\sigma_{f_x}$}& 
 	   \colhead{$ f_0$}& 
 	 \colhead{$\sigma_{f_0}$}\\ 
 	 & 
 	 \colhead{(\AA)} & 
 	 \colhead{($\kms$)}  & 
 	 \colhead{($\kms$)}  & 
 	 \colhead{($\kms$)}  & 
 	  \colhead{($\kms$)}  & 
 	  \multicolumn{4}{c}{($ 10^{-5} \pflux$) } 
 } 
 \startdata
         Fe XXV&       1.858&        \nodata&        \nodata&        \nodata&        \nodata&        0.47&        0.18&        0.48&        0.00 \\
    Ar XVII (i)&       3.968&       \nodata&        \nodata&        \nodata&        \nodata&          0.18&        0.07&        0.24&        0.00 \\
          S XVI&       4.730&        -506&         226&        1835&         666&        0.59&        0.14&        0.93&        0.01 \\
           S XV&       5.039&         115&          77&        1011&         198&        1.46&        0.17&        2.43&        0.03 \\
           S XV&       5.065&         \nodata&        \nodata&        \nodata&        \nodata&          0.23&        0.11&        0.38&        0.00 \\
           S XV&       5.102&         \nodata&        \nodata&        \nodata&        \nodata&        0.59&        0.11&        1.00&        0.01 \\
Si XIII $\beta$&       5.681&         \nodata&        \nodata&        \nodata&        \nodata&         0.34&        0.09&        0.70&        0.01 \\
         Si XIV&       6.183&         -91&          40&        1082&          97&        1.66&        0.11&        4.09&        0.09 \\
\enddata 
 \end{deluxetable} \label{tbl:Cyg_OB2-12}

\begin{deluxetable}{lrrccccccc} 
 	 \tablecaption{ Cyg~OB2-9} 
 	 \tablehead{ 
 	 \colhead{Ion}& 
 	 \colhead{$\lambda_0$}& 
 	 \colhead{$\delta v$}& 
 	 \colhead{$\sigma_{\delta v}$}& 
 	 \colhead{$ FWHM$}& 
 	 \colhead{$\sigma_{FW}$}& 
 	   \colhead{$ f_x$}& 
 	 \colhead{$\sigma_{f_x}$}& 
 	   \colhead{$ f_0$}& 
 	 \colhead{$\sigma_{f_0}$}\\ 
 	 & 
 	 \colhead{(\AA)} & 
 	 \colhead{($\kms$)}  & 
 	 \colhead{($\kms$)}  & 
 	 \colhead{($\kms$)}  & 
 	  \colhead{($\kms$)}  & 
 	  \multicolumn{4}{c}{($ 10^{-5} \pflux$) } 
 } 
 \startdata
   Ar XVII (f)&       3.994&        \nodata&           \nodata&         \nodata&       \nodata&        0.27&        0.11&        0.32&  \nodata \\
          S XV&       5.102&         \nodata&           \nodata&         \nodata&       \nodata&        0.78&        0.21&        1.07&  \nodata \\
        Si XIV&       6.183&         \nodata&           \nodata&         \nodata&       \nodata&        0.70&        0.14&        1.19&  \nodata \\
        Mg XII&       8.422&       -1048&         207&        3944&         922&        0.79&        0.15&        2.48&  \nodata \\
     Mg XI (r)&       9.169&       -1445&         209&        1702&         350&        0.30&        0.12&        1.29&  \nodata \\
     Mg XI (i)&       9.230&        \nodata&           \nodata&         \nodata&       \nodata&        0.44&        0.13&        1.91&  \nodata \\
          Ne X&       9.362&        \nodata&           \nodata&         \nodata&       \nodata&       0.21&        0.09&        0.97&  \nodata \\
Ne IX $\delta$&       9.481&        \nodata&           \nodata&         \nodata&       \nodata&      0.26&        0.10&        1.15&  \nodata \\
          Ne X&      12.135&        -287&         317&        1685&         414&        0.62&        0.23&       11.13&  \nodata \\
       Fe XVII&      12.266&         \nodata&           \nodata&         \nodata&       \nodata&      0.39&        0.18&        7.60&  \nodata \\
\enddata 
 \end{deluxetable} \label{tbl:Cyg_OB2-9}

\begin{deluxetable}{lrrccccccc} 
 	 \tablecaption{ Cyg~OB2-8A} 
 	 \tablehead{ 
 	 \colhead{Ion}& 
 	 \colhead{$\lambda_0$}& 
 	 \colhead{$\delta v$}& 
 	 \colhead{$\sigma_{\delta v}$}& 
 	 \colhead{$ FWHM$}& 
 	 \colhead{$\sigma_{FW}$}& 
 	   \colhead{$ f_x$}& 
 	 \colhead{$\sigma_{f_x}$}& 
 	   \colhead{$ f_0$}& 
 	 \colhead{$\sigma_{f_0}$}\\ 
 	 & 
 	 \colhead{(\AA)} & 
 	 \colhead{($\kms$)}  & 
 	 \colhead{($\kms$)}  & 
 	 \colhead{($\kms$)}  & 
 	  \colhead{($\kms$)}  & 
 	  \multicolumn{4}{c}{($ 10^{-5} \pflux$) } 
 } 
 \startdata
         Fe XXV&       1.858&       \nodata&         \nodata&        \nodata&        \nodata&        0.78&        0.30&        0.80&  \nodata \\
    Ar XVII (r)&       3.949&         \nodata&         \nodata&        \nodata&        \nodata&       0.25&        0.12&        0.28&  \nodata \\
    Ar XVII (i)&       3.968&        \nodata&         \nodata&        \nodata&        \nodata&         0.39&        0.14&        0.43&  \nodata \\
          S XVI&       4.730&         -96&         249&        2567&         534&        1.56&        0.27&        1.88&  \nodata \\
           S XV&       5.039&        -171&         205&        2211&         367&        2.32&        0.49&        2.87&  \nodata \\
           S XV&       5.102&        \nodata&         \nodata&        \nodata&        \nodata&        1.68&        0.34&        2.09&  \nodata \\
Si XIII $\beta$&       5.681&        -468&         319&        2712&         662&        1.20&        0.26&        1.61&  \nodata \\
         Si XIV&       6.183&         -96&          66&        1954&         144&        3.81&        0.25&        5.54&  \nodata \\
\enddata 
 \end{deluxetable} \label{tbl:Cyg_OB2-8A}

\begin{deluxetable}{lrrccccccc} 
 	 \tablecaption{  HD~206267} 
 	 \tablehead{ 
 	 \colhead{Ion}& 
 	 \colhead{$\lambda_0$}& 
 	 \colhead{$\delta v$}& 
 	 \colhead{$\sigma_{\delta v}$}& 
 	 \colhead{$ FWHM$}& 
 	 \colhead{$\sigma_{FW}$}& 
 	   \colhead{$ f_x$}& 
 	 \colhead{$\sigma_{f_x}$}& 
 	   \colhead{$ f_0$}& 
 	 \colhead{$\sigma_{f_0}$}\\ 
 	 & 
 	 \colhead{(\AA)} & 
 	 \colhead{($\kms$)}  & 
 	 \colhead{($\kms$)}  & 
 	 \colhead{($\kms$)}  & 
 	  \colhead{($\kms$)}  & 
 	  \multicolumn{4}{c}{($ 10^{-5} \pflux$) } 
 } 
 \startdata
        Si XIV&       6.183&         \nodata&         \nodata&         \nodata&           \nodata&        0.02&        0.01&        0.03&  \nodata \\
       Si XIII&       6.648&        -130&         171&        1996&         350&        0.33&        0.06&        0.42&  \nodata \\
       Si XIII&       6.687&       \nodata&         \nodata&         \nodata&           \nodata&       0.12&        0.05&        0.15&  \nodata \\
       Si XIII&       6.740&        \nodata&         \nodata&         \nodata&           \nodata&       0.12&        0.03&        0.15&  \nodata \\
        Mg XII&       8.422&        -583&         309&        2778&         644&        0.17&        0.03&        0.25&  \nodata \\
     Mg XI (r)&       9.169&        -300&         140&        1997&         259&        0.75&        0.10&        1.26&  \nodata \\
     Mg XI (i)&       9.230&        \nodata&         \nodata&         \nodata&           \nodata&       0.48&        0.08&        0.82&  \nodata \\
     Mg XI (f)&       9.314&        \nodata&         \nodata&         \nodata&           \nodata&       0.18&        0.05&        0.30&  \nodata \\
\enddata 
 \end{deluxetable} \label{tbl:HD_206267}

\begin{deluxetable}{lrrccccccc} 
 	 \tablecaption{ $\gamma\,$Cas} 
 	 \tablehead{ 
 	 \colhead{Ion}& 
 	 \colhead{$\lambda_0$}& 
 	 \colhead{$\delta v$}& 
 	 \colhead{$\sigma_{\delta v}$}& 
 	 \colhead{$ FWHM$}& 
 	 \colhead{$\sigma_{FW}$}& 
 	   \colhead{$ f_x$}& 
 	 \colhead{$\sigma_{f_x}$}& 
 	   \colhead{$ f_0$}& 
 	 \colhead{$\sigma_{f_0}$}\\ 
 	 & 
 	 \colhead{(\AA)} & 
 	 \colhead{($\kms$)}  & 
 	 \colhead{($\kms$)}  & 
 	 \colhead{($\kms$)}  & 
 	  \colhead{($\kms$)}  & 
 	  \multicolumn{4}{c}{($ 10^{-5} \pflux$) } 
 } 
 \startdata
         S XVI&       4.730&          -9&         101&        1408&         236&        8.83&        1.03&        8.86&        0.01 \\
          S XV&       5.039&        1427&         386&        2356&         725&        3.74&        1.19&        3.75&        0.00 \\
        Si XIV&       6.183&         107&          50&        1176&         158&        8.99&        0.67&        9.04&        0.01 \\
       Si XIII&       6.648&          70&         106&         801&         244&        2.29&        0.47&        2.31&        0.00 \\
       Si XIII&       6.687&          \nodata&           \nodata&         \nodata&        \nodata&        0.99&        0.42&        1.00&        0.00 \\
       Si XIII&       6.740&          \nodata&           \nodata&         \nodata&        \nodata&        0.89&        0.39&        0.90&        0.00 \\
        Mg XII&       8.422&          40&          78&        1158&         228&        5.98&        0.75&        6.06&        0.02 \\
 Ne X $\gamma$&       9.708&        \nodata&           \nodata&         \nodata&        \nodata&         1.26&        0.40&        1.28&        0.01 \\
\enddata 
 \end{deluxetable} \label{tbl:gamma_Cas}

\begin{deluxetable}{lrrccccccc} 
 	 \tablecaption{ $\xi\,$Per} 
 	 \tablehead{ 
 	 \colhead{Ion}& 
 	 \colhead{$\lambda_0$}& 
 	 \colhead{$\delta v$}& 
 	 \colhead{$\sigma_{\delta v}$}& 
 	 \colhead{$ FWHM$}& 
 	 \colhead{$\sigma_{FW}$}& 
 	   \colhead{$ f_x$}& 
 	 \colhead{$\sigma_{f_x}$}& 
 	   \colhead{$ f_0$}& 
 	 \colhead{$\sigma_{f_0}$}\\ 
 	 & 
 	 \colhead{(\AA)} & 
 	 \colhead{($\kms$)}  & 
 	 \colhead{($\kms$)}  & 
 	 \colhead{($\kms$)}  & 
 	  \colhead{($\kms$)}  & 
 	  \multicolumn{4}{c}{($ 10^{-5} \pflux$) } 
 } 
 \startdata
         Si XIV&       6.183&         \nodata&         \nodata&         \nodata&            \nodata&        0.07&        0.03&        0.08&        0.00 \\
        Si XIII&       6.648&         108&         118&        1701&         203&        0.49&        0.07&        0.52&        0.01 \\
        Si XIII&       6.687&         \nodata&         \nodata&         \nodata&            \nodata&         0.26&        0.06&        0.27&        0.00 \\
        Si XIII&       6.740&          \nodata&         \nodata&         \nodata&            \nodata&        0.28&        0.04&        0.30&        0.00 \\
         Mg XII&       8.422&          35&         199&        1640&         357&        0.22&        0.05&        0.25&        0.00 \\
      Mg XI (r)&       9.169&        -377&         105&        1721&         160&        1.17&        0.12&        1.33&        0.03 \\
      Mg XI (i)&       9.230&         \nodata&         \nodata&         \nodata&            \nodata&         1.04&        0.11&        1.19&        0.03 \\
      Mg XI (f)&       9.314&        \nodata&         \nodata&         \nodata&            \nodata&        0.25&        0.12&        0.29&        0.01 \\
\enddata 
 \end{deluxetable} \label{tbl:xi_Per}

\begin{deluxetable}{lrrccccccc} 
 	 \tablecaption{ $15\,$Mon} 
 	 \tablehead{ 
 	 \colhead{Ion}& 
 	 \colhead{$\lambda_0$}& 
 	 \colhead{$\delta v$}& 
 	 \colhead{$\sigma_{\delta v}$}& 
 	 \colhead{$ FWHM$}& 
 	 \colhead{$\sigma_{FW}$}& 
 	   \colhead{$ f_x$}& 
 	 \colhead{$\sigma_{f_x}$}& 
 	   \colhead{$ f_0$}& 
 	 \colhead{$\sigma_{f_0}$}\\ 
 	 & 
 	 \colhead{(\AA)} & 
 	 \colhead{($\kms$)}  & 
 	 \colhead{($\kms$)}  & 
 	 \colhead{($\kms$)}  & 
 	  \colhead{($\kms$)}  & 
 	  \multicolumn{4}{c}{($ 10^{-5} \pflux$) } 
 } 
 \startdata
        Si XIII&       6.648&         \nodata&         \nodata&         \nodata&         \nodata&        0.10&        0.03&        0.10&        0.00 \\
      Mg XI (r)&       9.169&         -50&         169&        1415&         284&        0.30&        0.07&        0.31&        0.00 \\
      Mg XI (i)&       9.230&        \nodata&         \nodata&         \nodata&         \nodata&        0.17&        0.06&        0.17&        0.00 \\
           Ne X&      12.135&        -284&         191&        2451&         413&        1.47&        0.21&        1.55&        0.02 \\
        Fe XVII&      12.266&        \nodata&         \nodata&         \nodata&         \nodata&        0.41&        0.14&        0.43&        0.01 \\
      Ne IX (r)&      13.447&          49&         178&        2144&         258&        3.81&        0.75&        4.07&        0.06 \\
      Ne IX (i)&      13.552&         \nodata&         \nodata&         \nodata&         \nodata&       2.32&        0.57&        2.48&        0.04 \\
        Fe XVII&      15.014&         -48&         118&        2009&         290&        5.36&        0.74&        5.78&        0.10 \\
\enddata 
 \end{deluxetable} \label{tbl:15_Mon}

\begin{deluxetable}{lrrccccccc} 
 	 \tablecaption{ $\delta\,$Ori} 
 	 \tablehead{ 
 	 \colhead{Ion}& 
 	 \colhead{$\lambda_0$}& 
 	 \colhead{$\delta v$}& 
 	 \colhead{$\sigma_{\delta v}$}& 
 	 \colhead{$ FWHM$}& 
 	 \colhead{$\sigma_{FW}$}& 
 	   \colhead{$ f_x$}& 
 	 \colhead{$\sigma_{f_x}$}& 
 	   \colhead{$ f_0$}& 
 	 \colhead{$\sigma_{f_0}$}\\ 
 	 & 
 	 \colhead{(\AA)} & 
 	 \colhead{($\kms$)}  & 
 	 \colhead{($\kms$)}  & 
 	 \colhead{($\kms$)}  & 
 	  \colhead{($\kms$)}  & 
 	  \multicolumn{4}{c}{($ 10^{-5} \pflux$) } 
 } 
 \startdata
         Fe XXV&       1.858&        \nodata&        \nodata&        \nodata&           \nodata&        0.15&        0.04&        0.15&        0.00 \\
          S XVI&       4.730&        \nodata&        \nodata&        \nodata&           \nodata&        0.09&        0.03&        0.09&        0.00 \\
           S XV&       5.039&        \nodata&        \nodata&        \nodata&           \nodata&        0.36&        0.13&        0.36&        0.00 \\
           S XV&       5.102&       \nodata&        \nodata&        \nodata&           \nodata&        0.26&        0.04&        0.26&        0.00 \\
Si XIII $\beta$&       5.681&          84&         144&        1334&         393&        0.17&        0.03&        0.17&        0.00 \\
         Si XIV&       6.183&         -75&          50&        1191&         143&        0.38&        0.03&        0.38&        0.00 \\
        Si XIII&       6.648&           5&          20&        1169&          43&        1.38&        0.05&        1.39&        0.00 \\
        Si XIII&       6.687&           \nodata&        \nodata&        \nodata&           \nodata&       0.45&        0.03&        0.45&        0.00 \\
\enddata 
 \end{deluxetable} \label{tbl:delta_Ori}

\begin{deluxetable}{lrrccccccc} 
 	 \tablecaption{ $\epsilon\,$Ori} 
 	 \tablehead{ 
 	 \colhead{Ion}& 
 	 \colhead{$\lambda_0$}& 
 	 \colhead{$\delta v$}& 
 	 \colhead{$\sigma_{\delta v}$}& 
 	 \colhead{$ FWHM$}& 
 	 \colhead{$\sigma_{FW}$}& 
 	   \colhead{$ f_x$}& 
 	 \colhead{$\sigma_{f_x}$}& 
 	   \colhead{$ f_0$}& 
 	 \colhead{$\sigma_{f_0}$}\\ 
 	 & 
 	 \colhead{(\AA)} & 
 	 \colhead{($\kms$)}  & 
 	 \colhead{($\kms$)}  & 
 	 \colhead{($\kms$)}  & 
 	  \colhead{($\kms$)}  & 
 	  \multicolumn{4}{c}{($ 10^{-5} \pflux$) } 
 } 
 \startdata
           S XV&       5.039&        \nodata&        \nodata&        \nodata&           \nodata&        0.15&        0.07&        0.15&        0.00 \\
        Si XIII&       6.648&          46&          62&        1072&         137&        0.78&        0.09&        0.80&        0.00 \\
        Si XIII&       6.687&          \nodata&        \nodata&        \nodata&           \nodata&          0.21&        0.06&        0.21&        0.00 \\
        Si XIII&       6.740&          \nodata&        \nodata&        \nodata&           \nodata&         0.42&        0.06&        0.42&        0.00 \\
         Mg XII&       8.422&        -146&          96&        1063&         233&        0.41&        0.06&        0.42&        0.00 \\
      Mg XI (r)&       9.169&         -61&          60&        1364&         102&        2.15&        0.17&        2.23&        0.02 \\
      Mg XI (i)&       9.230&         \nodata&        \nodata&        \nodata&           \nodata&         0.73&        0.12&        0.76&        0.01 \\
           Ne X&       9.291&          \nodata&        \nodata&        \nodata&           \nodata&         0.35&        0.17&        0.37&        0.00 \\
\enddata 
 \end{deluxetable} \label{tbl:eps_Ori}

\begin{deluxetable}{lrrccccccc} 
 	 \tablecaption{   V640 Mon} 
 	 \tablehead{ 
 	 \colhead{Ion}& 
 	 \colhead{$\lambda_0$}& 
 	 \colhead{$\delta v$}& 
 	 \colhead{$\sigma_{\delta v}$}& 
 	 \colhead{$ FWHM$}& 
 	 \colhead{$\sigma_{FW}$}& 
 	   \colhead{$ f_x$}& 
 	 \colhead{$\sigma_{f_x}$}& 
 	   \colhead{$ f_0$}& 
 	 \colhead{$\sigma_{f_0}$}\\ 
 	 & 
 	 \colhead{(\AA)} & 
 	 \colhead{($\kms$)}  & 
 	 \colhead{($\kms$)}  & 
 	 \colhead{($\kms$)}  & 
 	  \colhead{($\kms$)}  & 
 	  \multicolumn{4}{c}{($ 10^{-5} \pflux$) } 
 } 
 \startdata
         Fe XXV&       1.858&        \nodata&         \nodata&        \nodata&        \nodata&        0.44&        0.10&        0.44&        0.00 \\
          S XVI&       4.730&        -226&         302&        2536&         914&        0.37&        0.08&        0.39&        0.00 \\
           S XV&       5.039&           7&         123&        1307&         242&        0.49&        0.09&        0.51&        0.00 \\
           S XV&       5.102&            \nodata&         \nodata&        \nodata&        \nodata&       0.33&        0.07&        0.35&        0.00 \\
Si XIII $\beta$&       5.681&          \nodata&         \nodata&        \nodata&        \nodata&        0.20&        0.06&        0.21&        0.00 \\
         Si XIV&       6.183&          73&          52&        1537&         119&        0.98&        0.06&        1.04&        0.02 \\
        Si XIII&       6.648&         -30&          56&        1711&         124&        1.23&        0.08&        1.33&        0.03 \\
        Si XIII&       6.687&          \nodata&         \nodata&        \nodata&        \nodata&        0.34&        0.07&        0.37&        0.01 \\
\enddata 
 \end{deluxetable} \label{tbl:V640_Mon}

\begin{deluxetable}{lrrccccccc} 
 	 \tablecaption{ $\zeta\,$Ori~A} 
 	 \tablehead{ 
 	 \colhead{Ion}& 
 	 \colhead{$\lambda_0$}& 
 	 \colhead{$\delta v$}& 
 	 \colhead{$\sigma_{\delta v}$}& 
 	 \colhead{$ FWHM$}& 
 	 \colhead{$\sigma_{FW}$}& 
 	   \colhead{$ f_x$}& 
 	 \colhead{$\sigma_{f_x}$}& 
 	   \colhead{$ f_0$}& 
 	 \colhead{$\sigma_{f_0}$}\\ 
 	 & 
 	 \colhead{(\AA)} & 
 	 \colhead{($\kms$)}  & 
 	 \colhead{($\kms$)}  & 
 	 \colhead{($\kms$)}  & 
 	  \colhead{($\kms$)}  & 
 	  \multicolumn{4}{c}{($ 10^{-5} \pflux$) } 
 } 
 \startdata
          S XVI&       4.730&        \nodata&         \nodata&        \nodata&       \nodata&       0.11&        0.04&        0.12&        0.00 \\
           S XV&       5.039&         133&         139&        1373&         238&        0.30&        0.05&        0.30&        0.00 \\
           S XV&       5.065&         \nodata&         \nodata&        \nodata&        \nodata&        0.13&        0.05&        0.13&        0.00 \\
           S XV&       5.102&         \nodata&         \nodata&        \nodata&        \nodata&       0.22&        0.05&        0.22&        0.00 \\
        Si XIII&       5.405&         \nodata&         \nodata&        \nodata&        \nodata&        0.12&        0.04&        0.12&        0.00 \\
         Si XIV&       6.183&        -223&          99&        1490&         242&        0.26&        0.03&        0.26&        0.00 \\
        Si XIII&       6.648&         -42&          27&        1457&          58&        1.62&        0.07&        1.64&        0.00 \\
        Si XIII&       6.687&           \nodata&         \nodata&        \nodata&        \nodata&        0.53&        0.05&        0.53&        0.00 \\
\enddata 
 \end{deluxetable} \label{tbl:zeta_Ori}

\begin{deluxetable}{lrrccccccc} 
 	 \tablecaption{ $\zeta\,$Ori~B } 
 	 \tablehead{ 
 	 \colhead{Ion}& 
 	 \colhead{$\lambda_0$}& 
 	 \colhead{$\delta v$}& 
 	 \colhead{$\sigma_{\delta v}$}& 
 	 \colhead{$ FWHM$}& 
 	 \colhead{$\sigma_{FW}$}& 
 	   \colhead{$ f_x$}& 
 	 \colhead{$\sigma_{f_x}$}& 
 	   \colhead{$ f_0$}& 
 	 \colhead{$\sigma_{f_0}$}\\ 
 	 & 
 	 \colhead{(\AA)} & 
 	 \colhead{($\kms$)}  & 
 	 \colhead{($\kms$)}  & 
 	 \colhead{($\kms$)}  & 
 	  \colhead{($\kms$)}  & 
 	  \multicolumn{4}{c}{($ 10^{-5} \pflux$) } 
 } 
 \startdata
       Si XIII&       6.648&          -9&         246&        1511&         534&        0.09&        0.03&        0.09&        0.00 \\
       Si XIII&       6.687&         \nodata&           \nodata&        \nodata&        \nodata&        0.05&        0.02&        0.05&        0.00 \\
       Si XIII&       6.740&          \nodata&           \nodata&        \nodata&        \nodata&        0.06&        0.02&        0.06&        0.00 \\
        Mg XII&       8.422&        -106&         154&         988&         325&        0.08&        0.02&        0.08&        0.00 \\
     Mg XI (r)&       9.169&         112&          76&         891&         155&        0.27&        0.04&        0.28&        0.00 \\
     Mg XI (i)&       9.230&         \nodata&           \nodata&        \nodata&        \nodata&        0.14&        0.03&        0.14&        0.00 \\
 Ne X $\gamma$&       9.708&         \nodata&           \nodata&        \nodata&        \nodata&       0.06&        0.03&        0.06&        0.00 \\
  Ne X $\beta$&      10.239&         -67&          78&          69&         107&        0.07&        0.03&        0.07&        0.00 \\
\enddata 
 \end{deluxetable} \label{tbl:zeta_Ori_B}

\begin{deluxetable}{lrrccccccc} 
 	 \tablecaption{ $\sigma\,$Ori~Aa} 
 	 \tablehead{ 
 	 \colhead{Ion}& 
 	 \colhead{$\lambda_0$}& 
 	 \colhead{$\delta v$}& 
 	 \colhead{$\sigma_{\delta v}$}& 
 	 \colhead{$ FWHM$}& 
 	 \colhead{$\sigma_{FW}$}& 
 	   \colhead{$ f_x$}& 
 	 \colhead{$\sigma_{f_x}$}& 
 	   \colhead{$ f_0$}& 
 	 \colhead{$\sigma_{f_0}$}\\ 
 	 & 
 	 \colhead{(\AA)} & 
 	 \colhead{($\kms$)}  & 
 	 \colhead{($\kms$)}  & 
 	 \colhead{($\kms$)}  & 
 	  \colhead{($\kms$)}  & 
 	  \multicolumn{4}{c}{($ 10^{-5} \pflux$) } 
 } 
 \startdata
        Si XIII&       6.648&         -49&          85&         592&         229&        0.13&        0.04&        0.14&        0.00 \\
        Si XIII&       6.687&         \nodata&           \nodata&         \nodata&        \nodata&        0.06&        0.03&        0.07&        0.00 \\
        Si XIII&       6.740&         \nodata&           \nodata&         \nodata&        \nodata&        0.07&        0.03&        0.07&        0.00 \\
      Mg XI (r)&       9.169&         -56&          47&         612&         122&        0.43&        0.07&        0.45&        0.00 \\
      Mg XI (i)&       9.230&        \nodata&           \nodata&         \nodata&        \nodata&         0.38&        0.06&        0.40&        0.00 \\
  Ne X $\gamma$&       9.708&         \nodata&           \nodata&         \nodata&        \nodata&         0.06&        0.00&        0.06&        0.00 \\
   Ne X $\beta$&      10.239&        -152&          70&          74&         139&        0.20&        0.06&        0.21&        0.00 \\
           Ne X&      12.135&        -114&          32&         567&          90&        2.32&        0.24&        2.52&        0.03 \\
\enddata 
 \end{deluxetable} \label{tbl:sig_Ori}

\begin{deluxetable}{lrrccccccc} 
 	 \tablecaption{ $\theta^1\,$Ori~C} 
 	 \tablehead{ 
 	 \colhead{Ion}& 
 	 \colhead{$\lambda_0$}& 
 	 \colhead{$\delta v$}& 
 	 \colhead{$\sigma_{\delta v}$}& 
 	 \colhead{$ FWHM$}& 
 	 \colhead{$\sigma_{FW}$}& 
 	   \colhead{$ f_x$}& 
 	 \colhead{$\sigma_{f_x}$}& 
 	   \colhead{$ f_0$}& 
 	 \colhead{$\sigma_{f_0}$}\\ 
 	 & 
 	 \colhead{(\AA)} & 
 	 \colhead{($\kms$)}  & 
 	 \colhead{($\kms$)}  & 
 	 \colhead{($\kms$)}  & 
 	  \colhead{($\kms$)}  & 
 	  \multicolumn{4}{c}{($ 10^{-5} \pflux$) } 
 } 
 \startdata
         Fe XXV&       1.858&         100&          43&        3466&         104&        8.13&        0.18&        8.17&        0.01 \\
    Ar XVII (r)&       3.949&         -67&          24&         879&          72&        1.48&        0.07&        1.54&        0.02 \\
    Ar XVII (i)&       3.968&         \nodata&           \nodata&         \nodata&       \nodata&        0.60&        0.06&        0.62&        0.01 \\
    Ar XVII (f)&       3.994&         \nodata&           \nodata&         \nodata&       \nodata&        1.71&        0.07&        1.79&        0.02 \\
          S XVI&       4.730&         -29&          12&         917&          38&        6.08&        0.11&        6.56&        0.13 \\
           S XV&       5.039&          27&          12&         805&          33&        5.19&        0.11&        5.66&        0.13 \\
           S XV&       5.065&         \nodata&           \nodata&         \nodata&       \nodata&        1.41&        0.08&        1.54&        0.03 \\
           S XV&       5.102&         \nodata&           \nodata&         \nodata&       \nodata&        2.40&        0.09&        2.63&        0.06 \\
\enddata 
 \end{deluxetable} \label{tbl:tet01_Ori_C}

\begin{deluxetable}{lrrccccccc} 
 	 \tablecaption{  $\iota\,$Ori} 
 	 \tablehead{ 
 	 \colhead{Ion}& 
 	 \colhead{$\lambda_0$}& 
 	 \colhead{$\delta v$}& 
 	 \colhead{$\sigma_{\delta v}$}& 
 	 \colhead{$ FWHM$}& 
 	 \colhead{$\sigma_{FW}$}& 
 	   \colhead{$ f_x$}& 
 	 \colhead{$\sigma_{f_x}$}& 
 	   \colhead{$ f_0$}& 
 	 \colhead{$\sigma_{f_0}$}\\ 
 	 & 
 	 \colhead{(\AA)} & 
 	 \colhead{($\kms$)}  & 
 	 \colhead{($\kms$)}  & 
 	 \colhead{($\kms$)}  & 
 	  \colhead{($\kms$)}  & 
 	  \multicolumn{4}{c}{($ 10^{-5} \pflux$) } 
 } 
 \startdata
         Si XIV&       6.183&        \nodata&        \nodata&        \nodata&           \nodata&        0.28&        0.07&        0.28&        0.00 \\
        Si XIII&       6.648&         146&         147&        2035&         277&        0.70&        0.13&        0.71&        0.00 \\
        Si XIII&       6.687&         \nodata&        \nodata&        \nodata&           \nodata&        0.30&        0.12&        0.30&        0.00 \\
        Si XIII&       6.740&         \nodata&        \nodata&        \nodata&           \nodata&       0.75&        0.12&        0.76&        0.00 \\
         Mg XII&       8.422&        -216&         336&        3016&        1059&        0.86&        0.17&        0.88&        0.00 \\
  Ne X $\gamma$&       9.708&         \nodata&        \nodata&        \nodata&           \nodata&       0.41&        0.12&        0.42&        0.00 \\
   Ne X $\beta$&      10.239&         149&         180&        1276&         480&        0.79&        0.24&        0.81&        0.00 \\
           Ne X&      12.135&         -57&          91&        2364&         161&        9.50&        0.69&        9.87&        0.09 \\
\enddata 
 \end{deluxetable} \label{tbl:iota_Ori}

\begin{deluxetable}{lrrccccccc} 
 	 \tablecaption{  $\kappa\,$Ori} 
 	 \tablehead{ 
 	 \colhead{Ion}& 
 	 \colhead{$\lambda_0$}& 
 	 \colhead{$\delta v$}& 
 	 \colhead{$\sigma_{\delta v}$}& 
 	 \colhead{$ FWHM$}& 
 	 \colhead{$\sigma_{FW}$}& 
 	   \colhead{$ f_x$}& 
 	 \colhead{$\sigma_{f_x}$}& 
 	   \colhead{$ f_0$}& 
 	 \colhead{$\sigma_{f_0}$}\\ 
 	 & 
 	 \colhead{(\AA)} & 
 	 \colhead{($\kms$)}  & 
 	 \colhead{($\kms$)}  & 
 	 \colhead{($\kms$)}  & 
 	  \colhead{($\kms$)}  & 
 	  \multicolumn{4}{c}{($ 10^{-5} \pflux$) } 
 } 
 \startdata
           S XV&       5.039&        \nodata&           \nodata&          \nodata&          \nodata&        0.07&        0.03&        0.07&        0.00 \\
         Si XIV&       6.183&        \nodata&           \nodata&          \nodata&          \nodata&       0.06&        0.02&        0.06&        0.00 \\
        Si XIII&       6.648&         125&          65&        1012&         145&        0.26&        0.03&        0.26&        0.00 \\
        Si XIII&       6.687&         \nodata&           \nodata&          \nodata&          \nodata&        0.06&        0.02&        0.06&        0.00 \\
        Si XIII&       6.740&         \nodata&           \nodata&          \nodata&          \nodata&       0.12&        0.02&        0.12&        0.00 \\
         Mg XII&       8.422&        -100&         146&         966&         342&        0.09&        0.02&        0.09&        0.00 \\
      Mg XI (r)&       9.169&         -31&          62&        1165&         119&        0.58&        0.06&        0.61&        0.01 \\
      Mg XI (i)&       9.230&       \nodata&           \nodata&          \nodata&          \nodata&        0.32&        0.05&        0.34&        0.00 \\
\enddata 
 \end{deluxetable} \label{tbl:kappa_Ori}

\begin{deluxetable}{lrrccccccc} 
 	 \tablecaption{  WR~6} 
 	 \tablehead{ 
 	 \colhead{Ion}& 
 	 \colhead{$\lambda_0$}& 
 	 \colhead{$\delta v$}& 
 	 \colhead{$\sigma_{\delta v}$}& 
 	 \colhead{$ FWHM$}& 
 	 \colhead{$\sigma_{FW}$}& 
 	   \colhead{$ f_x$}& 
 	 \colhead{$\sigma_{f_x}$}& 
 	   \colhead{$ f_0$}& 
 	 \colhead{$\sigma_{f_0}$}\\ 
 	 & 
 	 \colhead{(\AA)} & 
 	 \colhead{($\kms$)}  & 
 	 \colhead{($\kms$)}  & 
 	 \colhead{($\kms$)}  & 
 	  \colhead{($\kms$)}  & 
 	  \multicolumn{4}{c}{($ 10^{-5} \pflux$) } 
 } 
 \startdata
        Fe XXV&       1.858&         \nodata&        \nodata&        \nodata&           \nodata&        0.14&        0.05&        0.14&  \nodata \\
   Ar XVII (r)&       3.949&         \nodata&        \nodata&        \nodata&           \nodata&        0.12&        0.03&        0.12&  \nodata \\
         S XVI&       4.730&         \nodata&        \nodata&        \nodata&           \nodata&       0.16&        0.05&        0.16&  \nodata \\
        Si XIV&       6.183&        -324&         161&        2744&         385&        0.29&        0.03&        0.32&  \nodata \\
       Si XIII&       6.648&        -477&          73&        2793&         121&        1.02&        0.07&        1.12&  \nodata \\
       Si XIII&       6.687&         \nodata&        \nodata&        \nodata&           \nodata&        0.33&        0.07&        0.37&  \nodata \\
       Si XIII&       6.740&         \nodata&        \nodata&        \nodata&           \nodata&       0.91&        0.06&        0.99&  \nodata \\
        Mg XII&       7.106&         154&         298&        3513&         604&        0.18&        0.03&        0.20&  \nodata \\
\enddata 
 \end{deluxetable} \label{tbl:WR_6}

\begin{deluxetable}{lrrccccccc} 
 	 \tablecaption{ $\tau\,$CMa} 
 	 \tablehead{ 
 	 \colhead{Ion}& 
 	 \colhead{$\lambda_0$}& 
 	 \colhead{$\delta v$}& 
 	 \colhead{$\sigma_{\delta v}$}& 
 	 \colhead{$ FWHM$}& 
 	 \colhead{$\sigma_{FW}$}& 
 	   \colhead{$ f_x$}& 
 	 \colhead{$\sigma_{f_x}$}& 
 	   \colhead{$ f_0$}& 
 	 \colhead{$\sigma_{f_0}$}\\ 
 	 & 
 	 \colhead{(\AA)} & 
 	 \colhead{($\kms$)}  & 
 	 \colhead{($\kms$)}  & 
 	 \colhead{($\kms$)}  & 
 	  \colhead{($\kms$)}  & 
 	  \multicolumn{4}{c}{($ 10^{-5} \pflux$) } 
 } 
 \startdata
        Si XIII&       6.648&         609&         136&        1434&         259&        0.11&        0.02&        0.11&        0.00 \\
        Si XIII&       6.740&        \nodata&         \nodata&        \nodata&        \nodata&        0.09&        0.02&        0.09&        0.00 \\
         Mg XII&       8.422&        -475&         125&        1102&         289&        0.09&        0.02&        0.10&        0.00 \\
      Mg XI (r)&       9.169&        -429&         146&        1670&         228&        0.29&        0.04&        0.31&        0.00 \\
      Mg XI (i)&       9.230&         \nodata&         \nodata&        \nodata&        \nodata&        0.25&        0.04&        0.27&        0.00 \\
           Ne X&       9.362&         \nodata&         \nodata&        \nodata&        \nodata&       0.07&        0.03&        0.07&        0.00 \\
 Ne IX $\delta$&       9.481&        \nodata&         \nodata&        \nodata&        \nodata&        0.07&        0.02&        0.07&        0.00 \\
   Ne X $\beta$&      10.239&         -56&         230&        1830&         559&        0.16&        0.05&        0.17&        0.00 \\
\enddata 
 \end{deluxetable} \label{tbl:tau_CMa}

\begin{deluxetable}{lrrccccccc} 
 	 \tablecaption{   HD~42054} 
 	 \tablehead{ 
 	 \colhead{Ion}& 
 	 \colhead{$\lambda_0$}& 
 	 \colhead{$\delta v$}& 
 	 \colhead{$\sigma_{\delta v}$}& 
 	 \colhead{$ FWHM$}& 
 	 \colhead{$\sigma_{FW}$}& 
 	   \colhead{$ f_x$}& 
 	 \colhead{$\sigma_{f_x}$}& 
 	   \colhead{$ f_0$}& 
 	 \colhead{$\sigma_{f_0}$}\\ 
 	 & 
 	 \colhead{(\AA)} & 
 	 \colhead{($\kms$)}  & 
 	 \colhead{($\kms$)}  & 
 	 \colhead{($\kms$)}  & 
 	  \colhead{($\kms$)}  & 
 	  \multicolumn{4}{c}{($ 10^{-5} \pflux$) } 
 } 
 \startdata
      Fe XXV&       1.858&         \nodata&         \nodata&        \nodata&        \nodata&        1.21&        0.23&        1.21&  \nodata \\
 Ar XVII (f)&       3.994&        \nodata&         \nodata&        \nodata&        \nodata&         0.12&        0.05&        0.12&  \nodata \\
       S XVI&       4.730&        \nodata&         \nodata&        \nodata&        \nodata&         0.24&        0.11&        0.24&  \nodata \\
      Si XIV&       6.183&         159&          62&         743&         181&        0.53&        0.06&        0.53&  \nodata \\
     Si XIII&       6.648&         \nodata&         \nodata&        \nodata&        \nodata&         0.17&        0.05&        0.17&  \nodata \\
      Mg XII&       8.422&          85&          61&         503&         148&        0.29&        0.05&        0.29&  \nodata \\
Ne X $\beta$&      10.239&         246&         154&         582&         312&        0.14&        0.06&        0.14&  \nodata \\
        Ne X&      12.135&         131&         159&        1684&         557&        1.02&        0.20&        1.02&  \nodata \\
\enddata 
 \end{deluxetable} \label{tbl:HD_42054}

\begin{deluxetable}{lrrccccccc} 
 	 \tablecaption{  $\zeta\,$Pup} 
 	 \tablehead{ 
 	 \colhead{Ion}& 
 	 \colhead{$\lambda_0$}& 
 	 \colhead{$\delta v$}& 
 	 \colhead{$\sigma_{\delta v}$}& 
 	 \colhead{$ FWHM$}& 
 	 \colhead{$\sigma_{FW}$}& 
 	   \colhead{$ f_x$}& 
 	 \colhead{$\sigma_{f_x}$}& 
 	   \colhead{$ f_0$}& 
 	 \colhead{$\sigma_{f_0}$}\\ 
 	 & 
 	 \colhead{(\AA)} & 
 	 \colhead{($\kms$)}  & 
 	 \colhead{($\kms$)}  & 
 	 \colhead{($\kms$)}  & 
 	  \colhead{($\kms$)}  & 
 	  \multicolumn{4}{c}{($ 10^{-5} \pflux$) } 
 } 
 \startdata
    Ar XVII (r)&       3.949&        -129&         199&        1558&         274&        0.09&        0.03&        0.09&        0.00 \\
    Ar XVII (i)&       3.968&        \nodata&           \nodata&        \nodata&         \nodata&        0.07&        0.03&        0.07&        0.00 \\
    Ar XVII (f)&       3.994&         \nodata&           \nodata&        \nodata&         \nodata&       0.13&        0.03&        0.13&        0.00 \\
           S XV&       4.088&        \nodata&           \nodata&        \nodata&         \nodata&       0.06&        0.02&        0.06&        0.00 \\
           S XV&       5.039&        -184&          43&        1672&         115&        1.60&        0.10&        1.60&        0.00 \\
           S XV&       5.065&         \nodata&           \nodata&        \nodata&         \nodata&       0.67&        0.09&        0.67&        0.00 \\
           S XV&       5.102&         \nodata&           \nodata&        \nodata&         \nodata&       0.97&        0.07&        0.97&        0.00 \\
Si XIII $\beta$&       5.681&        -369&          45&        1720&         155&        0.88&        0.06&        0.88&        0.00 \\
\enddata 
 \end{deluxetable} \label{tbl:zeta_Pup}

\begin{deluxetable}{lrrccccccc} 
 	 \tablecaption{ $\gamma^2\,$Vel} 
 	 \tablehead{ 
 	 \colhead{Ion}& 
 	 \colhead{$\lambda_0$}& 
 	 \colhead{$\delta v$}& 
 	 \colhead{$\sigma_{\delta v}$}& 
 	 \colhead{$ FWHM$}& 
 	 \colhead{$\sigma_{FW}$}& 
 	   \colhead{$ f_x$}& 
 	 \colhead{$\sigma_{f_x}$}& 
 	   \colhead{$ f_0$}& 
 	 \colhead{$\sigma_{f_0}$}\\ 
 	 & 
 	 \colhead{(\AA)} & 
 	 \colhead{($\kms$)}  & 
 	 \colhead{($\kms$)}  & 
 	 \colhead{($\kms$)}  & 
 	  \colhead{($\kms$)}  & 
 	  \multicolumn{4}{c}{($ 10^{-5} \pflux$) } 
 } 
 \startdata
         Fe XXV&       1.858&         673&         385&        3367&         963&        2.35&        0.48&        2.35&        0.00 \\
    Ar XVII (r)&       3.949&         \nodata&          \nodata&         \nodata&         \nodata&        1.27&        0.26&        1.28&        0.00 \\
    Ar XVII (i)&       3.968&          \nodata&          \nodata&         \nodata&         \nodata&        0.67&        0.23&        0.67&        0.00 \\
    Ar XVII (f)&       3.994&          \nodata&          \nodata&         \nodata&         \nodata&       1.19&        0.27&        1.19&        0.00 \\
          S XVI&       4.730&          82&         107&        1433&         305&        3.37&        0.42&        3.37&        0.00 \\
           S XV&       5.039&         \nodata&          \nodata&         \nodata&         \nodata&        8.90&        0.73&        8.91&        0.00 \\
           S XV&       5.065&           \nodata&          \nodata&         \nodata&         \nodata&       3.04&        0.60&        3.04&        0.00 \\
           S XV&       5.102&           \nodata&          \nodata&         \nodata&         \nodata&       4.86&        0.49&        4.87&        0.00 \\
\enddata 
 \end{deluxetable} \label{tbl:gamma_Vel}

\begin{deluxetable}{lrrccccccc} 
 	 \tablecaption{   RCW 38 IRS 2} 
 	 \tablehead{ 
 	 \colhead{Ion}& 
 	 \colhead{$\lambda_0$}& 
 	 \colhead{$\delta v$}& 
 	 \colhead{$\sigma_{\delta v}$}& 
 	 \colhead{$ FWHM$}& 
 	 \colhead{$\sigma_{FW}$}& 
 	   \colhead{$ f_x$}& 
 	 \colhead{$\sigma_{f_x}$}& 
 	   \colhead{$ f_0$}& 
 	 \colhead{$\sigma_{f_0}$}\\ 
 	 & 
 	 \colhead{(\AA)} & 
 	 \colhead{($\kms$)}  & 
 	 \colhead{($\kms$)}  & 
 	 \colhead{($\kms$)}  & 
 	  \colhead{($\kms$)}  & 
 	  \multicolumn{4}{c}{($ 10^{-5} \pflux$) } 
 } 
 \startdata
           S XV&       5.039&         819&         223&        1715&         573&        0.37&        0.10&        0.53&       -0.06 \\
Si XIII $\beta$&       5.681&        \nodata&         \nodata&         \nodata&         \nodata&        0.15&        0.06&        0.24&       -0.04 \\
         Si XIV&       6.183&        -250&         237&        2819&         611&        0.41&        0.06&        0.76&       -0.14 \\
        Si XIII&       6.648&         224&         417&        3431&         527&        0.80&        0.25&        1.70&       -0.36 \\
        Si XIII&       6.740&         \nodata&         \nodata&         \nodata&         \nodata&      0.24&        0.08&        0.50&       -0.10 \\
         Mg XII&       8.422&        -620&         282&        2982&         681&        0.25&        0.05&        0.93&       -0.26 \\
      Mg XI (i)&       9.230&        \nodata&         \nodata&         \nodata&         \nodata&     0.33&        0.09&        1.77&       -0.53 \\
           Ne X&      12.135&        -665&         336&        3462&        1144&        0.24&        0.09&        6.65&       -1.95 \\
\enddata 
 \end{deluxetable} \label{tbl:RCW_38}

\begin{deluxetable}{lrrccccccc} 
 	 \tablecaption{ HD~93129A} 
 	 \tablehead{ 
 	 \colhead{Ion}& 
 	 \colhead{$\lambda_0$}& 
 	 \colhead{$\delta v$}& 
 	 \colhead{$\sigma_{\delta v}$}& 
 	 \colhead{$ FWHM$}& 
 	 \colhead{$\sigma_{FW}$}& 
 	   \colhead{$ f_x$}& 
 	 \colhead{$\sigma_{f_x}$}& 
 	   \colhead{$ f_0$}& 
 	 \colhead{$\sigma_{f_0}$}\\ 
 	 & 
 	 \colhead{(\AA)} & 
 	 \colhead{($\kms$)}  & 
 	 \colhead{($\kms$)}  & 
 	 \colhead{($\kms$)}  & 
 	  \colhead{($\kms$)}  & 
 	  \multicolumn{4}{c}{($ 10^{-5} \pflux$) } 
 } 
 \startdata
           S XV&       5.102&       \nodata&            \nodata&            \nodata&        \nodata&        0.15&        0.06&        0.16&  \nodata \\
Si XIII $\beta$&       5.681&         \nodata&            \nodata&            \nodata&        \nodata&       0.12&        0.05&        0.13&  \nodata \\
        Si XIII&       6.648&        -600&         200&        2676&         343&        0.82&        0.15&        0.95&  \nodata \\
        Si XIII&       6.687&         \nodata&            \nodata&            \nodata&        \nodata&       0.62&        0.13&        0.72&  \nodata \\
        Si XIII&       6.740&        \nodata&            \nodata&            \nodata&        \nodata&     0.57&        0.10&        0.66&  \nodata \\
         Mg XII&       8.422&        -873&         209&        3386&         774&        0.53&        0.11&        0.68&  \nodata \\
      Mg XI (r)&       9.169&        -860&         213&        2881&         446&        1.02&        0.22&        1.40&  \nodata \\
      Mg XI (i)&       9.230&         \nodata&            \nodata&            \nodata&        \nodata&       0.67&        0.21&        0.93&  \nodata \\
\enddata 
 \end{deluxetable}%
 \label{tbl:tau_Sco}

\clearpage

\clearpage
\section{Spectral Plots for Each Star}\label{sec:specplots}

We show each spectrum as luminosity density on two scales.  Panel A in
each pair uses a logarithmic $y$-axis scaling which is useful to
convey the overall shape of the spectrum.  All logarithmic $y$-ranges
span $3\,\mathrm{dex}$.  Panel B of the pair has a linear scale to
emphasize the emission lines, auto-scaled from zero to the maximum
flux in the range plotted.  In each plot, the black curve is
uncorrected for interstellar absorption, and the gray curve is the
corrected version. Some of these are apparently over-corrected with
the absorption values adopted from the literature (see
Table~\ref{stellar-parameters}). Plots are in the same order as given
in Table~\ref{stellar-parameters}. The plot captions give the
observed fluxes (ucorrected for absorption, corresponding to the black
curve), and the absorption corrected luminosity (corresponding to the
gray curves), along with the $N_\mathrm{H}$ and distances from
Table~\ref{stellar-parameters}.

\clearpage

\begin{figure}[t]
    \centering
   \includegraphics*[width=0.820\columnwidth, viewport= 0 0 535 377]{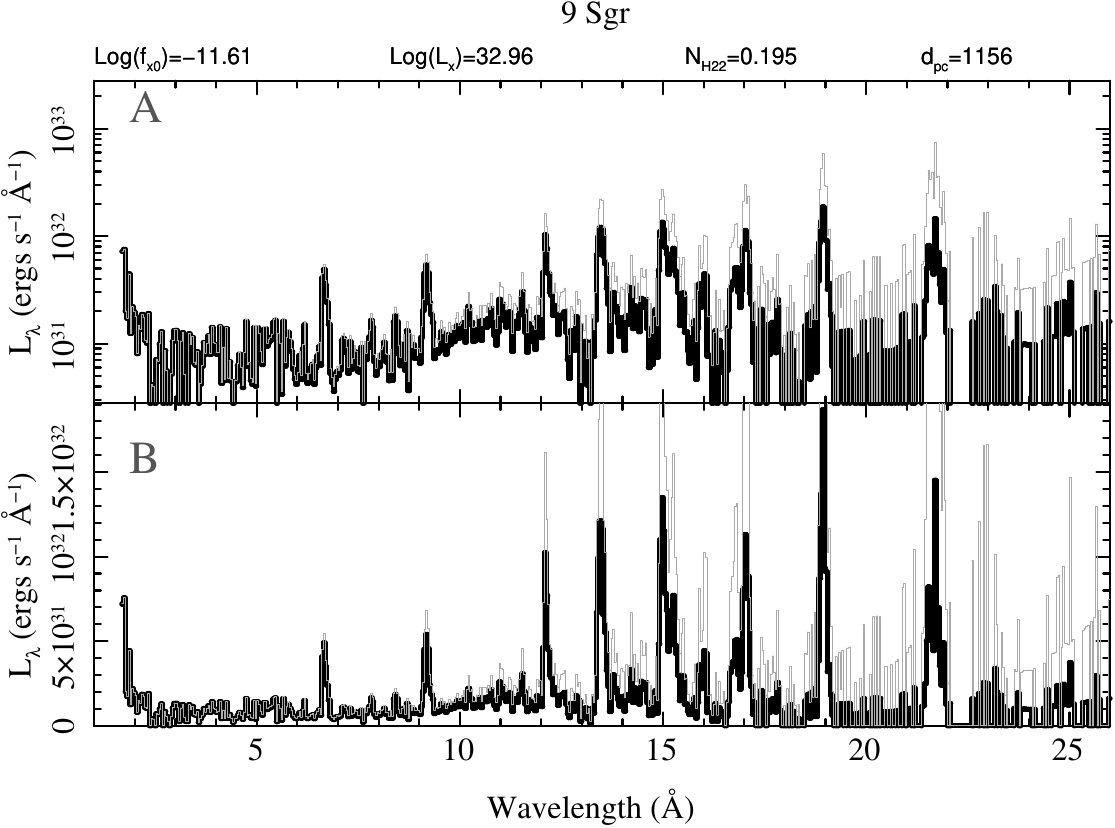}
   \caption{$9\,$Sgr}
   \label{fig:lx_9_Sgr}
\end{figure}

\begin{figure}[b]
    \centering
   \includegraphics*[width=0.820\columnwidth, viewport= 0 0 535 377]{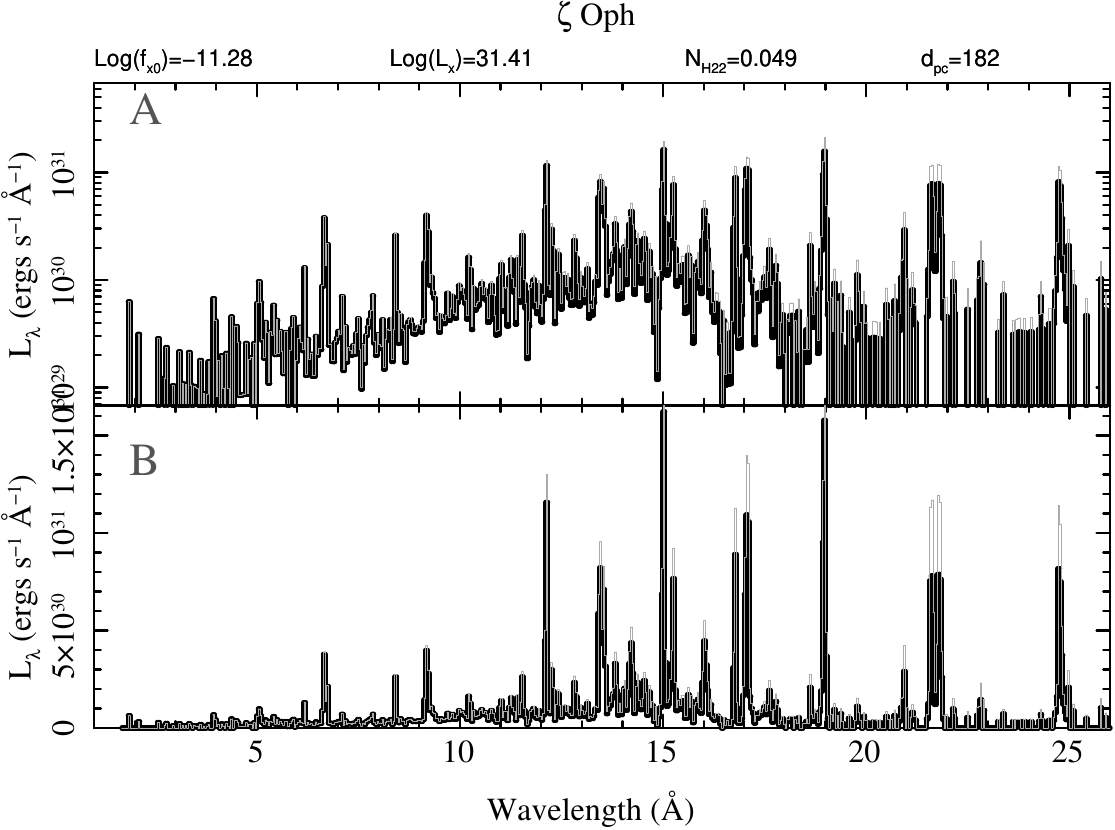}
   \caption{$\zeta\,$Oph}
   \label{fig:lx_zeta_Oph}
\end{figure}

\begin{figure}[tb]
    \centering
   \includegraphics*[width=0.820\columnwidth, viewport= 0 0 535 377]{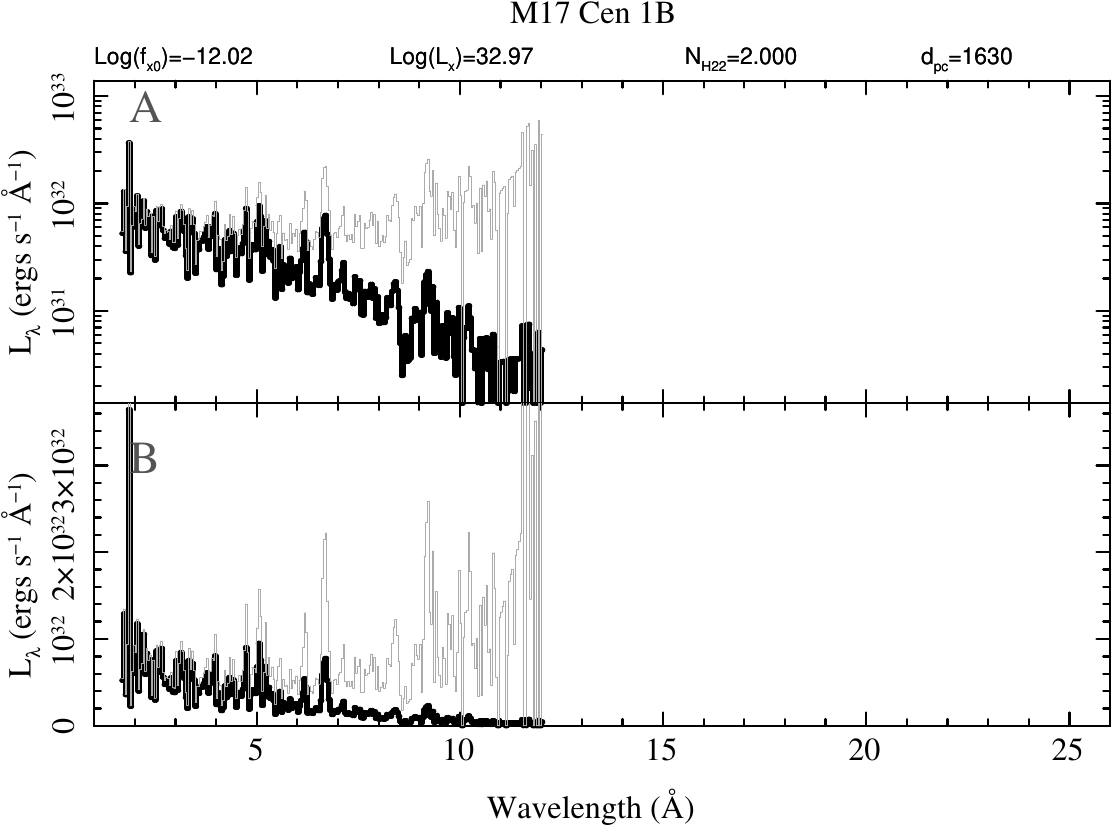}
   \caption{M17 Cen 1B}
   \label{fig:lx_M17_Cen_1B}
\end{figure}

\begin{figure}[tb]
    \centering
   \includegraphics*[width=0.820\columnwidth, viewport= 0 0 535 377]{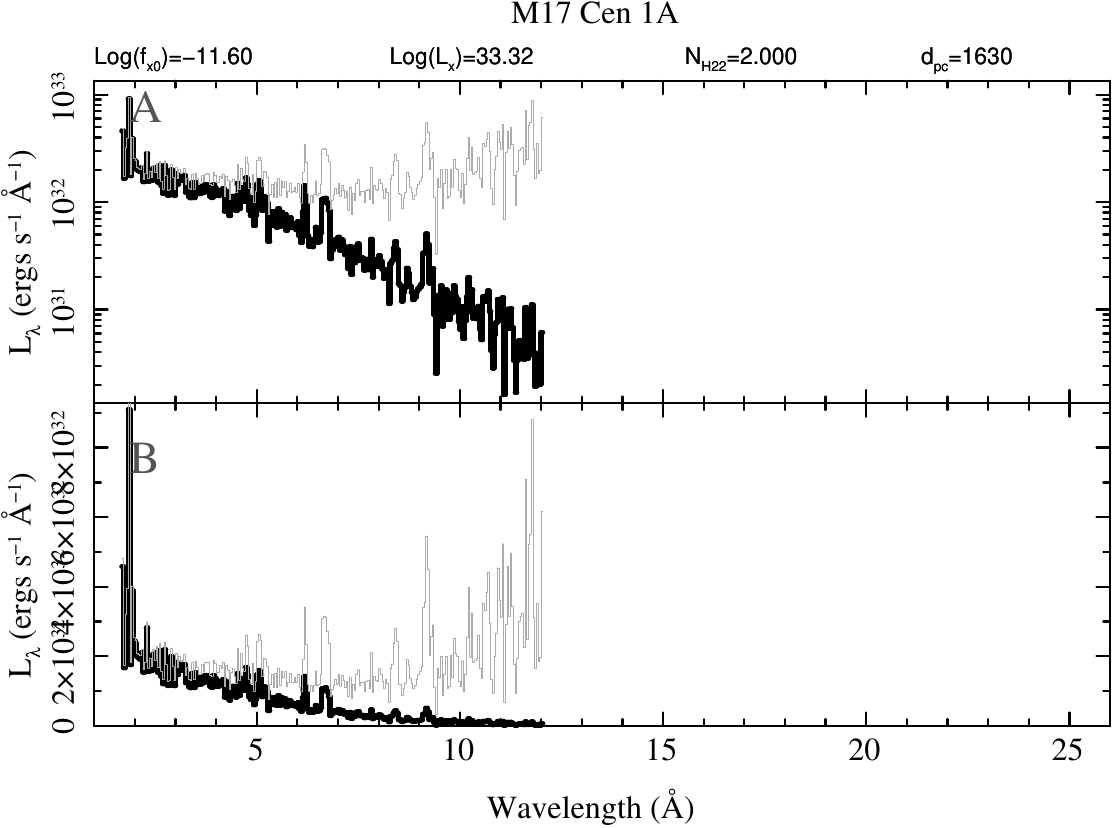}
   \caption{M17 Cen 1A}
   \label{fig:lx_M17_Cen_1A}
\end{figure}

\begin{figure}[tb]
    \centering
   \includegraphics*[width=0.820\columnwidth, viewport= 0 0 535 377]{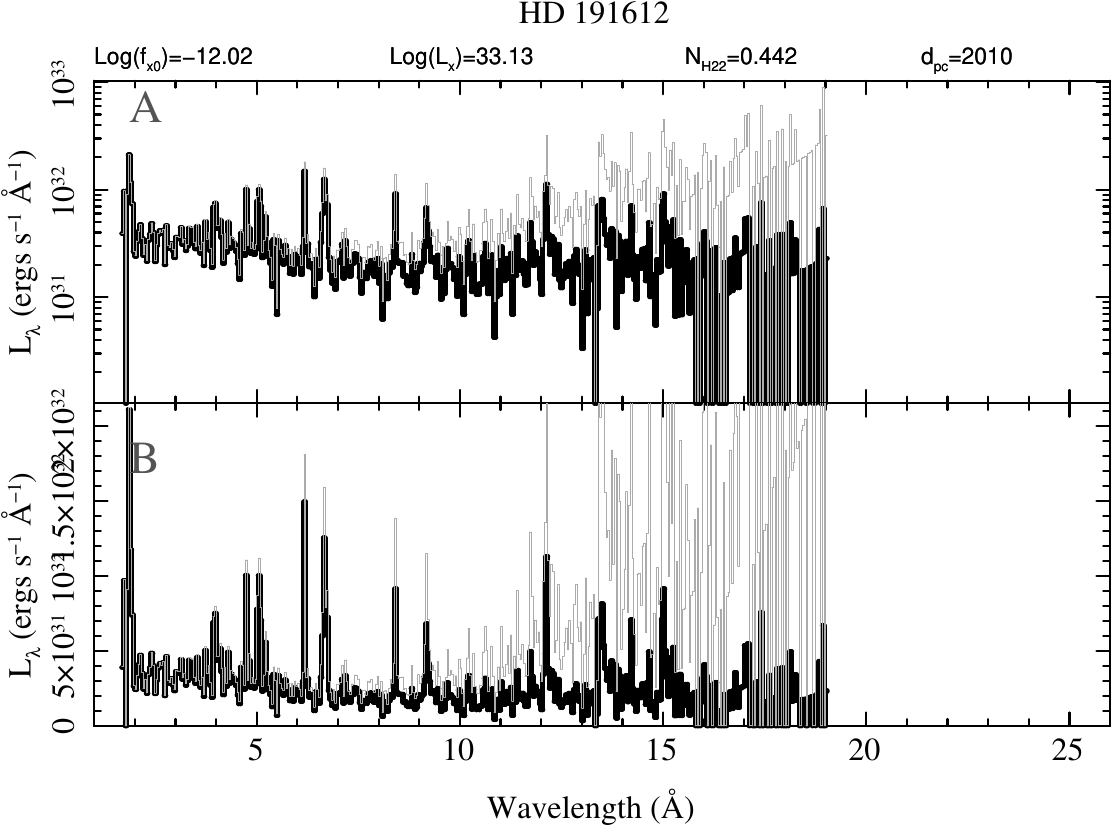}
   \caption{HD~191612}
   \label{fig:lx_HD_191612}
\end{figure}

\begin{figure}[tb]
    \centering
   \includegraphics*[width=0.820\columnwidth, viewport= 0 0 535 377]{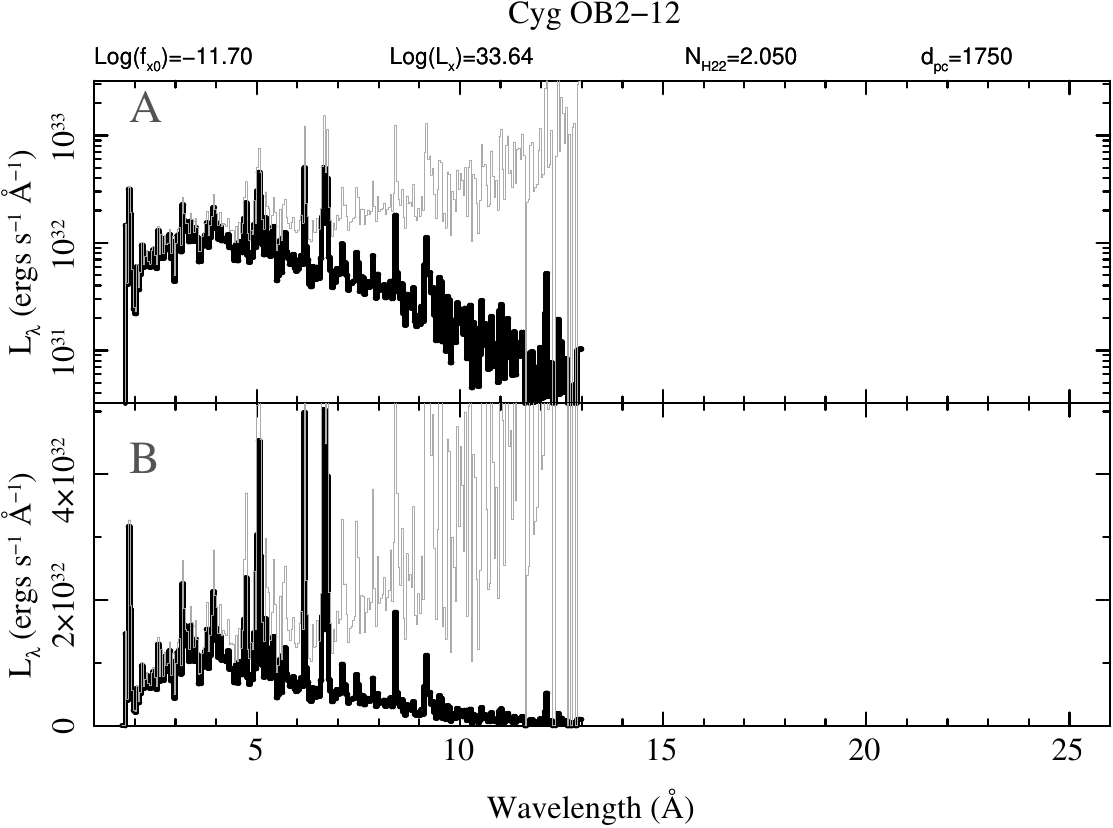}
   \caption{Cyg~OB2-12}
   \label{fig:lx_Cyg_OB2-12}
\end{figure}

\begin{figure}[tb]
    \centering
   \includegraphics*[width=0.820\columnwidth, viewport= 0 0 535 377]{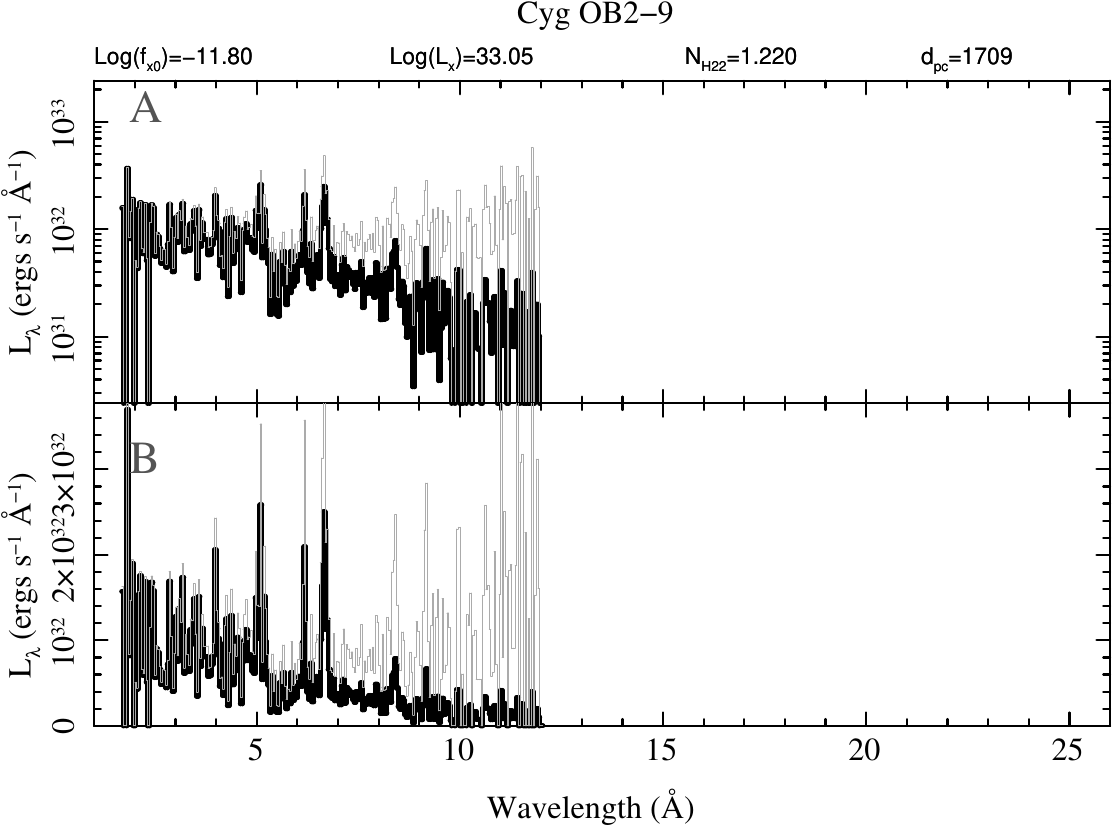}
   \caption{Cyg~OB2-9}
   \label{fig:lx_Cyg_OB2-9}
\end{figure}

\begin{figure}[tb]
    \centering
   \includegraphics*[width=0.820\columnwidth, viewport= 0 0 535 377]{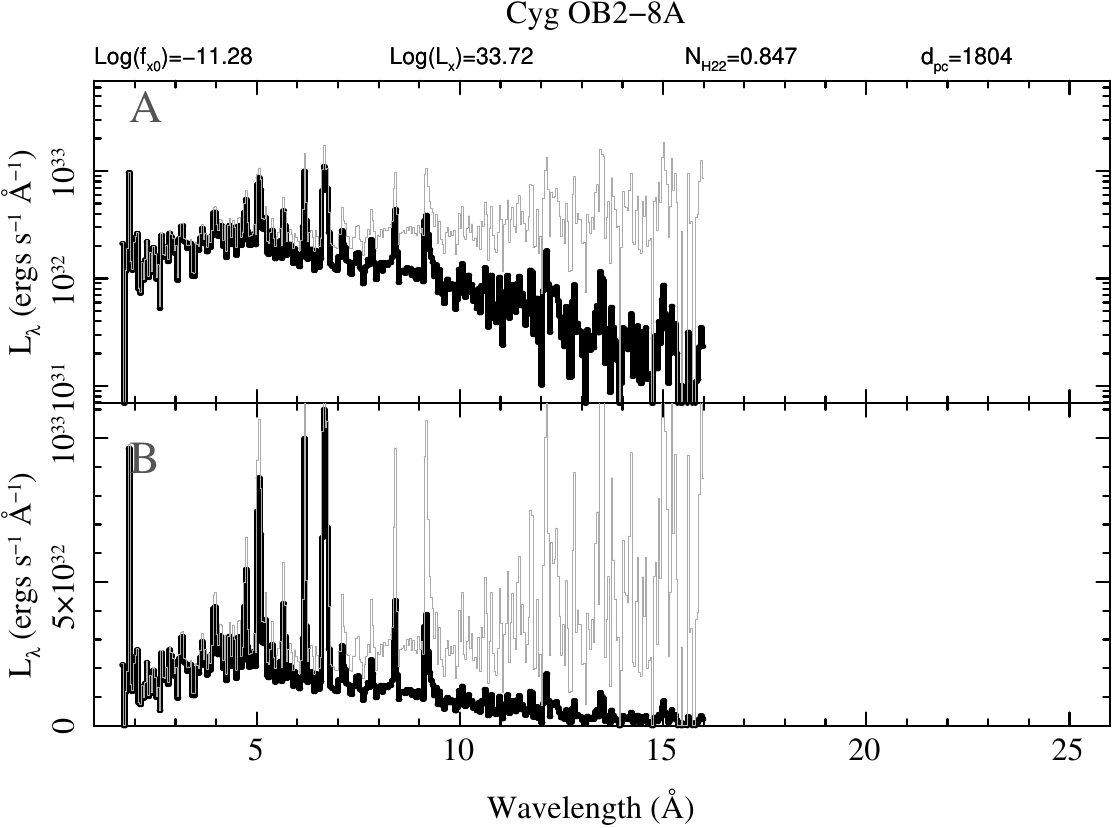}
   \caption{Cyg~OB2-8A}
   \label{fig:lx_Cyg_OB2-8A}
\end{figure}

\begin{figure}[tb]
    \centering
   \includegraphics*[width=0.820\columnwidth, viewport= 0 0 535 377]{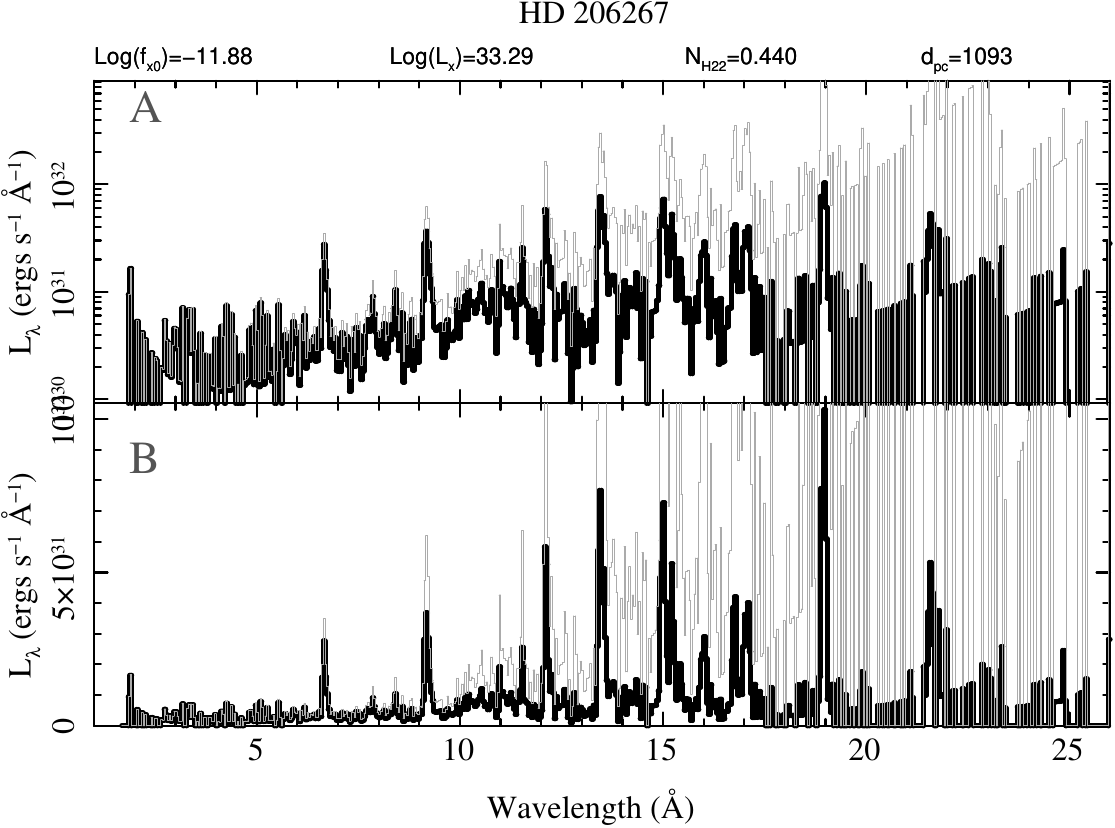}
   \caption{HD~206267}
   \label{fig:lx_HD_206267}
\end{figure}

\begin{figure}[tb]
    \centering
   \includegraphics*[width=0.820\columnwidth, viewport= 0 0 535 377]{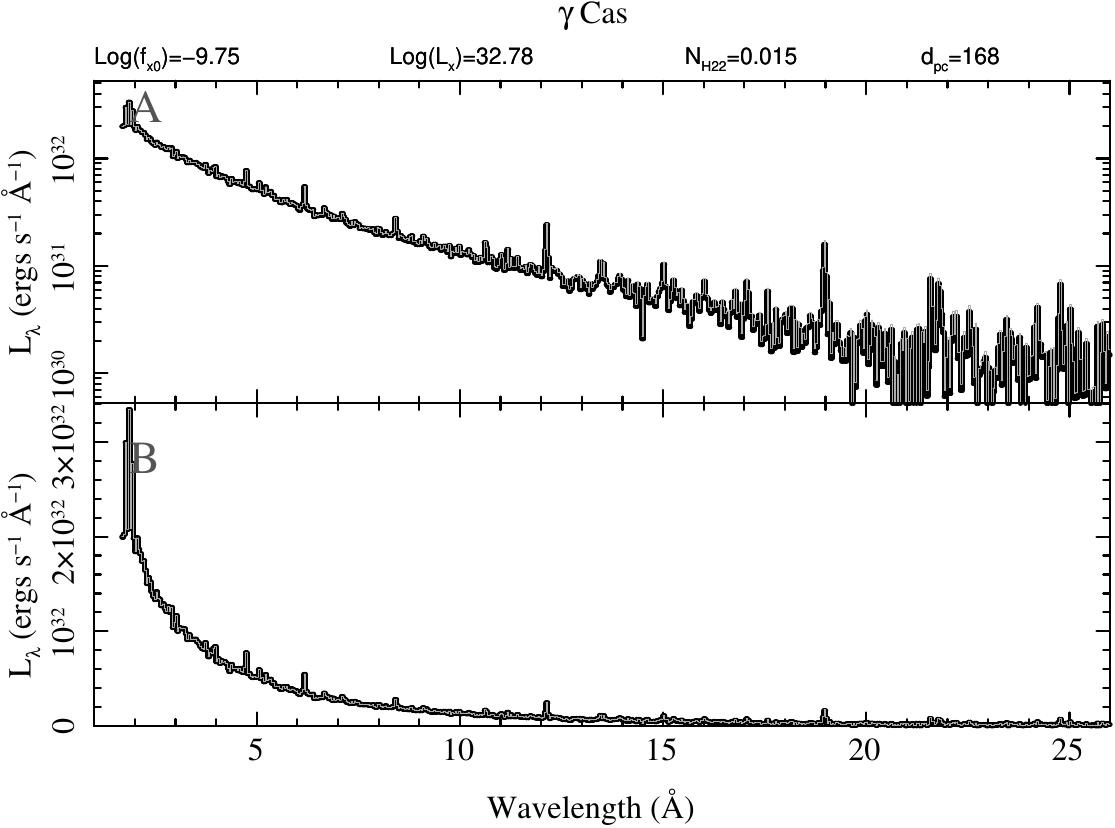}
   \caption{$\gamma\,$Cas}
   \label{fig:lx_gamma_Cas}
\end{figure}

\begin{figure}[tb]
    \centering
   \includegraphics*[width=0.820\columnwidth, viewport= 0 0 535 377]{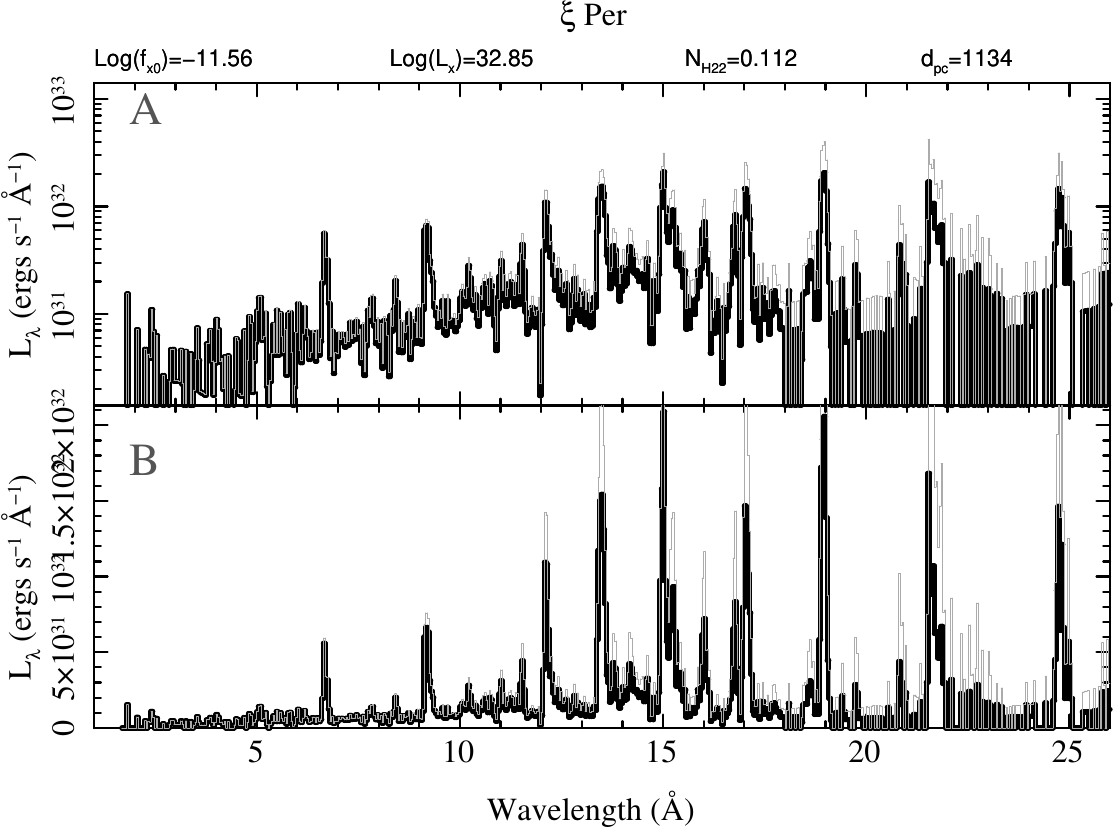}
   \caption{$\xi\,$Per}
   \label{fig:lx_xi_Per}
\end{figure}

\begin{figure}[tb]
    \centering
   \includegraphics*[width=0.820\columnwidth, viewport= 0 0 535 377]{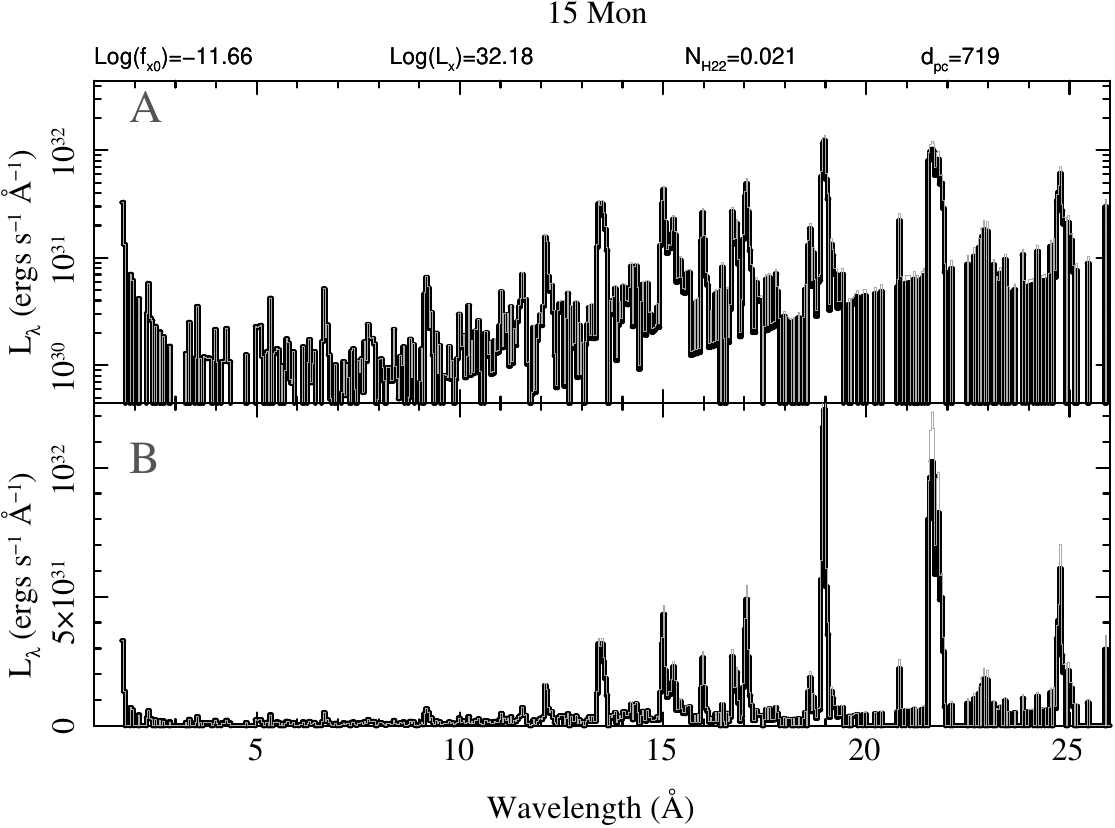}
   \caption{$15\,$Mon}
   \label{fig:lx_15_Mon}
\end{figure}

\begin{figure}[tb]
    \centering
   \includegraphics*[width=0.820\columnwidth, viewport= 0 0 535 377]{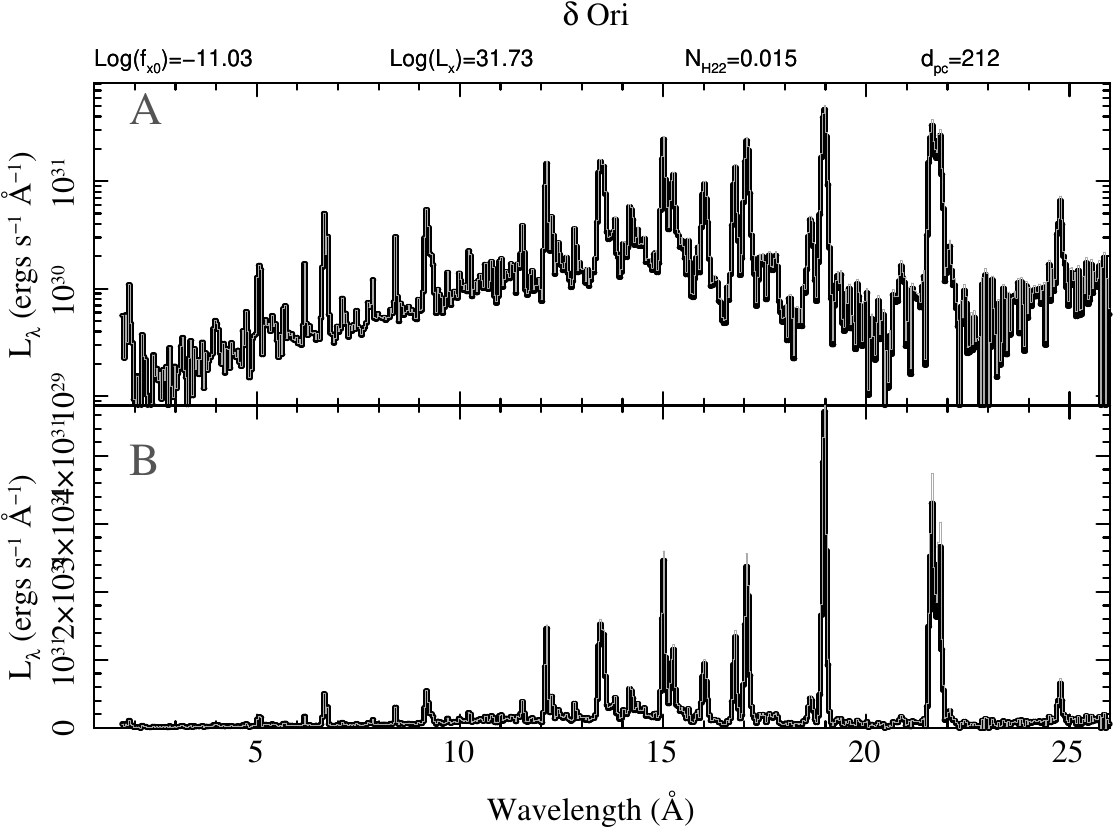}
   \caption{$\delta\,$Ori}
   \label{fig:lx_delta_Ori}
\end{figure}

\begin{figure}[tb]
    \centering
   \includegraphics*[width=0.820\columnwidth, viewport= 0 0 535 377]{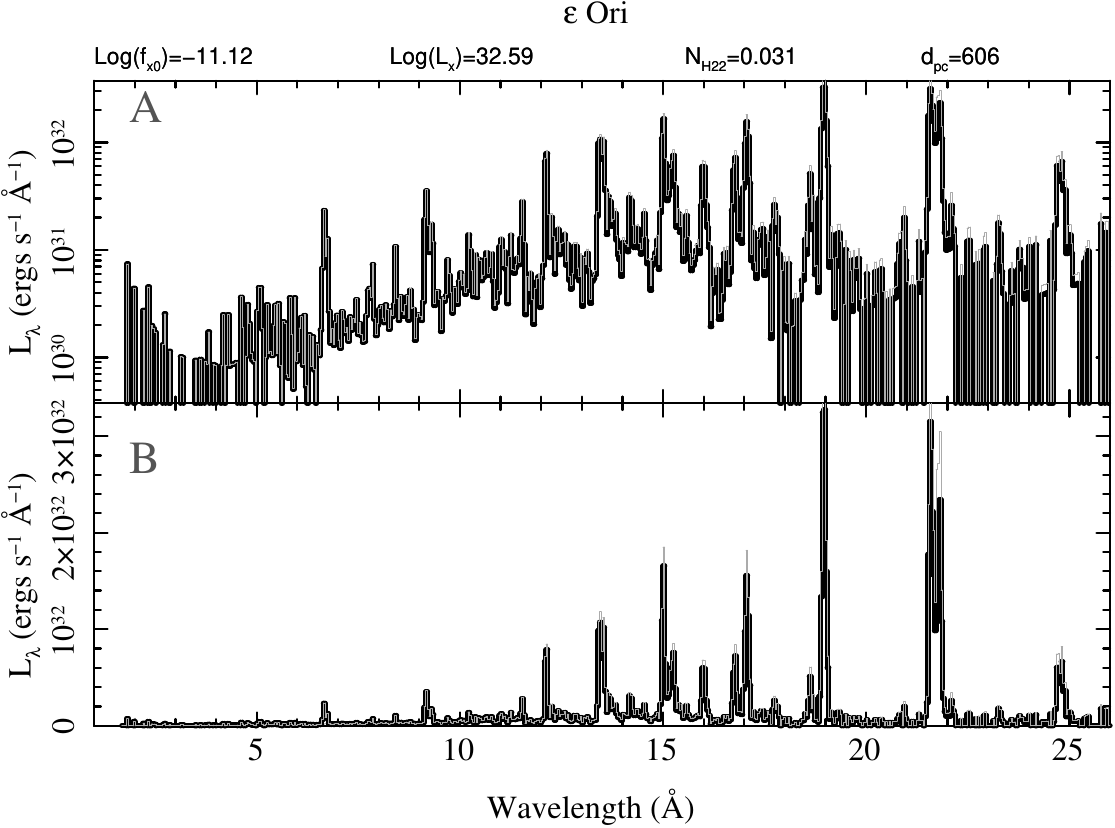}
   \caption{$\epsilon\,$Ori}
   \label{fig:lx_eps_Ori}
\end{figure}

\begin{figure}[tb]
    \centering
   \includegraphics*[width=0.820\columnwidth, viewport= 0 0 535 377]{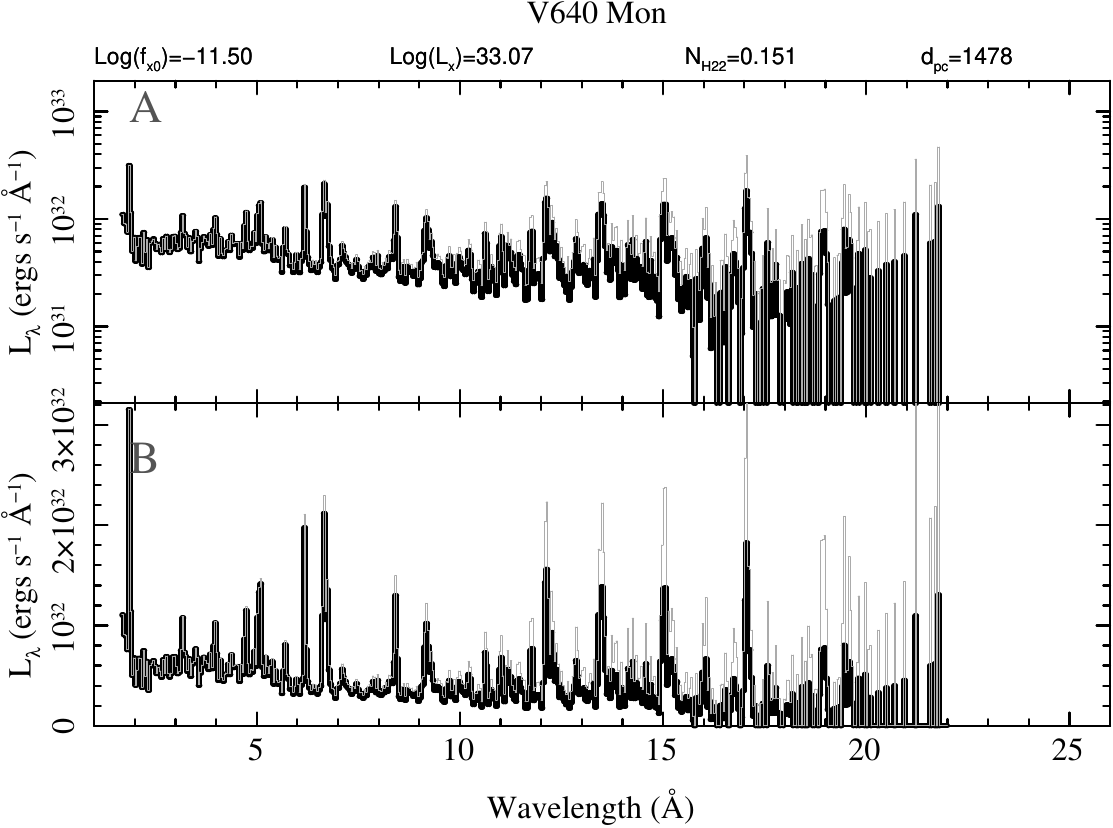}
   \caption{V640 Mon}
   \label{fig:lx_V640_Mon}
\end{figure}

\begin{figure}[tb]
    \centering
   \includegraphics*[width=0.820\columnwidth, viewport= 0 0 535 377]{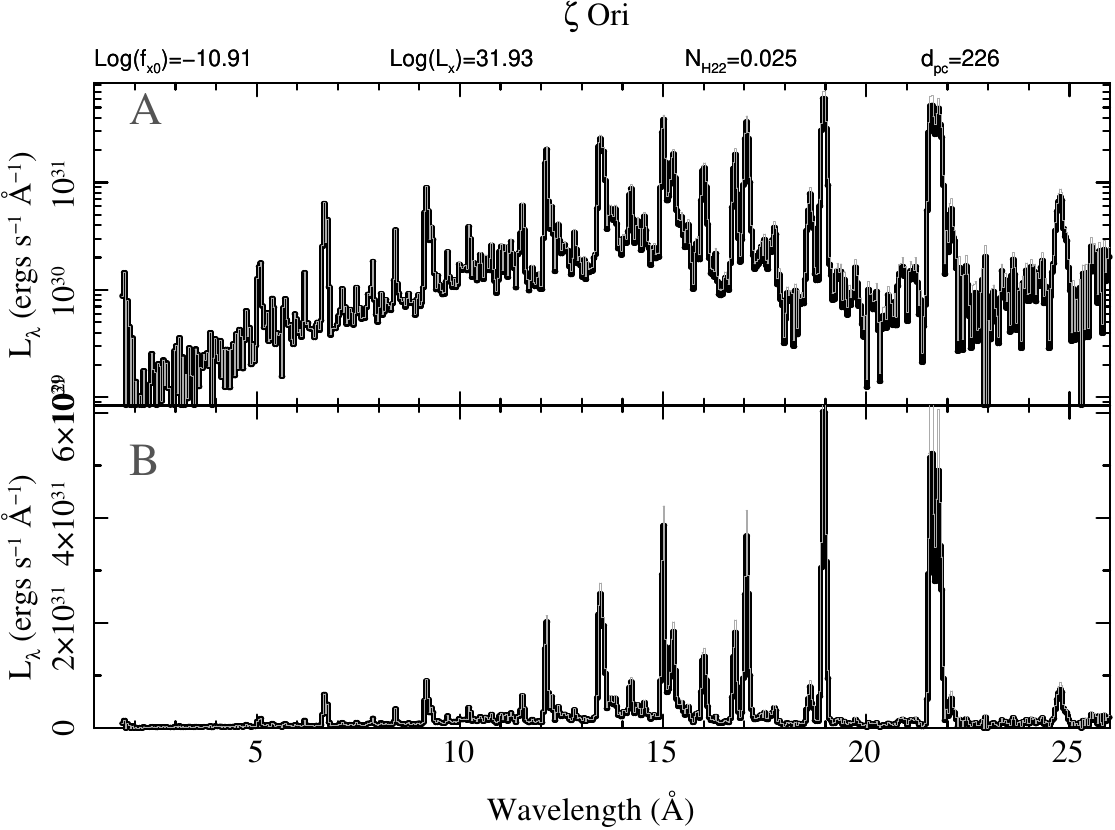}
   \caption{$\zeta\,$Ori}
   \label{fig:lx_zeta_Ori}
\end{figure}

\begin{figure}[tb]
    \centering
   \includegraphics*[width=0.820\columnwidth, viewport= 0 0 535 377]{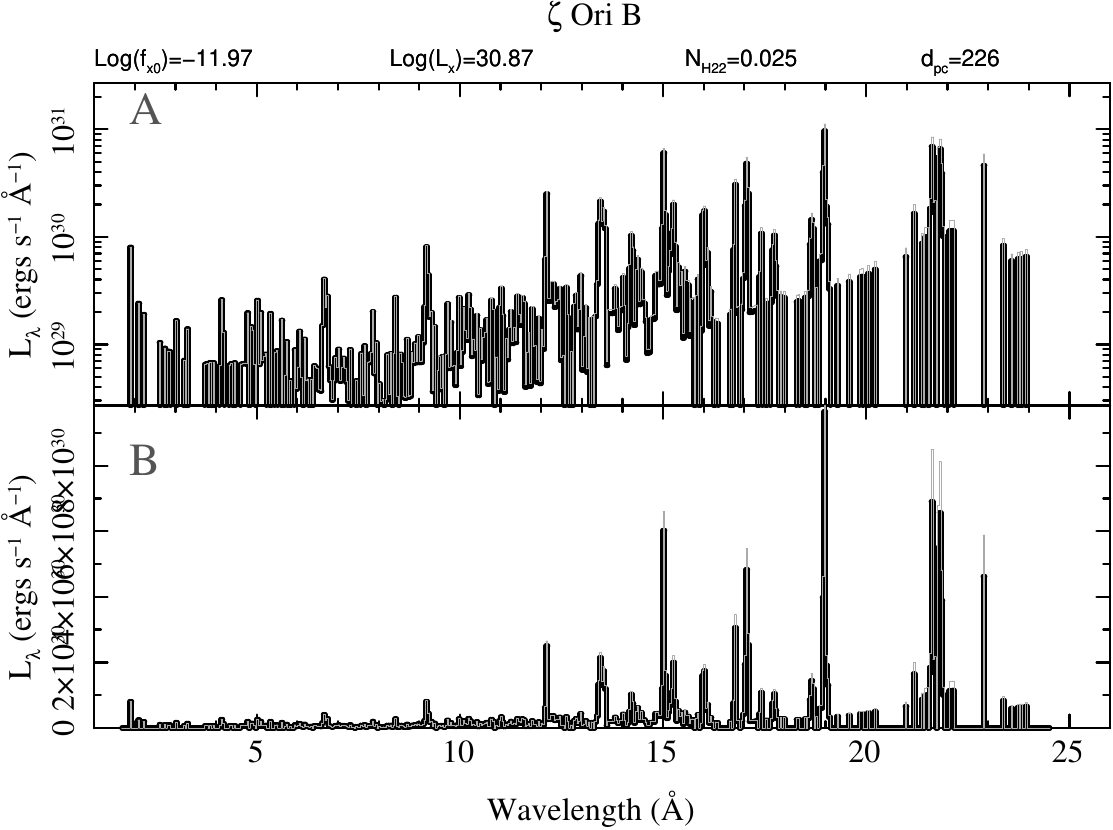}
   \caption{$\zeta\,$Ori~B}
   \label{fig:lx_zeta_Ori_B}
\end{figure}

\begin{figure}[tb]
    \centering
   \includegraphics*[width=0.820\columnwidth, viewport= 0 0 535 377]{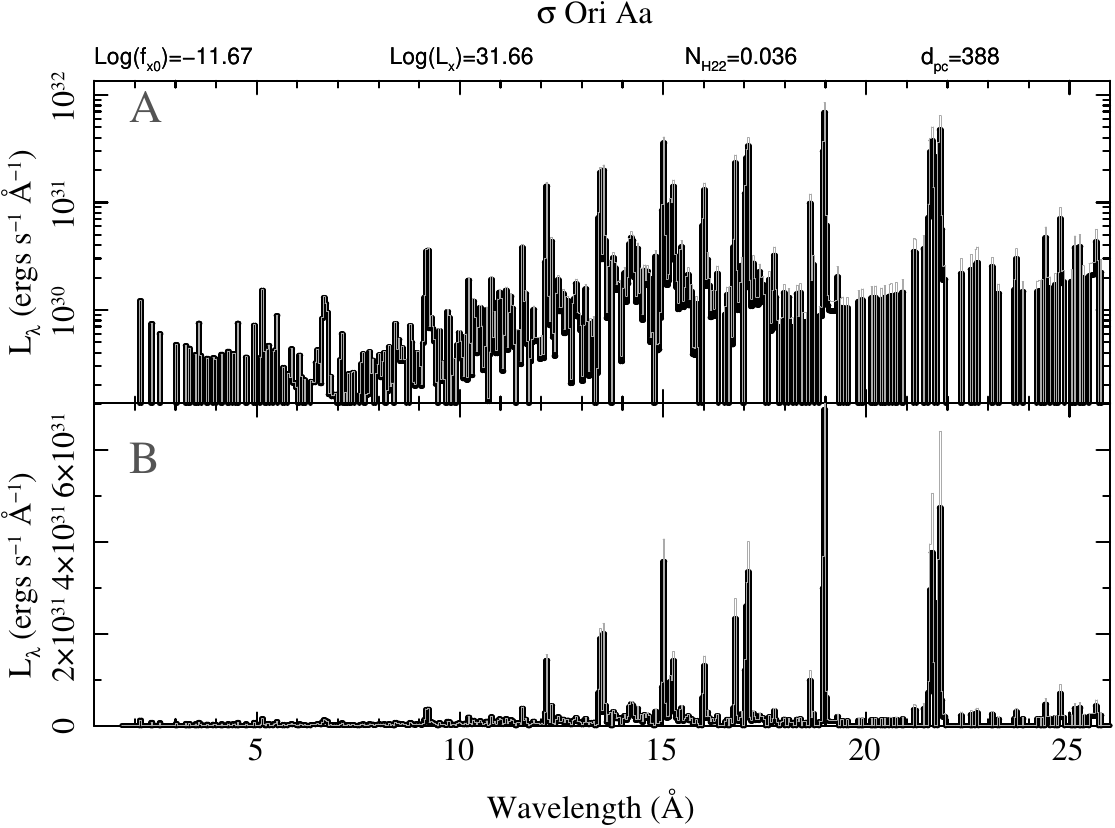}
   \caption{$\sigma\,$Ori~Aa}
   \label{fig:lx_sig_Ori}
\end{figure}

\begin{figure}[tb]
    \centering
   \includegraphics*[width=0.820\columnwidth, viewport= 0 0 535 377]{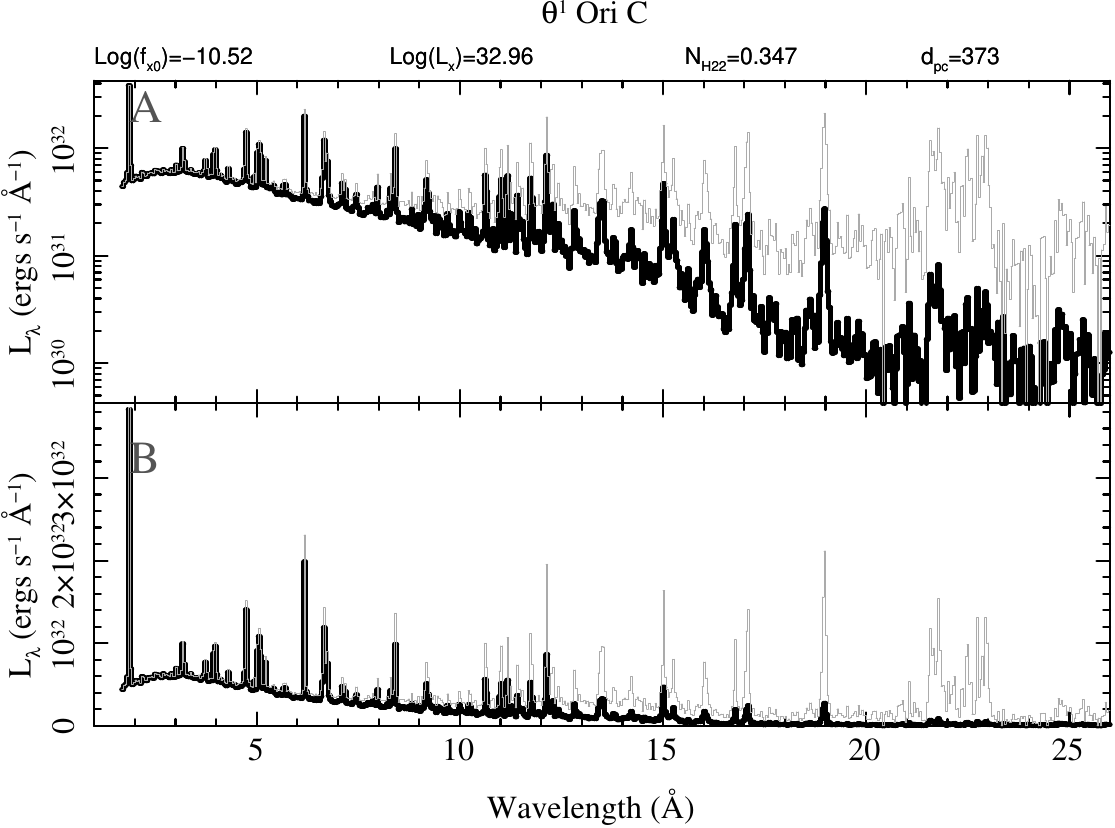}
   \caption{$\theta^1\,$Ori~C}
   \label{fig:lx_tet01_Ori_C}
\end{figure}

\begin{figure}[tb]
    \centering
   \includegraphics*[width=0.820\columnwidth, viewport= 0 0 535 377]{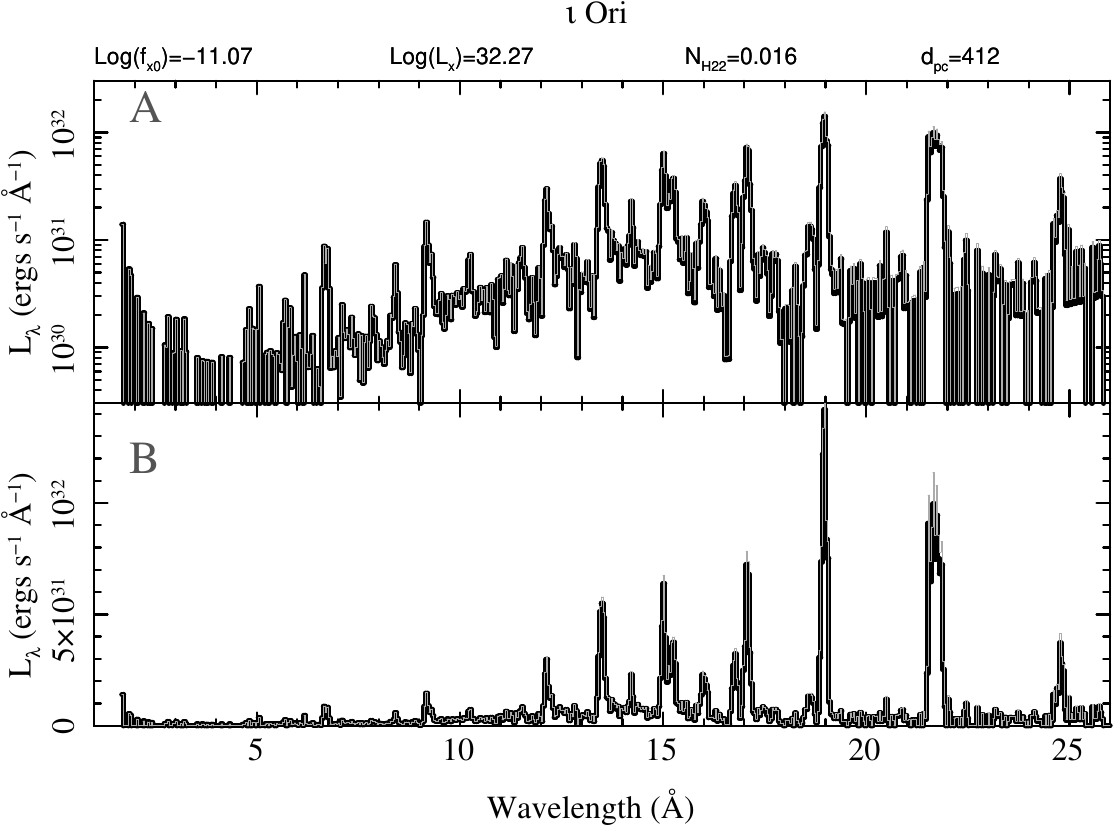}
   \caption{$\iota\,$Ori}
   \label{fig:lx_iota_Ori}
\end{figure}

\begin{figure}[tb]
    \centering
   \includegraphics*[width=0.820\columnwidth, viewport= 0 0 535 377]{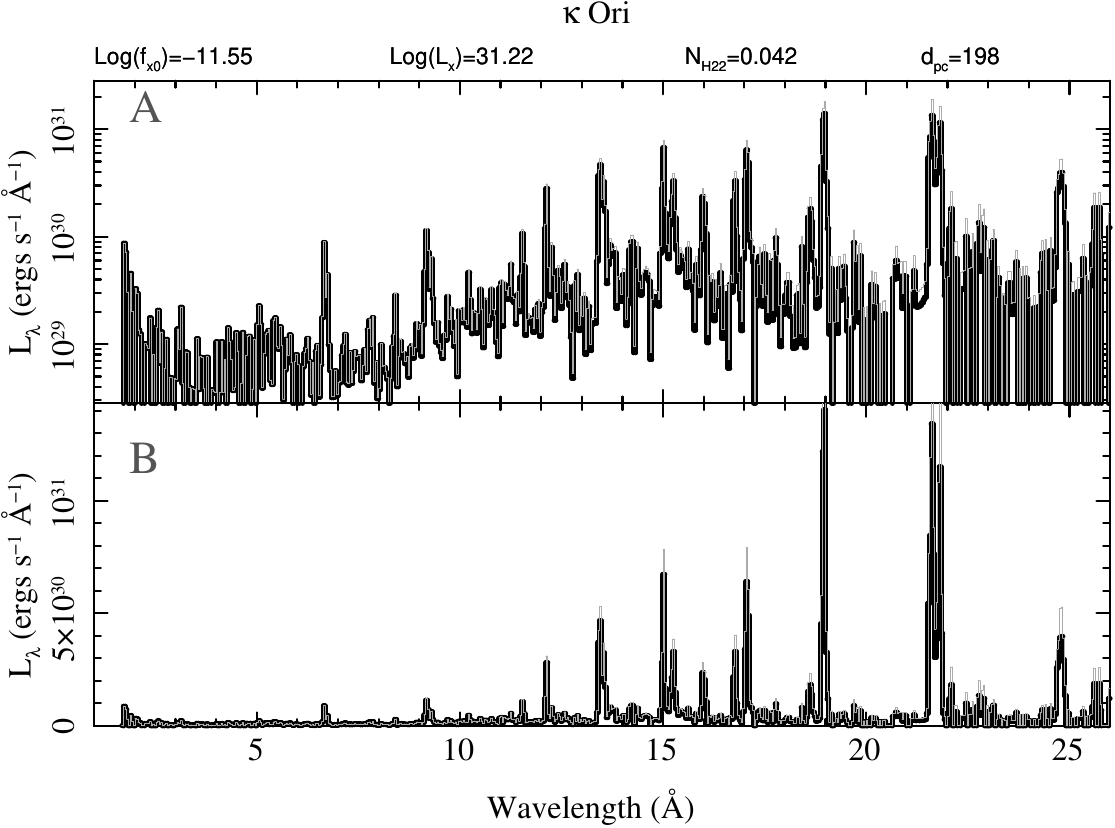}
   \caption{$\kappa\,$Ori}
   \label{fig:lx_kappa_Ori}
\end{figure}

\begin{figure}[tb]
    \centering
   \includegraphics*[width=0.820\columnwidth, viewport= 0 0 535 377]{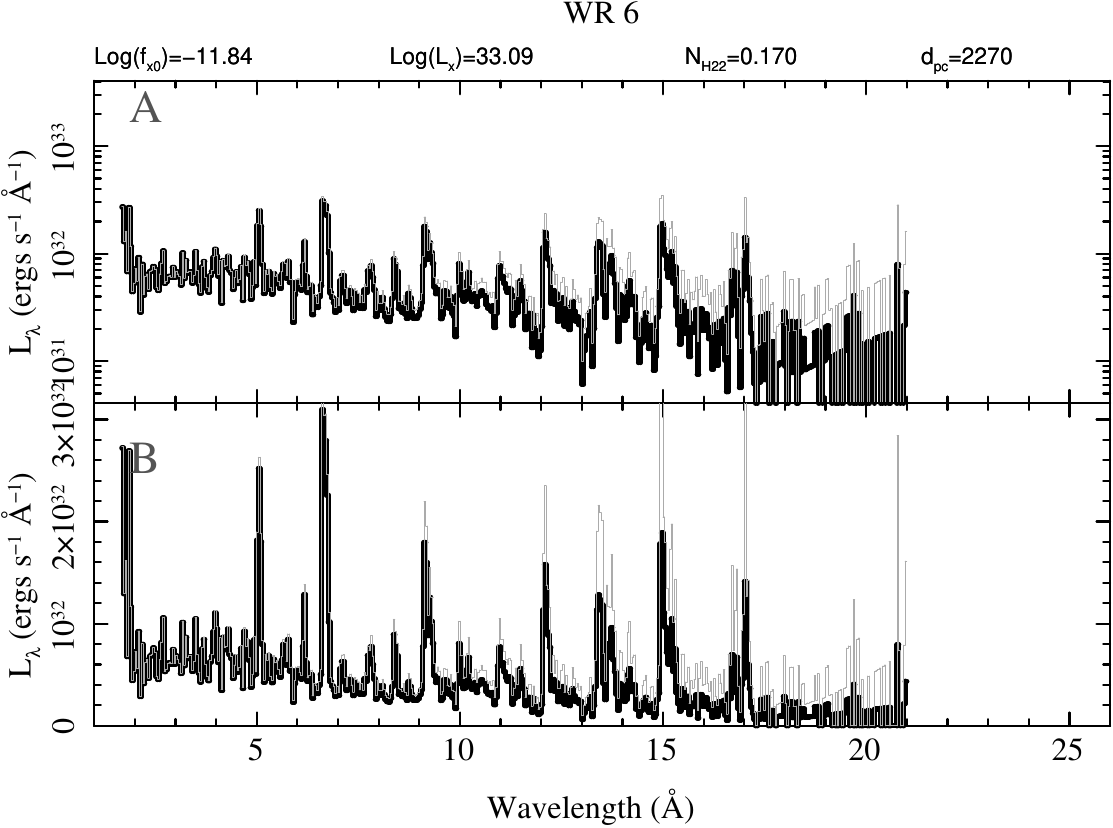}
   \caption{WR~6}
   \label{fig:lx_WR_6}
\end{figure}

\begin{figure}[tb]
    \centering
   \includegraphics*[width=0.820\columnwidth, viewport= 0 0 535 377]{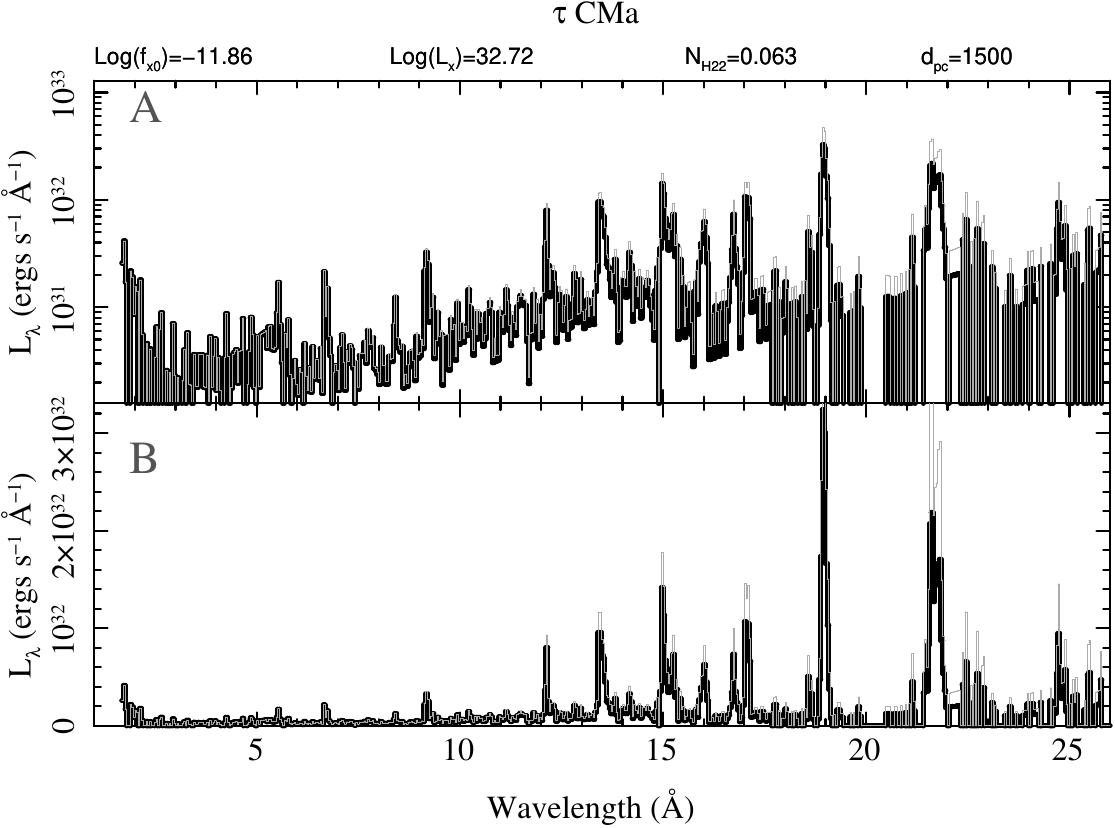}
   \caption{$\tau\,$CMa}
   \label{fig:lx_tau_CMa}
\end{figure}

\begin{figure}[tb]
    \centering
   \includegraphics*[width=0.820\columnwidth, viewport= 0 0 535 377]{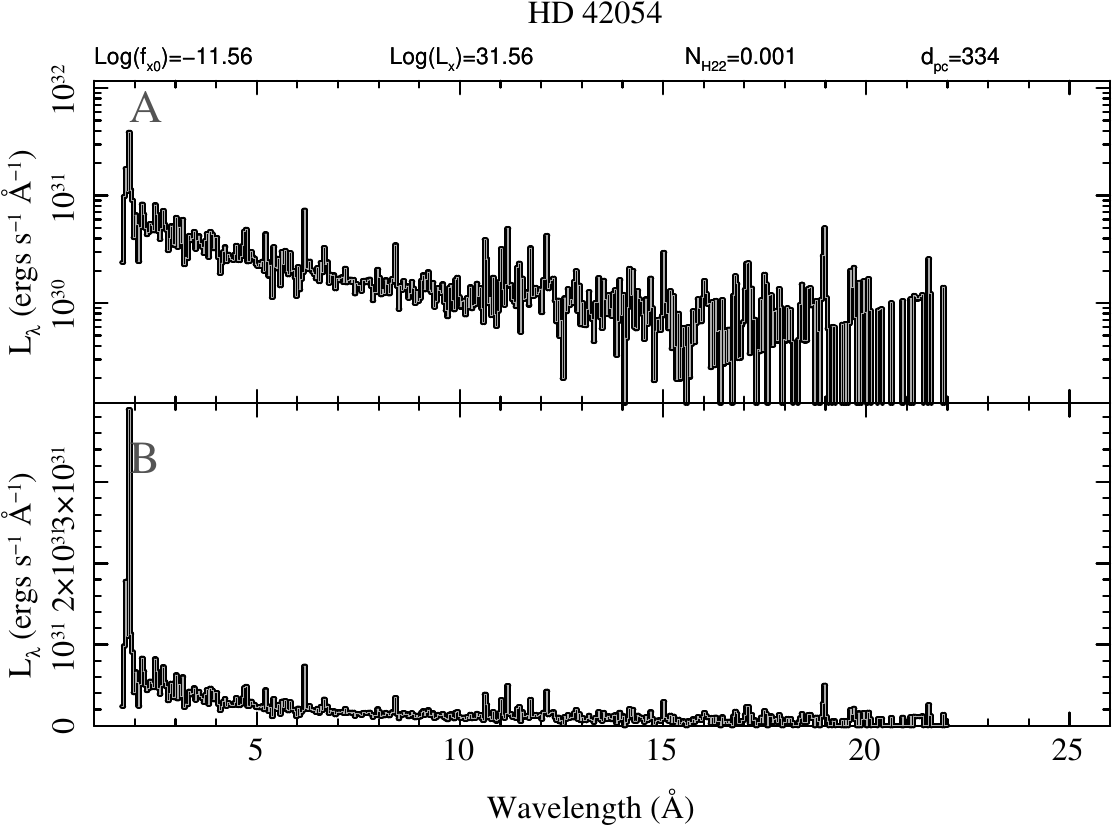}
   \caption{HD~42054}
   \label{fig:lx_HD_42054}
\end{figure}

\begin{figure}[tb]
    \centering
   \includegraphics*[width=0.820\columnwidth, viewport= 0 0 535 377]{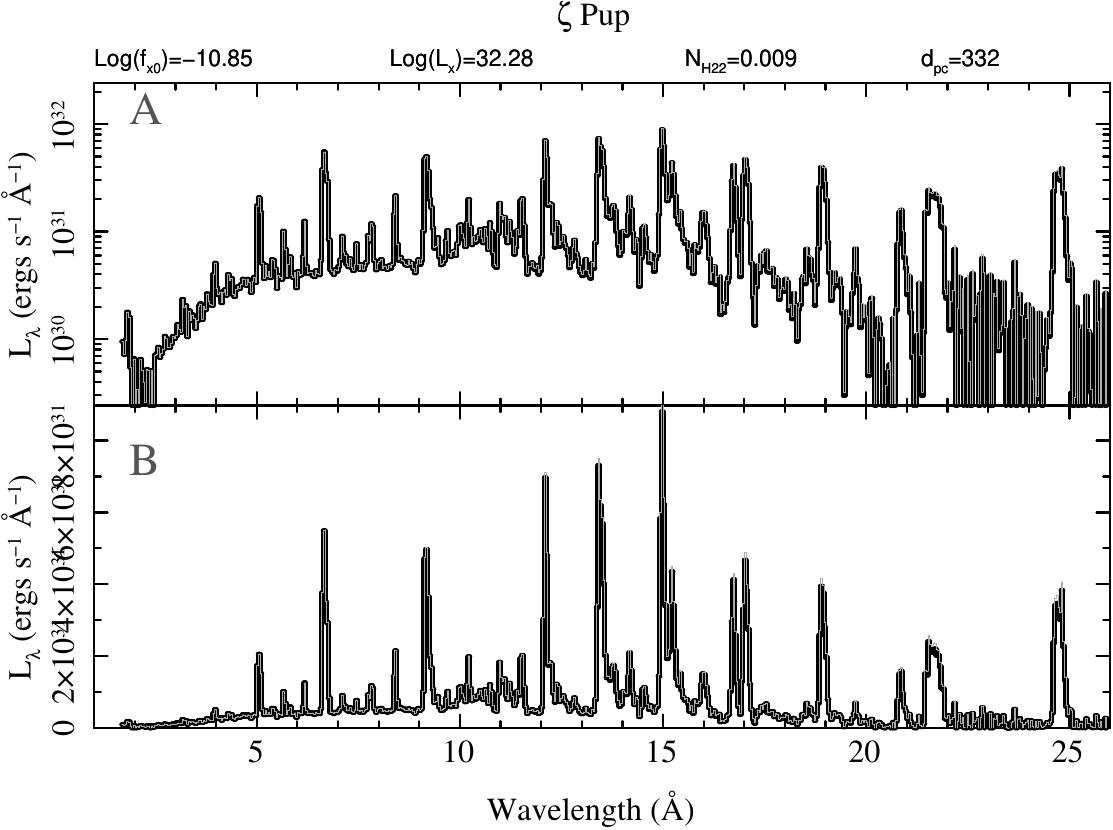}
   \caption{$\zeta\,$Pup}
   \label{fig:lx_zeta_Pup}
\end{figure}

\begin{figure}[tb]
    \centering
   \includegraphics*[width=0.820\columnwidth, viewport= 0 0 535 377]{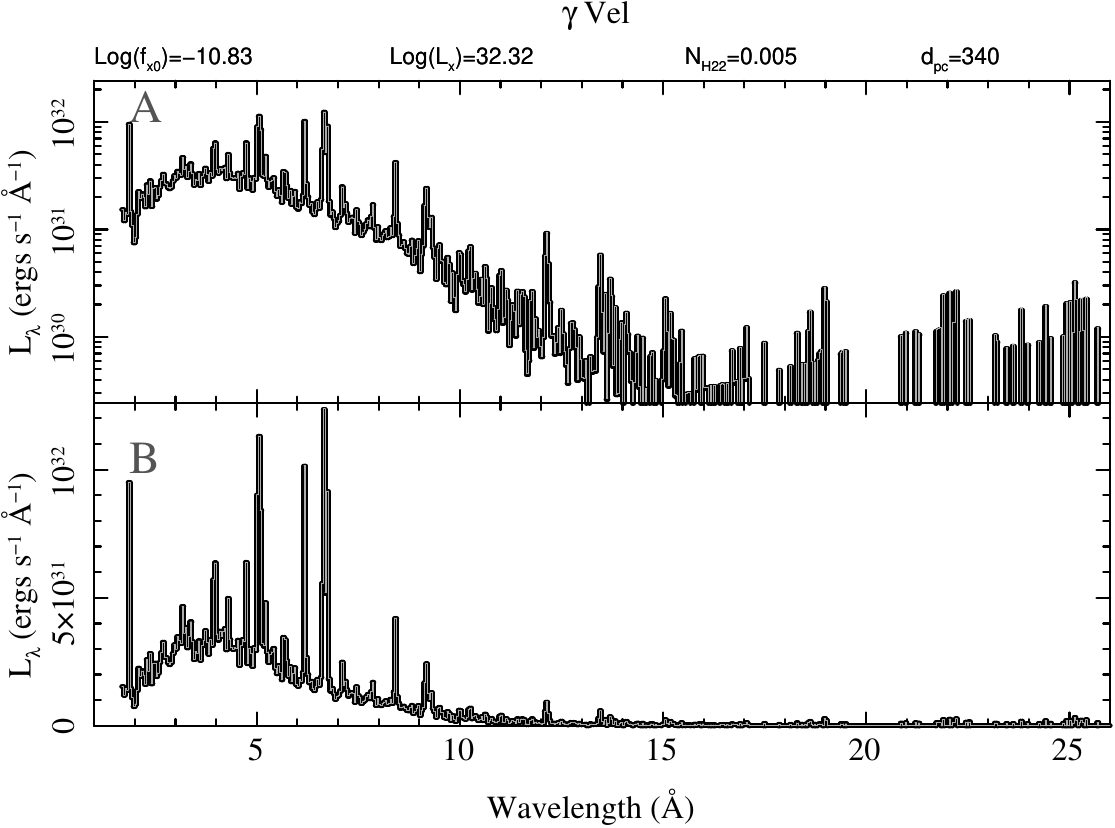}
   \caption{$\gamma^2\,$Vel}
   \label{fig:lx_gamma_Vel}
\end{figure}

\begin{figure}[tb]
    \centering
   \includegraphics*[width=0.820\columnwidth, viewport= 0 0 535 377]{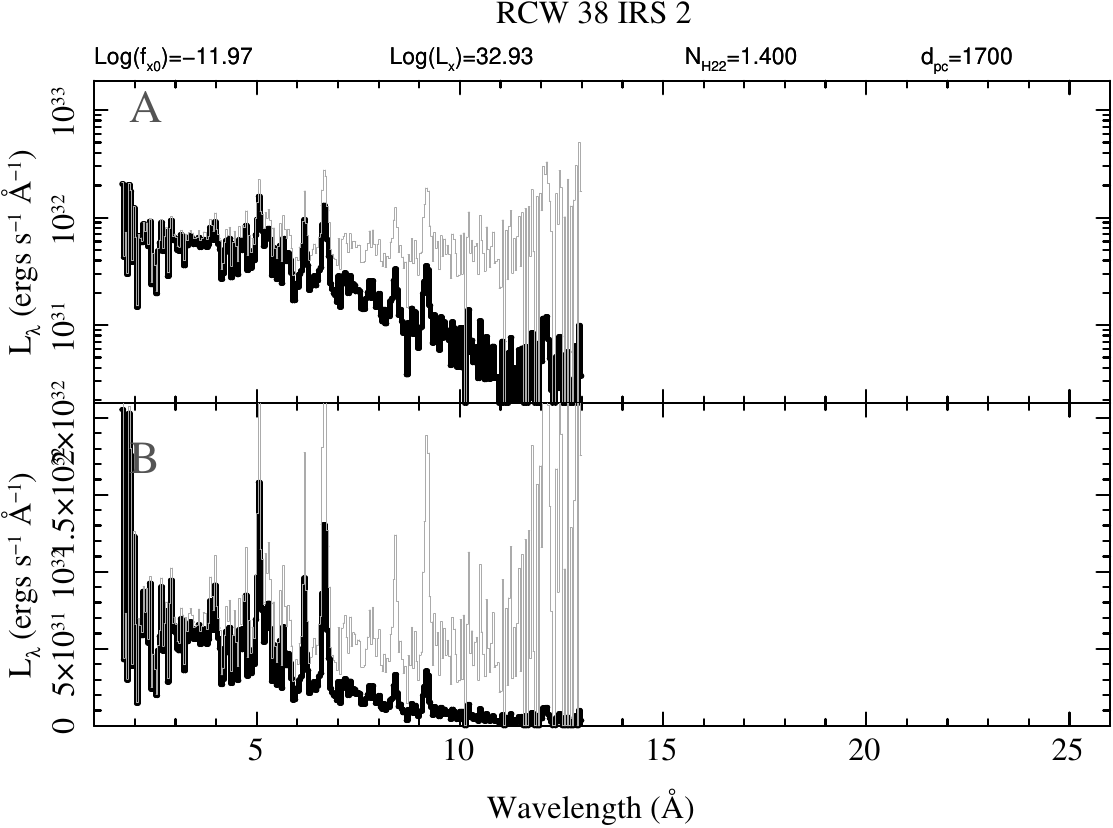}
   \caption{RCW 38 IRS 2}
   \label{fig:lx_RCW_38}
\end{figure}

\begin{figure}[tb]
    \centering
   \includegraphics*[width=0.820\columnwidth, viewport= 0 0 535 377]{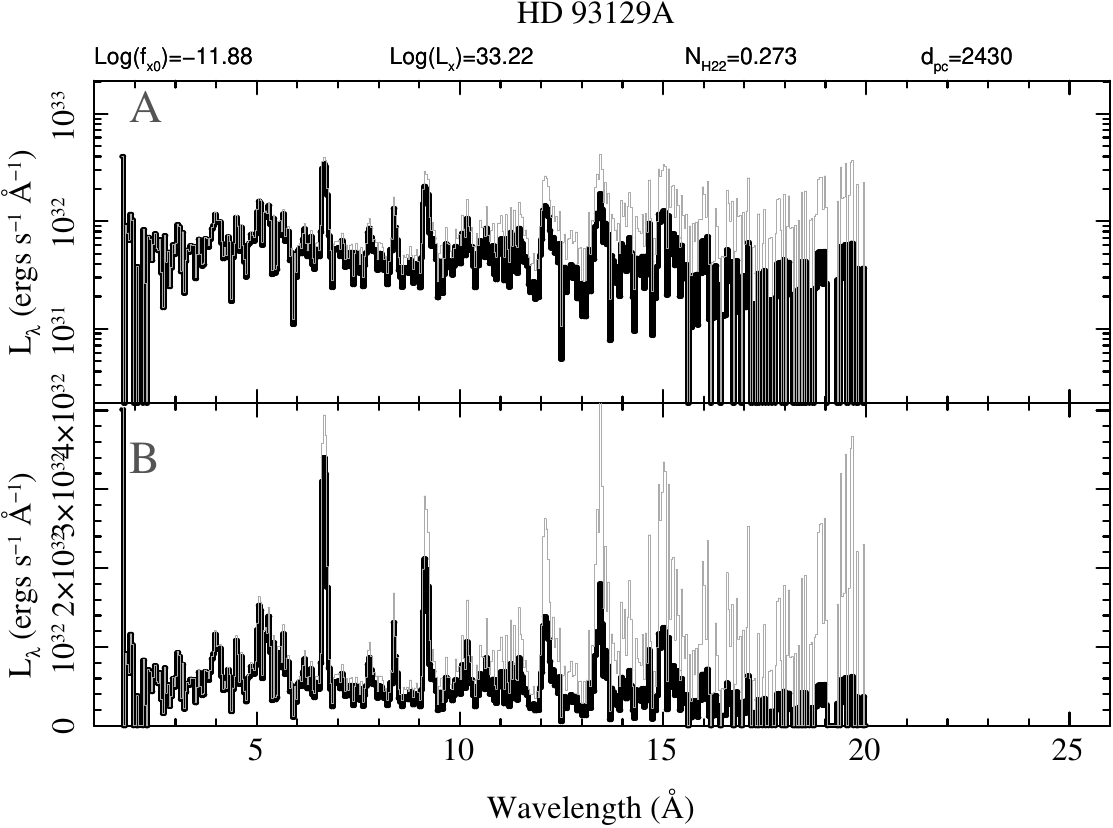}
   \caption{HD~93129A}
   \label{fig:lx_HD_93129A}
\end{figure}

\begin{figure}[tb]
    \centering
   \includegraphics*[width=0.820\columnwidth, viewport= 0 0 535 377]{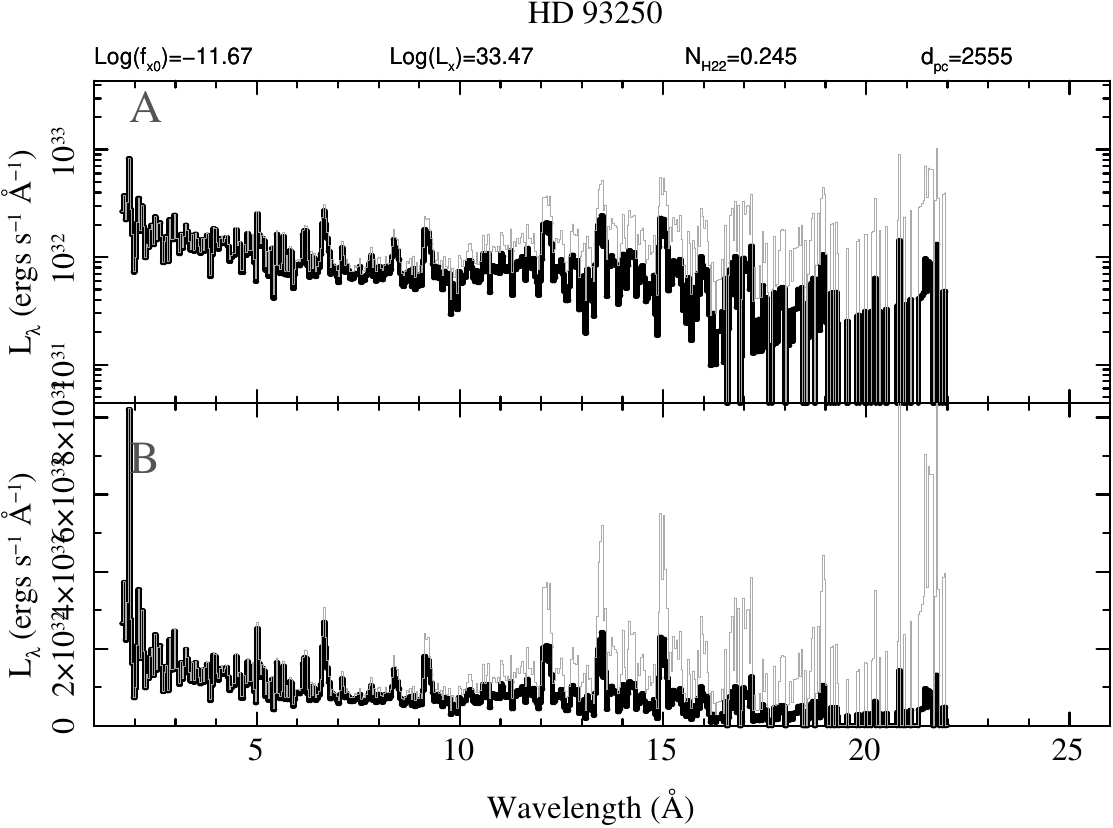}
   \caption{HD~93250}
   \label{fig:lx_HD_93250}
\end{figure}

\begin{figure}[tb]
    \centering
   \includegraphics*[width=0.820\columnwidth, viewport= 0 0 535 377]{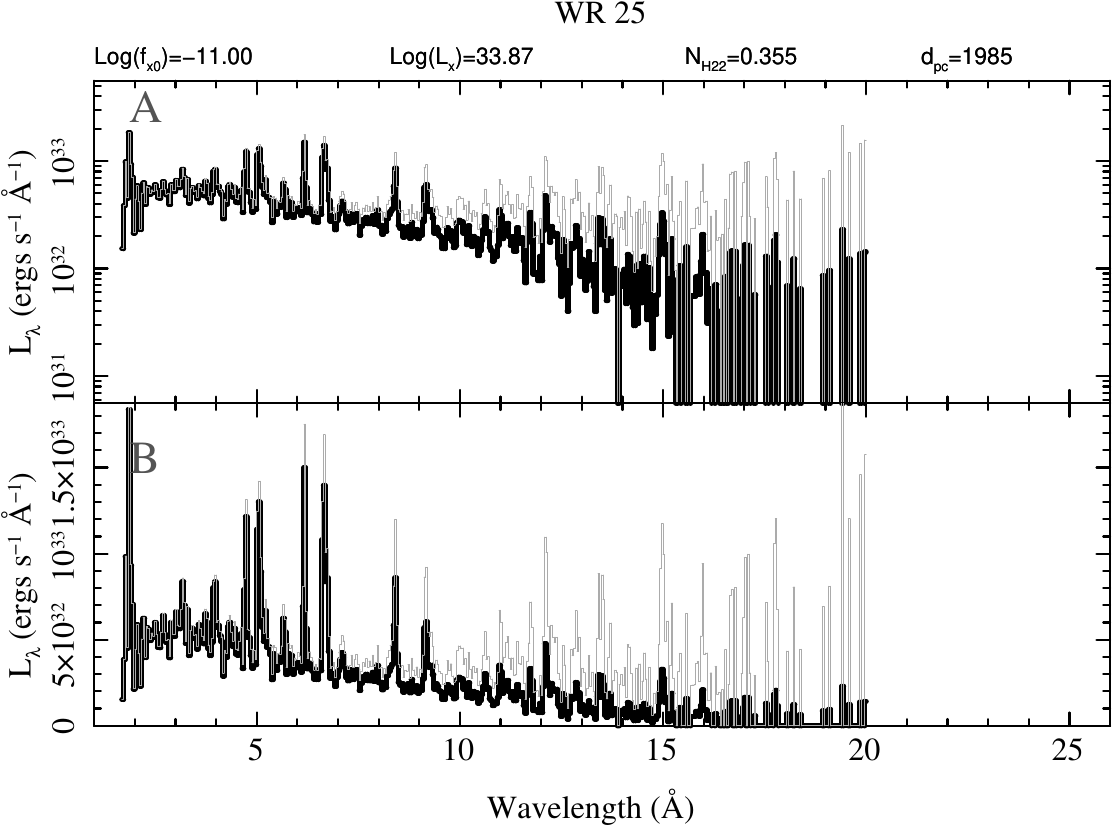}
   \caption{WR~25}
   \label{fig:lx_WR_25}
\end{figure}

\begin{figure}[tb]
    \centering
   \includegraphics*[width=0.820\columnwidth, viewport= 0 0 535 377]{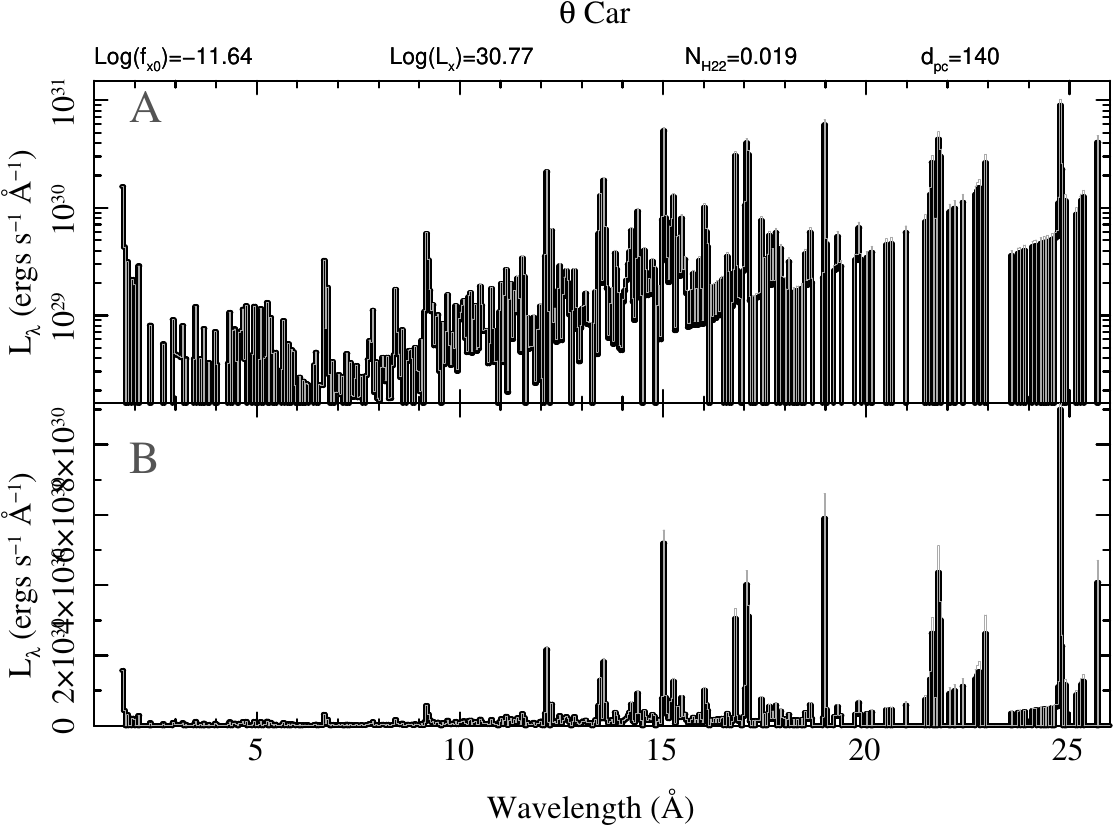}
   \caption{$\theta\,$Car}
   \label{fig:lx_HD_93030}
\end{figure}

\begin{figure}[tb]
    \centering
   \includegraphics*[width=0.820\columnwidth, viewport= 0 0 535 377]{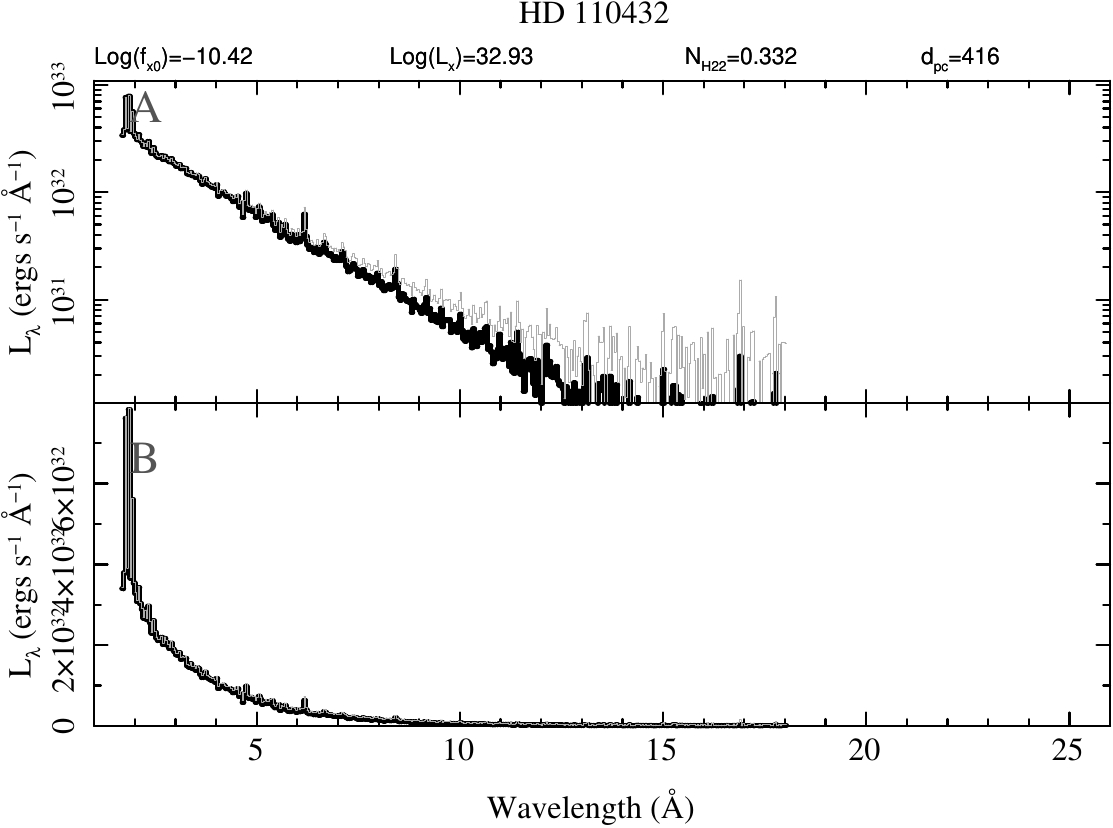}
   \caption{HD 110432}
   \label{fig:lx_HD_110432}
\end{figure}

\begin{figure}[tb]
    \centering
   \includegraphics*[width=0.820\columnwidth, viewport= 0 0 535 377]{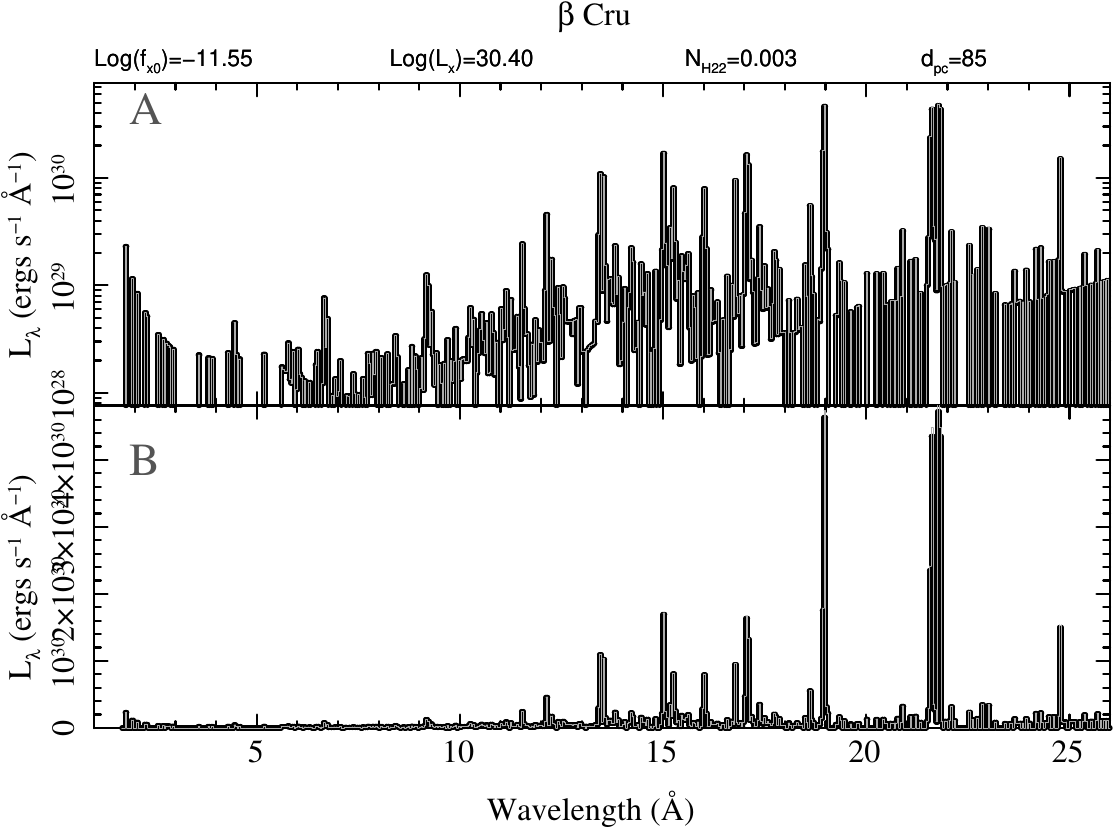}
   \caption{$\beta\,$Cru}
   \label{fig:lx_beta_Cru}
\end{figure}

\begin{figure}[tb]
    \centering
   \includegraphics*[width=0.820\columnwidth, viewport= 0 0 535 377]{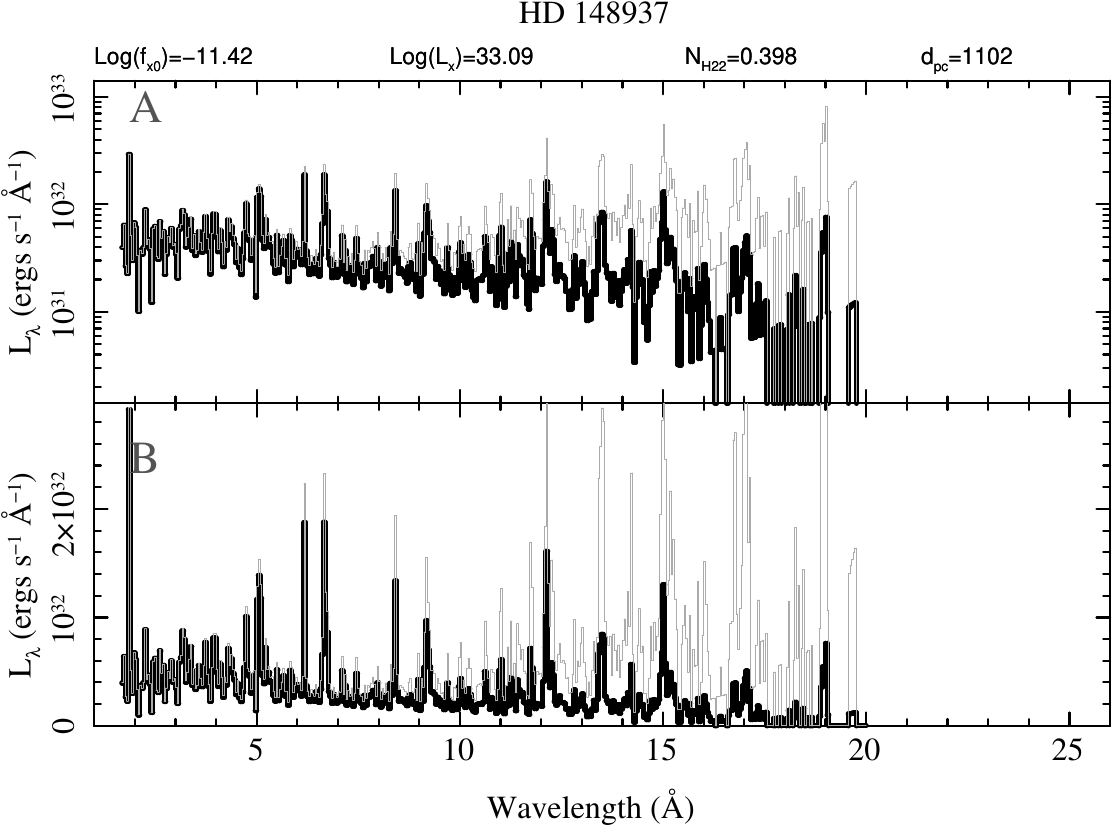}
   \caption{HD~148937}
   \label{fig:lx_HD_148937}
\end{figure}

\begin{figure}[tb]
    \centering
   \includegraphics*[width=0.820\columnwidth, viewport= 0 0 535 377]{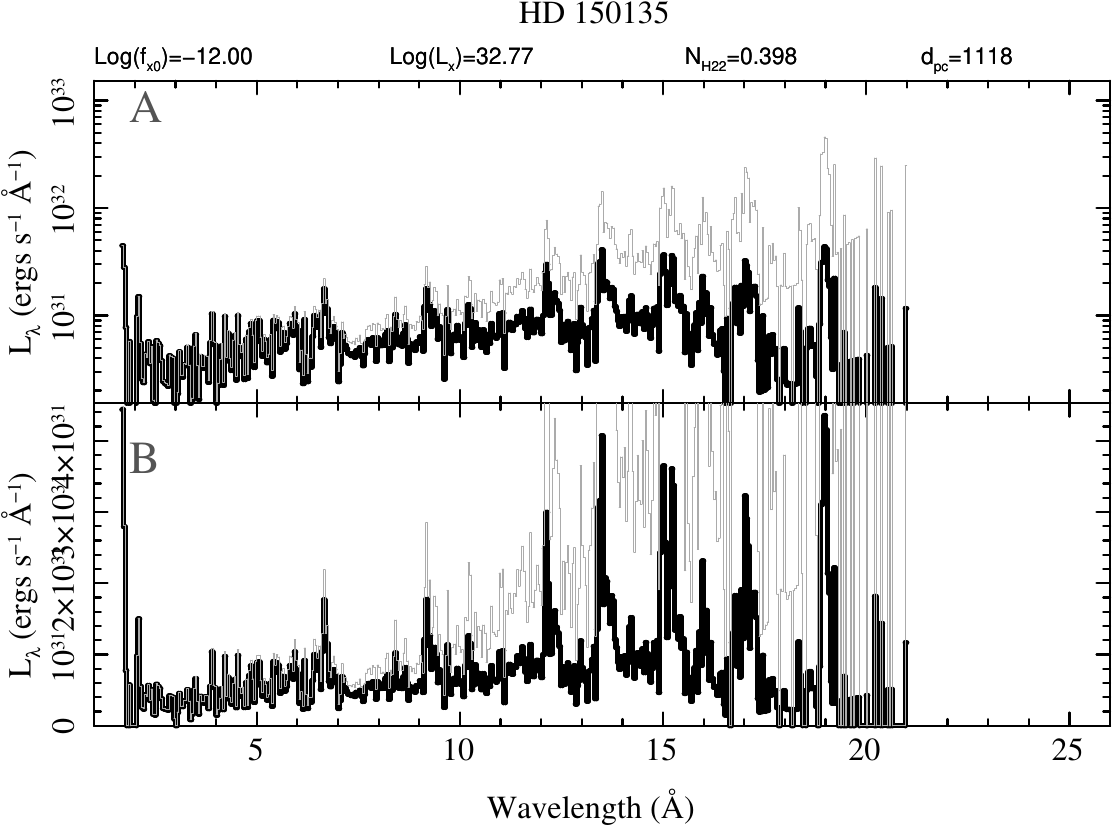}
   \caption{HD~150135}
   \label{fig:lx_HD_150135}
\end{figure}

\begin{figure}[tb]
    \centering
   \includegraphics*[width=0.820\columnwidth, viewport= 0 0 535 377]{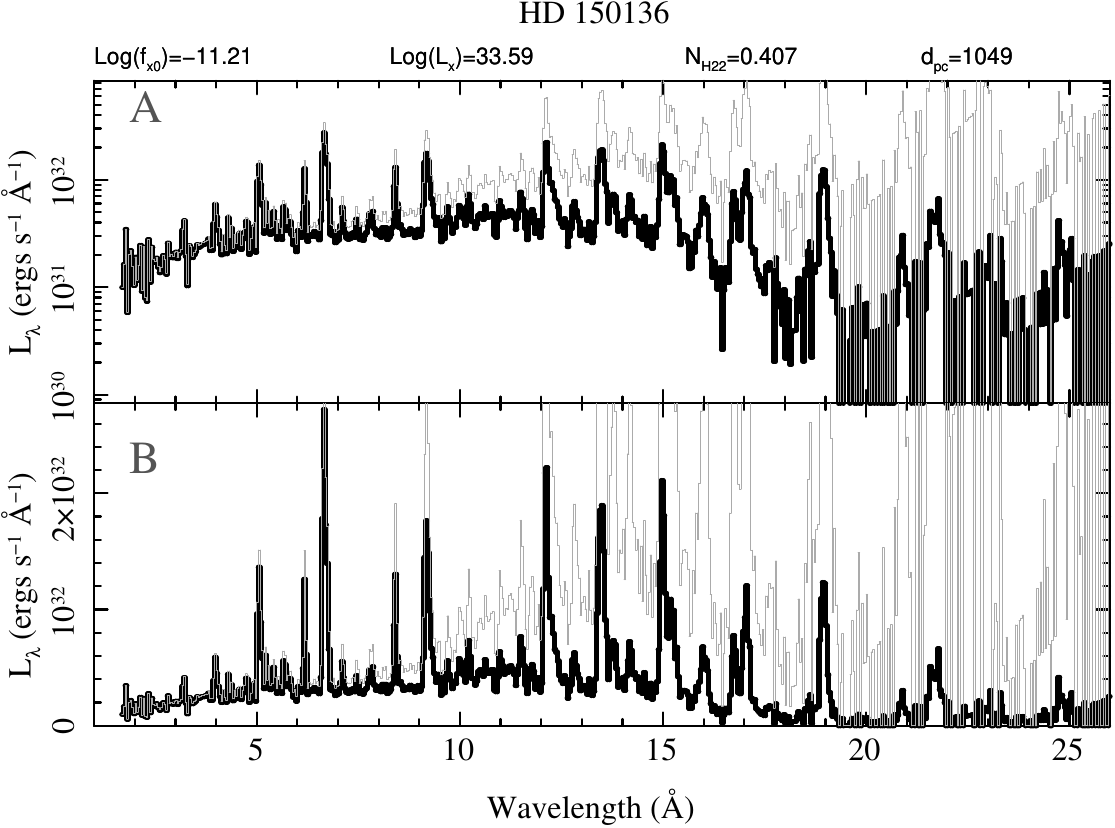}
   \caption{HD~150136}
   \label{fig:lx_HD_150136}
\end{figure}

\begin{figure}[tb]
    \centering
   \includegraphics*[width=0.820\columnwidth, viewport= 0 0 535 377]{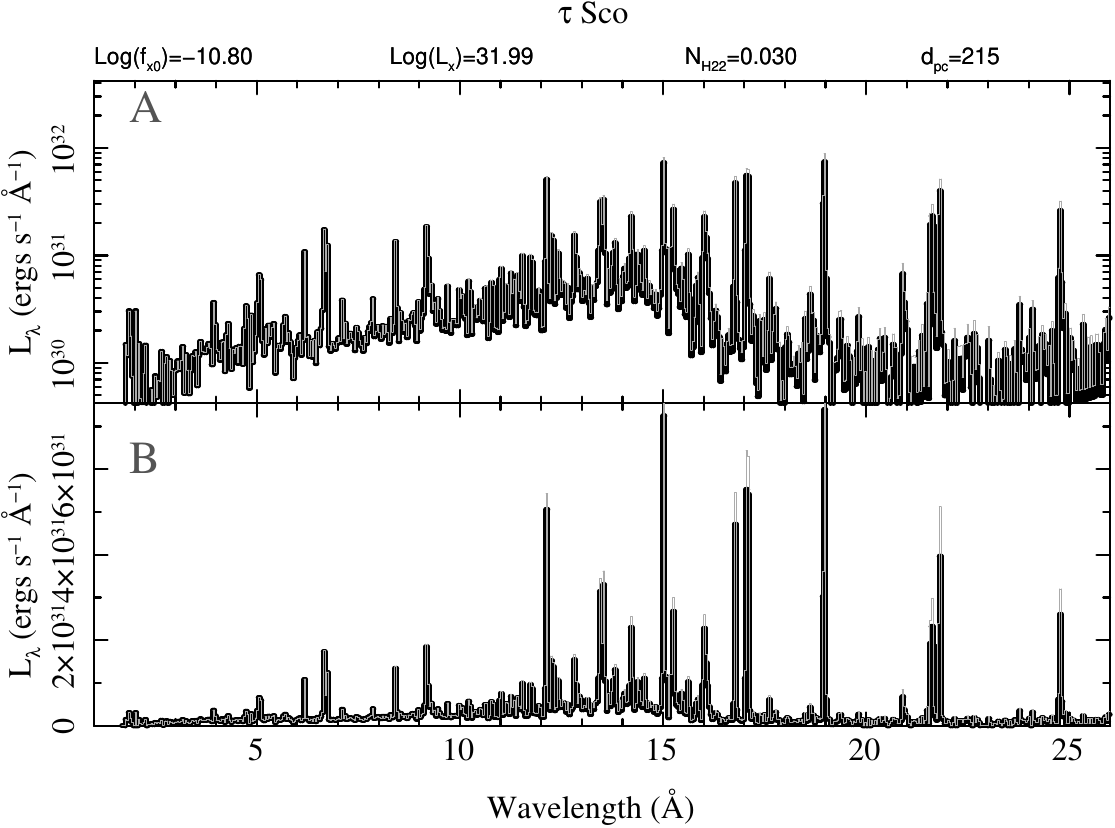}
   \caption{$\tau\,$Sco}
   \label{fig:lx_tau_Sco}
\end{figure}

\clearpage
\bibliographystyle{aasjournal} 
\bibliography{accepted} 


\end{document}